\documentclass[twocolumn]{aastex63}
\usepackage{amsmath,color,textcomp,url,graphicx,subfigure}
\usepackage{listings}
\newcolumntype{L}{>{\raggedright\arraybackslash}p{15.8cm}}

\newcommand{\assofull}{$\mu$~Tau}
\newcommand{\asso}{MUTA}
\newcommand{\assoage}{$61 \pm 5$\,Myr}
\newcommand{\isoage}{$62 \pm 7$\,Myr}
\newcommand{\wdage}{$60_{-6}^{+8}$\,Myr}
\newcommand{\assologage}{$7.79 \pm 0.05$}
\newcommand{\nstarsinitfirst}{37}
\newcommand{\nstarsinit}{33}
\newcommand{\nmodelmembers}{25}
\newcommand{\nwhitedwarfs}{12}
\newcommand{\ngaiasearch}{503}
\newcommand{\ninternalcomovers}{52}
\newcommand{\ninternalsystems}{26}
\newcommand{\nalonecomovers}{2}
\newcommand{\npartialcomovers}{21}
\newcommand{\nvisualcomovers}{12}
\newcommand{\npartialsystems}{15}

\newcommand{\majasso}{$\mu$~TAU}
\newcommand{\gaia}{\emph{Gaia}~DR2}
\newcommand{\gaiagr}{$G - G_{\rm RP}$}
\newcommand{\gaiar}{$G_{\rm RP}$}

\newcommand{\kms}{\hbox{km\,s$^{-1}$}}
\newcommand{\mjup}{$M_{\mathrm{Jup}}$}

\newcommand{\msol}{$M_{\odot}$}
\newcommand{\rsol}{$R_{\odot}$}
\newcommand{\masyr}{$\mathrm{mas}\,\mathrm{yr}^{-1}$}
\newcommand{\teff}{$T_{\rm eff}$}

\submitjournal{ApJ}

\shorttitle{The \asso\ Association: A 60 Myr-old Coeval Group at 150 pc from the Sun}
\shortauthors{Gagn\'e et al.}

\begin{document}

\title{THE \majasso\ ASSOCIATION: A 60 MYR-OLD COEVAL GROUP AT 150 \lowercase{pc} FROM THE SUN}

\author[0000-0002-2592-9612]{Jonathan Gagn\'e}
\affiliation{Plan\'etarium Rio Tinto Alcan, Espace pour la Vie, 4801 av. Pierre-de Coubertin, Montr\'eal, Qu\'ebec, Canada}
\affiliation{Institute for Research on Exoplanets, Universit\'e de Montr\'eal, D\'epartement de Physique, C.P.~6128 Succ. Centre-ville, Montr\'eal, QC H3C~3J7, Canada}
\email{gagne@astro.umontreal.ca}
\author[0000-0001-6534-6246]{Trevor J.\ David}
\affiliation{Center for Computational Astrophysics, Flatiron Institute, New York, NY 10010, USA}
\affiliation{Jet Propulsion Laboratory, California Institute of Technology, 4800 Oak Grove Drive, Pasadena, CA 91109, USA}
\author[0000-0003-2008-1488]{Eric E. Mamajek}
\affiliation{Jet Propulsion Laboratory, California Institute of Technology, 4800 Oak Grove Drive, Pasadena, CA 91109, USA}
\affiliation{Department of Physics \& Astronomy, University of Rochester, Rochester, NY 14627, USA}
\author[0000-0003-3654-1602]{Andrew W. Mann}
\affiliation{Department of Physics and Astronomy, University of North Carolina at Chapel Hill, Chapel Hill, NC 27599-3255, USA}
\author[0000-0001-6251-0573]{Jacqueline K. Faherty}
\affiliation{Department of Astrophysics, American Museum of Natural History, Central Park West at 79th St., New York, NY 10024, USA}
\author[0000-0002-2384-1326]{Antoine B\'edard}
\affiliation{D\'epartement de Physique, Universit\'e de Montr\'eal, C.P.~6128 Succ. Centre-ville, Montr\'eal, QC H3C~3J7, Canada}

\begin{abstract}

We present an analysis of the newly identified \assofull\ Association (MUTA) of young stars at $\simeq$\,150\,pc from the Sun that is part of the large Cas-Tau structure, coeval and co-moving with the $\alpha$~Persei cluster. This association is also located in the vicinity of the Taurus-Auriga star-forming region and the Pleiades association, although it is unrelated to them. We identify more than 500 candidate members of \asso\ using \gaia\ data and the BANYAN $\Sigma$ tool \citep{2018ApJ...856...23G} and we determine an age of \isoage\ for its population based on an empirical comparison of its color-magnitude diagram sequence with those of other nearby young associations. The \asso\ association is related to the Theia~160 group of \cite{2019AJ....158..122K} and corresponds to the e~Tau group of \cite{2020AJ....159..105L}. It is also part of the Cas-Tau group of \cite{1956ApJ...123..408B}. As part of this analysis, we introduce an iterative method based on spectral templates to perform an accurate correction of interstellar extinction of \gaia\ photometry, needed because of its wide photometric bandpasses. We show that the members of \asso\ display an expected increased rate of stellar activity and faster rotation rates compared with older stars, and that literature measurements of the lithium equivalent width of nine G0 to K3-type members are consistent with our age determination. We show that the present-day mass function of \asso\ is consistent with other known nearby young associations. We identify WD~0340+103 as a hot, massive white dwarf remnant of a B2 member that left its planetary nebula phase only 270,000~years ago, posing an independent age constraint of \wdage\ for \asso, consistent with our isochrone age. This relatively large collection of co-moving young stars near the Sun indicates that more work is required to unveil the full kinematic structure of the complex of young stars surrounding $\alpha$~Persei and Cas-Tau.

\end{abstract}


\section{INTRODUCTION}\label{sec:intro}

Young stellar associations in the Solar neighborhood ($\lesssim$\,200\,pc) are valuable laboratories to study stellar evolution and refine our age-dating methods because they contain groups of stars with many different masses that formed coevally from the same molecular cloud (e.g., \citealp{2004ARAA..42..685Z,2008hsf2.book..757T}). Their proximity is valuable because their members appear brighter, but it also causes them to be spread over larger areas of the sky, which makes their initial identification less straightforward. Obtaining credible lists of members with low contamination by unrelated field stars is challenging and typically requires measuring the six-dimensional position and space velocity of each member. As these stars formed from a single molecular cloud, they share the same velocities typically within $\simeq$\,2--4\,\kms, allowing us to distinguish them from most field stars.

Until recently, trigonometric distance measurements were only available for a limited set of bright stars (e.g., \citealt{1997AA...323L..49P}), and radial velocity measurements of stars in the Solar neighborhood were even more limited to small-scale samples (e.g., see \citealp{2006AstL...32..759G,2007AJ....133.2524W}). This led to the identification of co-moving and coeval massive stars that represented only the tip of the iceberg of each young association of stars in our neighborhood \citep{2004ARAA..42..685Z,2008hsf2.book..757T}. Efforts have been made to identify the lower-mass population based on various methods that can assign membership probabilities with missing parts of the 6-dimensional space and velocity, including the convergent point method \citep{2005ApJ...634.1385M,2006AA...460..695T} and various other flavors of selection cuts in space-velocity and/or photometry \citep{2004ARAA..42..685Z,2014AJ....147..146K,2017AJ....153...95R,2017AJ....154...69S} as well as methods based on Bayesian statistics \citep{2013ApJ...762...88M,2014ApJ...783..121G,2018ApJ...856...23G}.

The second data release of the \emph{Gaia} mission (\emph{Gaia}~DR2 hereafter; \added{\citealt{GaiaCollaboration:2018io,Lindegren:2018gy}})\footnote{\added{See also \cite{Luri:2018eu}, \cite{Mignard:2018bj}, \cite{Babusiaux:2018di},\cite{Sartoretti:2018jm}, \cite{Soubiran:2018fz}, \cite{Cropper:2018jx}, \cite{Evans:2018cj}, \cite{Hambly:2018gr}, and \cite{Riello:2018bo} for relevant calibration.}} changed this landscape completely in April of 2018 by providing trigonometric distance measurements for $\simeq$\,1.3 billion stars with an unprecedented precision, as well as radial velocities for more than 7.2 million bright stars. This allowed us to complete the 6-dimensional kinematics for a number of stars on a completely new scale, which led to a plethora of scientific discoveries that quickly unveiled the spatial and kinematic structure of the Solar neighborhood as well as the Milky Way in general. Some of these discoveries include many new associations of stars \citep{2017AJ....153..257O,2018ApJ...863...91F,2018ApJ...865..136G,2019AJ....158..122K,2019AA...622L..13M}, a large number of new M-type members of known associations \citep{2018ApJ...860...43G,2018ApJ...862..138G,2018AJ....156..271L,2018MNRAS.477.3197R,2019ApJ...870...27Z,2019ApJ...877...12T}, as well as the discovery of tidal disruption tails around three older, nearby clusters: the Hyades \citep{2019AA...621L...2R}, Praesepe \citep{2019AA...627A...4R} and Coma~Ber \citep{2019ApJ...877...12T}.

\begin{figure}
 	\centering
 	\includegraphics[width=0.465\textwidth]{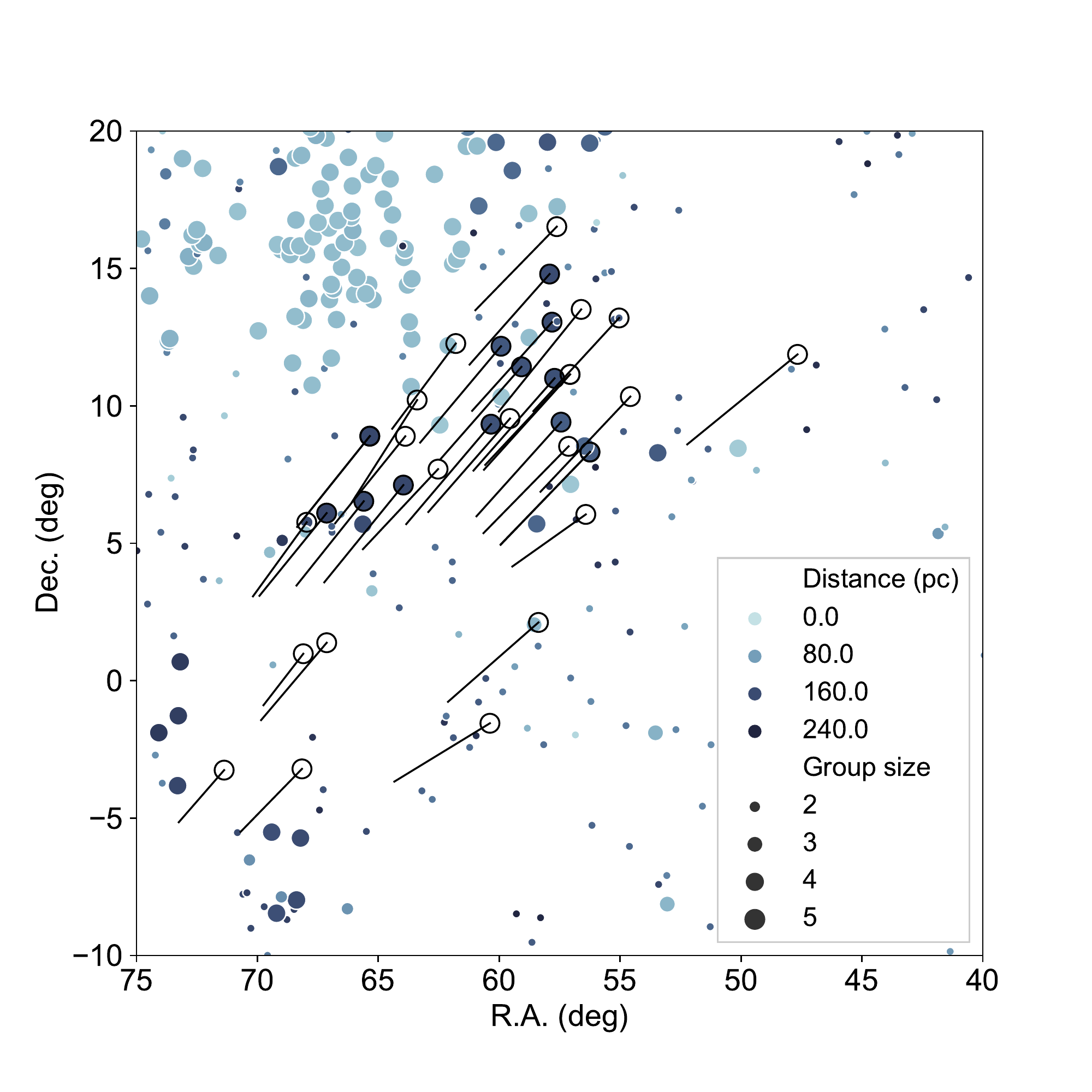}
 	\caption{Sky position and proper motions of \asso\ members (empty circles with proper motion arrows), compared with nearby co-moving systems recovered by \citeauthor{2017AJ....153..257O} (\citeyear{2017AJ....153..257O}; filled blue circles). The larger circles belong to \cite{2017AJ....153..257O} co-moving systems with more members (the maximum symbol size indicates 5 or more members), and the darker-shaded circles correspond to objects further away from the Sun. The tip of the Taurus star-forming region can be seen as large, dark blue circles at R.A. $\simeq$ 55--60\textdegree, Dec. $\simeq$\,20\textdegree, and part of the foreground Hyades cluster can be seen as large, pale blue circles at R.A. $\simeq$ 65--75\textdegree, Dec. $\simeq$\,10--20\textdegree. See Section~\ref{sec:initsample} for more details.}
 	\label{fig:ohfig}
\end{figure}

This paper presents the discovery and characterization the \asso\ association, based on an initial list of massive co-moving and coeval members that had been discovered in historical surveys but never before published. The advent of \gaia\ allowed us to complete this list and characterize \asso\ such that it will become yet another important laboratory for the investigation of stellar evolution and the grounds for discovery of age-calibrated brown dwarfs and exoplanets. In Section~\ref{sec:initsample}, we present the initial list of \asso\ members, which we use to build a spatial-kinematic model (Section~\ref{sec:kinmodel}) to search for additional members with the BANYAN~$\Sigma$ Bayesian identification tool \citep{2018ApJ...856...23G} in Section~\ref{sec:newmembers}. In Section~\ref{sec:ext}, we present an iterative method to correct interstellar extinction in \gaia\ color-magnitude diagrams, required because the photometric bandpasses are wider than usual. We discuss the properties of \asso\ as a whole and its individual members in Section~\ref{sec:discussion}, including their present-day mass function and stellar activity indicators, and a comparison with the Galactic kinematic structures recently unveiled by \cite{2019AJ....158..122K}. We summarize and conclude this work in Section~\ref{sec:conclusion}.

\section{INITIAL SAMPLE OF MEMBERS}\label{sec:initsample}

The existence of a distinct group of co-moving young stars in the vicinity of the Taurus-Auriga (\citealp{2008hsf1.book..405K}) star-forming region first appeared in a spatial distribution of Cas-Tau OB-type stars assembled by \cite{1956ApJ...123..408B}. Cas-Tau was identified by \cite{1956ApJ...123..408B} as an extended group of co-moving stars with an expansion age of $\simeq$\,50\,Myr that seems to be on the way to being dissolved. They noted that Cas-Tau may share a common origin with an extended stream of stars around the $\alpha$~Persei cluster (e.g., see \citealp{1958ZA.....45..243H,2019AA...628A..66L}) identified by \cite{1921MeLuS..26....3R}. An over-density in the Cas-Tau stars seemed to be located at Galactic coordinates $\left(\ell,b\right) = \left(190^{\circ},-10^{\circ}\right)$, and was recovered as part of the \cite{1999AJ....117..354D} census of nearby OB associations (see their Fig. 19). This over-density overlaps with subgroup~5 of Cas-Tau defined by \cite{1956ApJ...123..408B}, with five B-type stars in common (29~Tau, 30~Tau, 35~Eri, $\mu$~Tau, $\mu$~Eri) and one additional star (40~Tau) not in common that seems to be an un related background star. Combining this list of 12 early-type stars assembled by \cite{1999AJ....117..354D} to other co-moving B, A and F-type stars in the range $\ell$ from 170\textdegree\ to 205\textdegree\ and $b$ from $-$40\textdegree\ to $-$27\textdegree\ as well as nearby ROSAT entries \citep{2016AA...588A.103B} in the same region yielded a total set of 35 stars that appeared to be young and co-moving within 15\,\masyr\ of the average proper motions of the \cite{1999AJ....117..354D} list ($\mu_\alpha\cos\delta$ = 21.0\,\masyr, $\mu_\delta$ = $-$20.5\,\masyr). Four of these \nstarsinitfirst\ stars are clear outliers either in $XYZ$ (HD~23110, TYC~657--794--2, and HD~28796) or $UVW$ (HIP~18778) and were excluded from our initial list. The resulting \nstarsinit\ stars are listed in Table~\ref{tab:initialmembers} with their properties. We tentatively named this group \assofull\ Association (\asso) after one of its brightest members. We assigned initial members with MUTA identification numbers (from 1 to 30) in order of decreasing $V$-band brightness. We assigned the same MUTA ID to binaries with separations below 15$^{\prime\prime}$.

In a more recent analysis or the \emph{Gaia}~Data Release 1 (DR1), \cite{2017AJ....153..257O} recovered about a third of the stars in Table~\ref{tab:initialmembers} as three broken up groups of co-moving systems, which they named Groups 43 (6 matches), 52 (3 matches) and 60 (4 matches). The overlap between our initial list of \asso\ members and the \cite{2017AJ....153..257O} sample are shown in Figure~\ref{fig:ohfig}, where part of the Taurus star-forming region can be seen at a similar distance from the Sun (see e.g. \citealt{2000AA...359..181W}), and the Hyades cluster \citep{1998AA...331...81P} also appears in the foreground. The method that \cite{2017AJ....153..257O} used to identify systems of co-moving stars works directly in proper motion and parallax space, which tends to recover spatially large moving groups only as broken parts, explaining why the spatially extended \asso\ was broken up in three groups, similarly to other nearby young moving groups \citep{2018ApJ...863...91F}.

\begin{figure}
 	\centering
 	\includegraphics[width=0.465\textwidth]{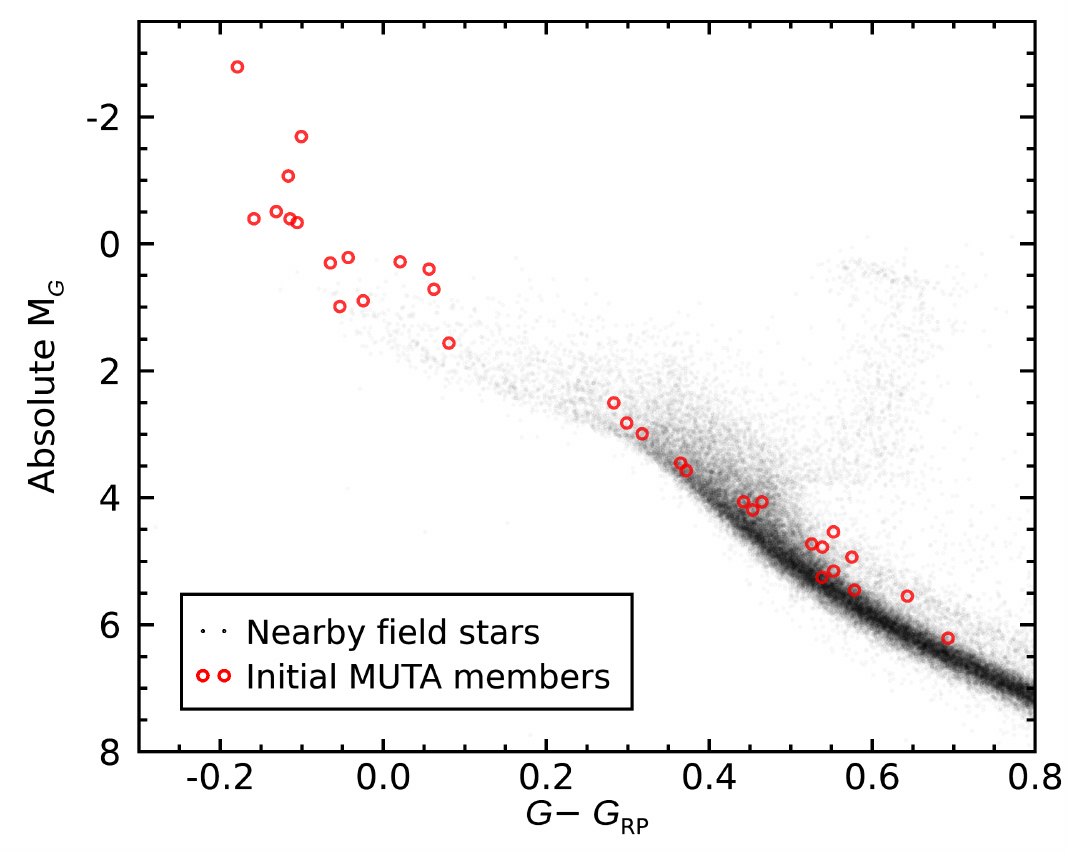}
 	\caption{\gaia\ Color-magnitude diagram of our initial list of \asso\ members (red circles), compared with field stars within 100\,pc of the Sun (black dots). This list of \asso\ members contains several OBA-type stars ($G - G_{\rm RP} < 0.2$) indicative of its young age, as well as later-type stars ($G - G_{\rm RP} > 0.2$) that constitute a narrow sequence. The \gaia\ photometry was not corrected for interstellar extinction. 2MASS~J04212444+0853488 is outside the range of this figure at $G - G_{\rm RP} = 1.12$. See Section~\ref{sec:initsample} for more details.}
 	\label{fig:initcmd}
\end{figure}

We cross-matched our initial list of \asso\ members with \gaia\ data to build a color-magnitude sequence shown in Figure~\ref{fig:initcmd} to demonstrate they constitute massive OBA-type stars ($G - G_{\rm RP} < 0.2$) and a well-defined sequence of later-type stars ($G - G_{\rm RP} > 0.2$), providing further evidence that they are coeval and young.

\startlongtable
\tablewidth{0.985\textwidth}
\begin{deluxetable*}{lllccccl}
\clearpage
\pagebreak
\tablecolumns{8}
\tablecaption{Initial members of \asso.\label{tab:initialmembers}}
\tablehead{\colhead{\asso} & \colhead{} & \colhead{Spectral} & \colhead{R.A.} & \colhead{Decl.} & \colhead{Distance\tablenotemark{a}} & \colhead{\gaia} & \colhead{}\\
\colhead{ID} & \colhead{Name} & \colhead{Type} & \colhead{(hh:mm:ss.sss)} & \colhead{(dd:mm:ss.ss)} & \colhead{(pc)} & \colhead{$G$ mag} & \colhead{Ref.\tablenotemark{b}} }
\startdata
1 & $\mu$~Eri & B3+A3 & 04:45:30.167 & --03:15:16.97 & $160 \pm 5$ & $3.931 \pm 0.004$ & 16\\
2 & $\mu$~Tau & B3IV & 04:15:32.079 & +08:53:32.14 & $149 \pm 7$ & $4.183 \pm 0.003$ & 5\\
3~A & 30~Tau & B3V & 03:48:16.292 & +11:08:35.52 & $129 \pm 3$ & $5.040 \pm 0.002$ & 5\\
3~B & TYC~661--1404--1 & F5+F5 & 03:48:16.835 & +11:08:40.16 & $138 \pm 1$ & $9.2693 \pm 0.0002$ & 2\\
4 & 35~Eri & B5V & 04:01:32.077 & --01:32:59.02 & $133 \pm 3$ & $5.230 \pm 0.002$ & 5\\
5 & 29~Tau & B3+A7 & 03:45:40.466 & +06:02:59.78 & $187 \pm 9$ & $5.295 \pm 0.001$ & 4\\
6 & HD~28375 & B5V & 04:28:32.142 & +01:22:50.65 & $146 \pm 4$ & $5.491 \pm 0.001$ & 14\\
7 & HD~28843 & B9III & 04:32:37.573 & --03:12:34.60 & $169 \pm 3$ & $5.740 \pm 0.002$ & 15\\
8 & HD~19698 & B8V & 03:10:38.828 & +11:52:21.07 & $134 \pm 2$ & $5.9439 \pm 0.0008$ & 1\\
9 & HR~1307 & B8V & 04:13:34.588 & +10:12:44.52 & $144 \pm 3$ & $6.1900 \pm 0.0006$ & 13\\
10 & V766~Tau & B9 & 03:51:15.896 & +13:02:45.52 & $161 \pm 2$ & $6.247 \pm 0.001$ & 9\\
11 & HD~28715 & B9 & 04:31:50.463 & +05:45:51.74 & $187 \pm 4$ & $6.6396 \pm 0.0004$ & 3\\
12 & HD~24456 & B9.5V & 03:53:30.257 & +02:07:08.57 & $138.7 \pm 0.9$ & $6.6983 \pm 0.0004$ & 10\\
13 & HD~23990 & B9.5V & 03:49:46.521 & +09:24:26.60 & $147 \pm 1$ & $6.7410 \pm 0.0004$ & 6\\
14 & HD~23538 & A0 & 03:46:26.278 & +13:30:32.46 & $168 \pm 2$ & $6.8479 \pm 0.0003$ & 3\\
15 & HD~25978 & B9V & 04:07:11.204 & +12:16:05.10 & $166 \pm 2$ & $7.6661 \pm 0.0003$ & 12\\
16 & HD~26323 & A2V & 04:10:06.873 & +07:41:52.12 & $161 \pm 2$ & $8.5401 \pm 0.0005$ & 10\\
17 & HD~27687 & A3 & 04:22:24.213 & +06:31:45.14 & $165 \pm 1$ & $8.9125 \pm 0.0004$ & 3\\
18 & HD~28356 & A3 & 04:28:32.733 & +06:05:52.07 & $157 \pm 2$ & $8.9675 \pm 0.0004$ & 3\\
19~A & HD~23376 & G5 & 03:44:58.957 & +08:19:10.09 & $145 \pm 1$ & $9.2549 \pm 0.0003$ & 3\\
19~B & TYC~658--1007--2 & $\cdots$ & 03:44:59.048 & +08:19:13.81 & $142 \pm 1$ & $10.493 \pm 0.002$ & --\\
20 & HIP~17133 & A0 & 03:40:09.988 & +13:11:55.07 & $150 \pm 1$ & $9.949 \pm 0.001$ & 3\\
21 & HD~286374 & F5 & 03:56:19.224 & +11:25:10.84 & $152 \pm 2$ & $9.9776 \pm 0.0005$ & 11\\
22 & PPM~119410 & F8 & 03:50:50.558 & +11:00:05.12 & $151 \pm 1$ & $10.0929 \pm 0.0006$ & 8\\
23 & {[LH98]}~108 & G5IV & 03:50:28.436 & +16:31:14.80 & $146 \pm 1$ & $10.364 \pm 0.001$ & 7\\
24 & RX~J0348.5+0832 & G7 & 03:48:31.461 & +08:31:36.43 & $152 \pm 2$ & $10.841 \pm 0.002$ & 2\\
25 & TYC~80--202--1 & $\cdots$ & 04:15:51.119 & +07:07:03.76 & $167 \pm 1$ & $10.8894 \pm 0.0006$ & --\\
26 & TYC~662--217--1 & $\cdots$ & 03:59:42.158 & +12:10:08.14 & $148 \pm 1$ & $11.111 \pm 0.002$ & --\\
27 & RX~J0338.3+1020 & G9 & 03:38:18.266 & +10:20:16.32 & $146 \pm 1$ & $10.976 \pm 0.001$ & 2\\
28 & TYC~664--136--1 & $\cdots$ & 03:51:39.673 & +14:47:47.84 & $160 \pm 1$ & $11.566 \pm 0.002$ & --\\
29 & RX~J0358.2+0932 & K3 & 03:58:12.749 & +09:32:21.97 & $146.8 \pm 0.9$ & $12.045 \pm 0.001$ & 2\\
30~A & TYC~668--737--1 & $\cdots$ & 04:21:24.386 & +08:53:54.34 & $151 \pm 1$ & $11.356 \pm 0.002$ & --\\
30~B & 2MASS~J04212444+0853488 & $\cdots$ & 04:21:24.473 & +08:53:48.52 & $151 \pm 1$ & $14.7603 \pm 0.0007$ & --
\enddata
\tablenotetext{a}{\gaia\ distances assuming a 0.029\,mas zero point \citep{Lindegren:2018gy}.}
\tablenotetext{b}{References for spectral types.}
\tablecomments{See section~\ref{sec:initsample} for more details.}
\tablerefs{(1)~\citealt{1969AJ.....74..375C}; (2)~\citealt{1997AAS..124..449M}; (3)~\citealt{1993yCat.3135....0C}; (4)~\citealt{1980ApJS...44..489B}; (5)~\citealt{1968ApJS...17..371L}; (6)~\citealt{2008ApJS..176..216A}; (7)~\citealt{2007AJ....133.2524W}; (8)~\citealt{2003AJ....125..359W}; (9)~\citealt{1968PASP...80..453C}; (10)~\citealt{1999AAS..137..451G}; (11)~\citealt{1995AAS..110..367N}; (12)~\citealt{1988PASP..100..828B}; (13)~\citealt{1972AJ.....77..750C}; (14)~\citealt{1972ApJ...175..453M}; (15)~\citealt{1980AAS...42..115J}; (16)~\citealt{2007AA...474..653V}.}
\end{deluxetable*}
\clearpage
\pagebreak
\startlongtable
\tablewidth{0.985\textwidth}
\begin{deluxetable*}{llccccl}
\tablecolumns{7}
\tablecaption{Core members of \asso\ used in the construction of a kinematic model.\label{tab:coremembers}}
\tablehead{\colhead{\asso} & \colhead{} & \colhead{$\mu_\alpha\cos\delta$} & \colhead{$\mu_\delta$} & \colhead{Parallax} & \colhead{RV} & \colhead{RV}\\
\colhead{ID} & \colhead{Name} & \colhead{(\masyr)} & \colhead{(\masyr)} & \colhead{(mas)} & \colhead{(\kms)} & \colhead{Ref.} }
\startdata
1 & $\mu$~Eri & $13.51 \pm 0.75$ & $-13.66 \pm 0.64$ & $6.3 \pm 0.2$ & $23 \pm 4$ & 1\\
2 & $\mu$~Tau & $20.88 \pm 0.62$ & $-22.79 \pm 0.52$ & $6.7 \pm 0.3$ & $16.3 \pm 0.6$ & 1\\
3~A & 30~Tau & $25.27 \pm 0.28$ & $-23.69 \pm 0.23$ & $7.7 \pm 0.2$ & $16.2 \pm 0.1$ & 1\\
4 & 35~Eri & $28.45 \pm 0.30$ & $-15.28 \pm 0.25$ & $7.5 \pm 0.2$ & $15.7 \pm 0.8$ & 1\\
5 & 29~Tau & $21.88 \pm 0.29$ & $-13.65 \pm 0.26$ & $5.3 \pm 0.2$ & $17 \pm 2$ & 3\\
6 & HD~28375 & $19.53 \pm 0.33$ & $-20.27 \pm 0.18$ & $6.8 \pm 0.2$ & $18 \pm 4$ & 1\\
7 & HD~28843 & $18.28 \pm 0.19$ & $-16.50 \pm 0.13$ & $5.9 \pm 0.1$ & $18 \pm 7$ & 6\\
8 & HD~19698 & $32.84 \pm 0.16$ & $-23.58 \pm 0.17$ & $7.4 \pm 0.1$ & $1 \pm 4$ & 1\\
9 & HR~1307 & $19.37 \pm 0.39$ & $-26.69 \pm 0.23$ & $6.9 \pm 0.1$ & $10 \pm 7$ & 6\\
10 & V766~Tau & $23.77 \pm 0.11$ & $-23.228 \pm 0.079$ & $6.19 \pm 0.06$ & $16 \pm 2$ & 5\\
12 & HD~24456 & $26.93 \pm 0.10$ & $-20.785 \pm 0.074$ & $7.18 \pm 0.05$ & $18 \pm 3$ & 1\\
15 & HD~25978 & $18.91 \pm 0.16$ & $-22.323 \pm 0.076$ & $5.99 \pm 0.07$ & $22 \pm 7$ & 6\\
16 & HD~26323 & $22.38 \pm 0.12$ & $-20.975 \pm 0.071$ & $6.18 \pm 0.06$ & $14 \pm 3$ & 1\\
18 & HD~28356 & $20.00 \pm 0.15$ & $-21.659 \pm 0.072$ & $6.36 \pm 0.07$ & $20.6 \pm 0.6$ & 2\\
19~A & HD~23376 & $26.61 \pm 0.11$ & $-24.306 \pm 0.066$ & $6.89 \pm 0.06$ & $16.5 \pm 0.5$ & 2\\
20 & HIP~17133 & $25.53 \pm 0.10$ & $-24.403 \pm 0.073$ & $6.63 \pm 0.05$ & $14 \pm 6$ & 2\\
21 & HD~286374 & $24.05 \pm 0.11$ & $-24.124 \pm 0.067$ & $6.54 \pm 0.07$ & $14 \pm 2$ & 2\\
22 & PPM~119410 & $24.14 \pm 0.10$ & $-24.167 \pm 0.068$ & $6.58 \pm 0.05$ & $15.0 \pm 0.6$ & 2\\
23 & {[LH98]}~108 & $24.24 \pm 0.14$ & $-21.892 \pm 0.072$ & $6.80 \pm 0.05$ & $8.0 \pm 0.7$ & 4\\
24 & RX~J0348.5+0832 & $25.33 \pm 0.11$ & $-22.738 \pm 0.070$ & $6.56 \pm 0.08$ & $10 \pm 10$ & 2\\
25 & TYC~80--202--1 & $23.547 \pm 0.086$ & $-25.480 \pm 0.054$ & $5.96 \pm 0.05$ & $20.7 \pm 0.6$ & 2\\
26 & TYC~662--217--1 & $24.07 \pm 0.11$ & $-25.242 \pm 0.063$ & $6.71 \pm 0.05$ & $15.3 \pm 0.6$ & 2\\
27 & RX~J0338.3+1020 & $26.75 \pm 0.10$ & $-24.923 \pm 0.070$ & $6.82 \pm 0.06$ & $15 \pm 1$ & 2\\
29 & RX~J0358.2+0932 & $24.321 \pm 0.071$ & $-24.493 \pm 0.051$ & $6.78 \pm 0.04$ & $16 \pm 2$ & 2\\
30~A & TYC~668--737--1 & $21.501 \pm 0.085$ & $-23.632 \pm 0.056$ & $6.57 \pm 0.05$ & $20 \pm 7$ & 2\\
\enddata
\tablecomments{All proper motion and parallax measurements are from \gaia, except for the parallax of $\mu$~Eri, which is from Hipparcos \citep{2007AA...474..653V}. See section~\ref{sec:kinmodel} for more details.}
\tablerefs{(1)~\citealt{2006AstL...32..759G}; (2)~\citealt{GaiaCollaboration:2018io}; (3)~\citealt{1967IAUS...30...57E}; (4)~\citealt{2007AJ....133.2524W}; (5)~\citealt{1953GCRV..C......0W}; (6)~\citealt{2007AN....328..889K}.}
\end{deluxetable*}
\clearpage
\pagebreak

The earliest-type member in our initial list is 29~Tau (MUTA~5), a B3\,V-type star \citep{1980ApJS...44..489B}, which corresponds to a mass of $\simeq$\,5.4\,\msol\ \citep{2013ApJS..208....9P}. \cite{2010AN....331..349H} and \cite{2016AJ....152...40G} estimated the mass of 29~Tau based on evolutionary tracks and found respective values of $6.0 \pm 0.7$\,\msol\ and $5.4 \pm 0.6$\,\msol, consistent with the expected mass for a B3 star. Following the evolutionary tracks of \cite{2016ApJ...823..102C}, such a star has a main-sequence life of only $\simeq$\,80\,Myr, indicating that the \asso\ association is likely younger than the Pleiades.

We note that both $\mu$~Tau (MUTA~2) and $\tau^1$~Ari are known eclipsing binaries \citep{2013AN....334..860A}. While the first is part of our initial list of members, $\tau^1$~Ari was identified in an earlier parsing of \cite{1999AJ....117..354D} but was not included because of its discrepant $UVW$ motion (it is separated from the other stars by $\simeq$6.3\,\kms). A further analysis of their respective light curves might be useful for constraining models of stellar structure at young ages.

\section{A KINEMATIC MODEL OF MUTA MEMBERS}\label{sec:kinmodel}

The BANYAN~$\Sigma$ tool \citep{2018ApJ...856...23G} makes it possible to identify additional stars with similar Galactic positions $XYZ$ and space velocities $UVW$ compared to our initial list of \asso\ members, if we provide it a 6-dimensional multivariate Gaussian model for \asso\ in $XYZUVW$ space. One of the main benefits of BANYAN~$\Sigma$ is its ability to recover stars with only partial kinematics, often a consequence of missing radial velocity or parallax measurements. The BANYAN~$\Sigma$ tool currently includes kinematic models for 29 nearby young associations, which consist of the 27 associations described in \cite{2018ApJ...856...23G}, as well as the recently discovered Volans-Carina \citep{2018ApJ...865..136G} and the Argus associations \citep{2000MNRAS.317..289M}, whose census of members was recently revised by \cite{2019ApJ...870...27Z}.

We compiled literature radial velocity measurements for the stars listed in Table~\ref{tab:initialmembers} to identify a set of \nmodelmembers\ core members with complete kinematics (see Table~\ref{tab:coremembers}). This list excludes any gravitationally bound companion to avoid artificially giving each system more weight in the kinematic construction of the \asso\ model (consistent with the model construction method of \citealt{2018ApJ...856...23G}). HD~28715, HD~23990, HD~23538, and HD~27687 (MUTA~11, 13, 14, and 17, respectively) currently do not have radial velocity measurements and were not included in Table~\ref{tab:coremembers} although they are likely part of \asso\ based on their position in a color-magnitude diagram (see Figure~\ref{fig:initcmd}) and their common proper motion and parallax compared to the other members.

\begin{figure*}
 	\centering
 	\includegraphics[width=0.98\textwidth]{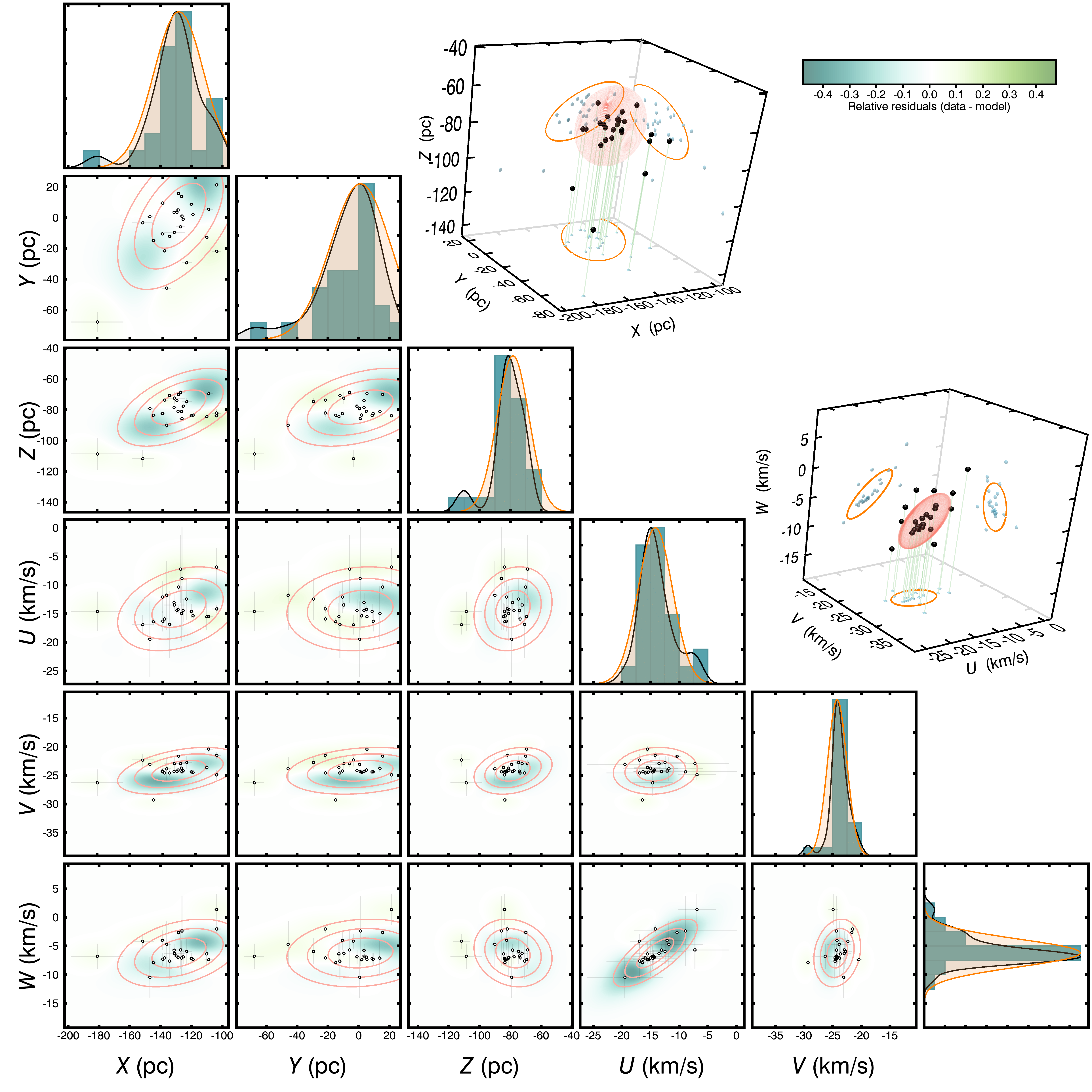}
 	\caption{Multivariate Gaussian model of \asso\ built for BANYAN~$\Sigma$. Orange lines show the 1, 2 and 3$\sigma$ projected contours of the modeled members distribution, and black points represent individual members. Blue and green shadings represent regions of over (green) and under (blue) density of actual members compared to the model, and therefore correspond to departures from a multivariate Gaussian distribution. One-dimensional distributions are displayed as green bars, and are compared with a kernel density estimate distribution of the members (black line) and the projected model (orange lines). A single 1$\sigma$ contour (orange surfaces) and individual members (black spheres) with their projections on the three axis planes are shown for the 3D model projections (upper right). See Section~\ref{sec:kinmodel} for more details.}
 	\label{fig:kinmodel}
\end{figure*}

The methodology described in \citeauthor{2018ApJ...856...23G} (\citeyear{2018ApJ...856...23G}; see their Section~5) was used to build a $XYZUVW$ multivariate Gaussian model of the stars listed in Table~\ref{tab:coremembers}. In summary, a 6-dimensional average vector and covariance matrix in $XYZUVW$ space were built by calculating the average, variance and covariances of the \nmodelmembers\ core members with full kinematics listed in Table~\ref{tab:coremembers}. When calculating the averages, variances and covariances, the individual measurements were weighted proportionally to the squared inverse of their individual error bars to minimize the impact of low quality measurements. The covariance matrix is then regularized to ensure its determinant is finite and positive with a singular value decomposition step. The resulting model is shown in Figure~\ref{fig:kinmodel}.

The multivariate Gaussian model in $XYZUVW$ space that was found to best represent \asso\ has the following central position $\bar x_0$ and covariance matrix $\bar{\bar\Sigma}$:
\begin{align}
	\bar x_0 &= \begin{bmatrix}
	-130.7 & 0.2 & -79.7 & -14.15 & -24.20 & -6.21
	\end{bmatrix},\notag
\end{align}
\begin{align}
	\bar{\bar\Sigma} &= \begin{bmatrix}
		478 & 286 & 196 & 16 & 11.9 & 15\\
		286 & 432 & 136 & 6.7 & 7.6 & 6.0\\
		196 & 136 & 155 & 5.2 & 4.7 & -3.9\\
		16 & 6.7 & 5.2 & 9.1 & 0.46 & 5.5\\
		12 & 7.6 & 4.7 & 0.46 & 2.8 & 0.76\\
		15 & 6.0 & -3.9 & 5.5 & 0.76 & 5.9\\
	\end{bmatrix}.\notag
\end{align}
\noindent both in units of pc and \kms.

The average sky position of \asso\ members is 04:01:29.54, $+$07:59:33.3 ($60.3731$\textdegree, $7.9926$\textdegree) with a standard deviation of 5\textdegree\ in both directions. The average galactic coordinates $\left(\ell,b\right)$ are ($182.4658$\textdegree, $-31.8645$\textdegree) with a standard deviation of ($9$\textdegree, $3$\textdegree). The total velocity $S_{\rm tot}$ of the members averages 28.3\,\kms\ with a standard deviation of 2.3\,\kms. The $UVW$ values we find correspond to a convergent point of $103.380$\textdegree, $-29.325$\textdegree\ in right ascension and declination (06:53:31, $-$29:19:30).

\section{A SEARCH FOR ADDITIONAL MEMBERS}\label{sec:newmembers}

The kinematic model described in Section~\ref{sec:kinmodel} was combined with the BANYAN~$\Sigma$ tool to identify candidate members of \asso\ in \gaia\ data. We pre-selected only \gaia\ entries with right ascensions in the range 10--150\textdegree, declinations in the range $-$20 to $+$40\textdegree\ and trigonometric distances within 300\,pc of the Sun. These limits are significantly wider than the ranges of sky positions (47\textdegree\ to 72\textdegree\ and $-$3.5\textdegree\ to 16.5\textdegree, respectively) and distances (all in the range 130--220\,pc) of the initial list of members. The sky positions, proper motions and parallaxes from \gaia\ were used to determine a membership probability, as well as the \gaia\ radial velocities when available. We selected only the stars with Bayesian membership probabilities above 90\% and a maximum likelihood separation of less than 5\,\kms\ from the core of our \asso\ kinematic model in $UVW$ space as new candidate members. The latter criterion avoids selecting stars that would fit all BANYAN~$\Sigma$ models poorly, including its model of the local Galactic neighborhood.

\begin{figure}
 	\centering
 	\includegraphics[width=0.465\textwidth]{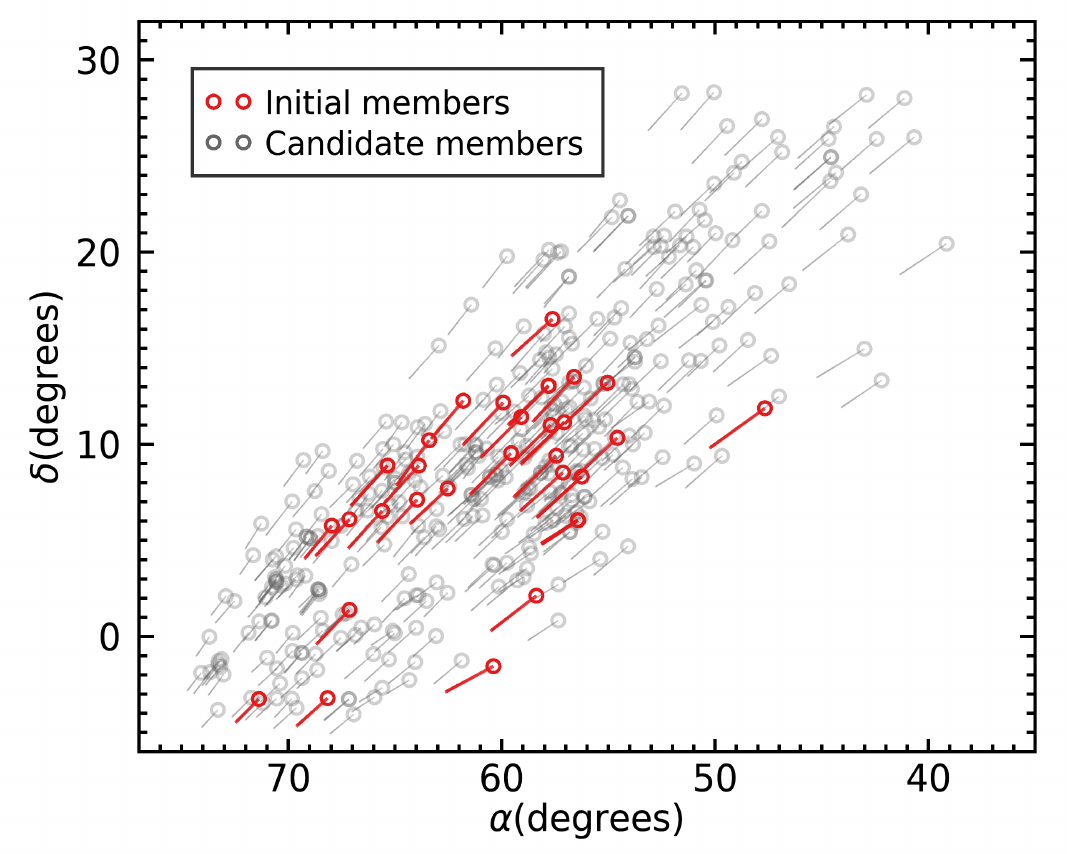}
 	\caption{Sky positions and proper motion vectors for initial members of \asso\ (red circles and lines) and additional candidate members recovered in this work (gray circles and lines). See Section~\ref{sec:newmembers} for more details.}
 	\label{fig:cand_pms}
\end{figure}

\begin{figure}
 	\centering
 	\includegraphics[width=0.465\textwidth]{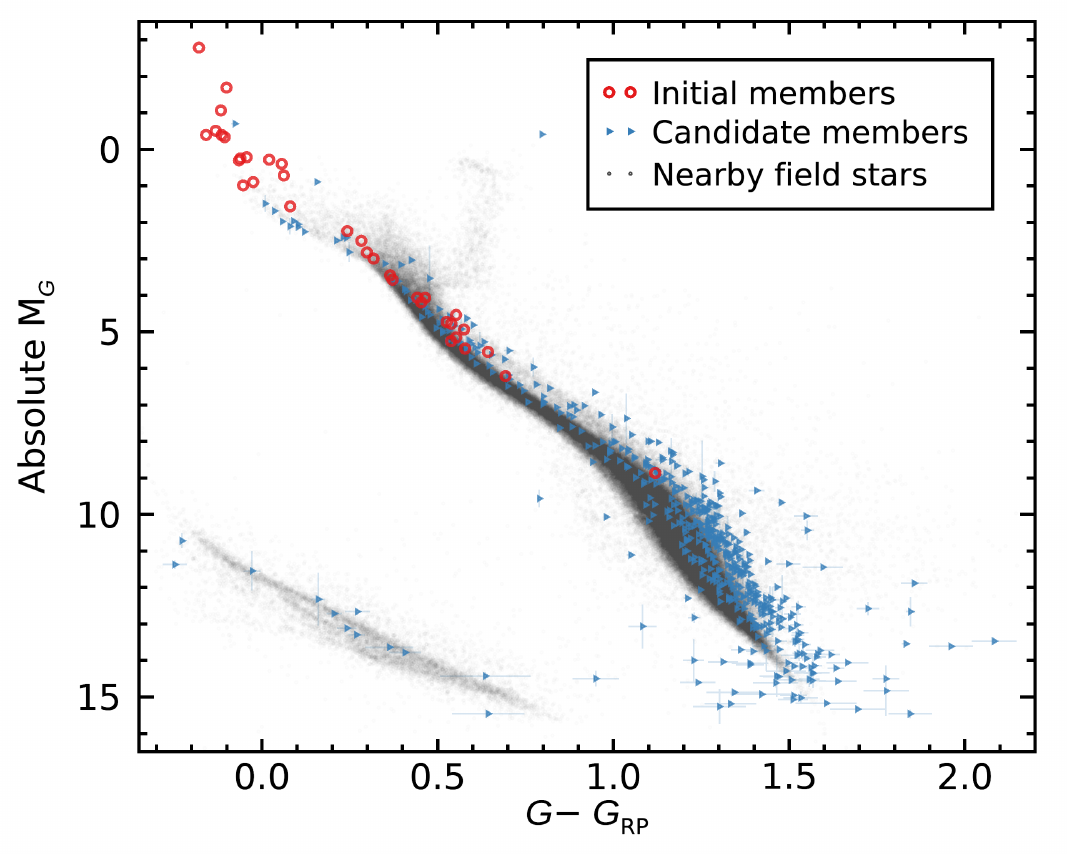}
 	\caption{\gaia\ color-magnitude diagram of initial \asso\ members (red circles) and candidate members (blue rightward triangles) recovered by BANYAN~$\Sigma$ based on their \gaia\ sky positions, proper motions, parallaxes and radial velocities when available. See Section~\ref{sec:newmembers} for more details.}
 	\label{fig:cand_cmd}
\end{figure}

These selection criteria resulted in a set of \ngaiasearch\ additional candidate members which are listed in Table~\ref{tab:allcandidates}. Their common proper motion is illustrated in Figure~\ref{fig:cand_pms} and their positions in a \gaia\ $G - G_{\rm RP}$ color versus absolute $G$ magnitude are shown in Figure~\ref{fig:cand_cmd}. Their sky positions are located in the range 37--74\textdegree\ and $-$4\textdegree\ to $+$29\textdegree\ in right ascension and declination, and their trigonometric distances are in the range 100--220\,pc, indicating that our initial filtering of \gaia\ entries was likely appropriate to encompass the full distribution of \asso\ members.

\subsection{A Search for Co-Moving Systems}\label{sec:comoving}

We complemented our search for \asso\ members with a subsequent search for stars co-moving with any one of the 540 members and candidate members. All \gaia\ entries within 180$''$ of each \asso\ candidate were inspected to find objects co-moving within 10\,\masyr\ and for which the proper motion difference is smaller than 5\% of the measurement. For most \gaia\ entries, a parallax measurement is also available: in these cases, we also required the trigonometric distance of the two objects to be within 5\,pc of each other\footnote{Throughout this work, we used a parallax zero-point of $-$0.029\,mas \citep{Lindegren:2018gy} to convert parallaxes to trigonometric distances. We determined trigonometric distances $\varpi = 1/\left(\pi+0.029\right)$ where $\pi$ is the parallax with a standard error propagation, which is accurate enough for the current purposes given the nearby distances of the stars under consideration.}, and we set a maximum parallax difference at 5\% of the parallax measurement.

This search identified \ninternalsystems\ co-moving systems (\ninternalcomovers\ components total) for which both components were already in the list of candidates, and \nalonecomovers\ stars (2MASS~J03424511+0754507 and 2MASS~J02581815+2456552) not already included in our list, each seemingly co-moving with a pair of stars in our list of candidates but failing to meet our membership selection criteria (i.e., their Bayesian membership probabilities are 49\% and 0\% respectively). In addition to those, we identified \npartialsystems\ systems (\npartialcomovers\ system components) for which only one component was in our list of candidates because the other component failed to pass our membership selection criteria. All objects were added to our list of low-likelihood candidates for completion, and all co-moving systems are listed in Table~\ref{tab:comovers}.

\startlongtable
\tabletypesize{\footnotesize}
\tablewidth{0.985\textwidth}
\clearpage
\pagebreak
\begin{longrotatetable}
\global\pdfpageattr\expandafter{\the\pdfpageattr/Rotate 90}
\begin{deluxetable*}{llccccccc}
\tablecolumns{9}
\tablecaption{List of new candidate members of \asso\ identified in \gaia\ with BANYAN~$\Sigma$.\label{tab:allcandidates}}
\tablehead{\colhead{\asso} & \colhead{Object} & \colhead{R.A.} & \colhead{Decl.} & \colhead{$\mu_\alpha\cos\delta$} & \colhead{$\mu_\delta$} & \colhead{Parallax} & \colhead{RV} & \colhead{\gaia}\\
\colhead{ID} & \colhead{Name} & \colhead{(hh:mm:ss.sss)} & \colhead{(dd:mm:ss.ss)} & \colhead{(\masyr)} & \colhead{(\masyr)} & \colhead{(mas)} & \colhead{(\kms)} & \colhead{$G$ mag}}
\startdata
31 & 2MASS~J02363660+2026331 & 02:36:36.648 & +20:26:32.84 & $40.3 \pm 1.4$ & $-26.7 \pm 1.1$ & $8.82 \pm 0.69$ & $\cdots$ & 15.289\\
32 & 2MASS~J02424085+2558585 & 02:42:40.905 & +25:58:58.08 & $38.60 \pm 0.16$ & $-31.61 \pm 0.14$ & $8.80 \pm 0.11$ & $\cdots$ & 16.090\\
33 & HD~17008 & 02:44:30.027 & +28:00:53.61 & $31.19 \pm 0.11$ & $-25.95 \pm 0.10$ & $7.372 \pm 0.066$ & $4 \pm 2$ & 7.909\\
34 & 2MASS~J02484851+1319378 & 02:48:48.553 & +13:19:37.66 & $34.627 \pm 0.062$ & $-23.355 \pm 0.054$ & $7.731 \pm 0.035$ & $\cdots$ & 14.077\\
35 & TYC~1785--155--1 & 02:49:43.813 & +25:53:11.25 & $30.146 \pm 0.087$ & $-26.174 \pm 0.077$ & $7.036 \pm 0.044$ & $7.2 \pm 0.3$ & 11.679\\
36 & 2MASS~J02513636+2811000 & 02:51:36.407 & +28:10:59.60 & $32.002 \pm 0.073$ & $-25.265 \pm 0.067$ & $7.866 \pm 0.036$ & $\cdots$ & 13.629\\
37 & 2MASS~J02515956+1458162 & 02:51:59.618 & +14:58:15.78 & $41.0 \pm 1.3$ & $-24.72 \pm 0.99$ & $8.68 \pm 0.78$ & $\cdots$ & 19.815\\
38 & 2MASS~J02523886+2300093 & 02:52:38.911 & +23:00:08.53 & $35.50 \pm 0.63$ & $-30.31 \pm 0.43$ & $8.30 \pm 0.32$ & $\cdots$ & 18.548\\
\enddata
\tablecomments{The MUTA identifiers listed in this table are defined in Section~\ref{sec:ruwe}. Some identifiers listed here contain only one component of a binary system (either A or B) because the other component was recovered in either comover searches of sections \ref{sec:comoving} and \ref{sec:ruwe}. Only a portion of the table is shown here. The full table is available as online-only additional material. Targets without a \asso\ ID number were flagged as problematic (i.e., low-likelihood candidates). See section~\ref{sec:newmembers} for more details.}
\end{deluxetable*}
\end{longrotatetable}
\clearpage
\pagebreak
\startlongtable
\tabletypesize{\scriptsize}
\tablewidth{0.985\textwidth}
\begin{longrotatetable}
\begin{deluxetable*}{llcccccccc}
\tablecolumns{10}
\tablecaption{Wide multiple systems recovered in \asso.\label{tab:comovers}}
\tablehead{\colhead{\asso} & \colhead{} & \colhead{R.A.} & \colhead{Decl.} & \colhead{$\mu_\alpha\cos\delta$} & \colhead{$\mu_\delta$} & \colhead{Parallax} & \colhead{\gaia} & \colhead{Sep.} & \colhead{Pos. Ang.}\\
\colhead{ID} & \colhead{Name} & \colhead{(hh:mm:ss.sss)} & \colhead{(dd:mm:ss.ss)} & \colhead{(\masyr)} & \colhead{(\masyr)} & \colhead{(mas)} & \colhead{$G$ mag} & \colhead{($''$)} & \colhead{(\textdegree)}}
\startdata
5 & 29~Tau & 03:45:40.466 & +06:02:59.78 & $21.88 \pm 0.29$ & $-13.65 \pm 0.26$ & $5.3 \pm 0.2$ & $5.295 \pm 0.001$ & $\cdots$ & $\cdots$\\
137 & 2MASS~J03454104+0602349 & 03:45:41.066 & +06:02:34.59 & $21.27 \pm 0.21$ & $-14.11 \pm 0.17$ & $5.5 \pm 0.1$ & $16.660 \pm 0.002$ & $26.7278 \pm 0.0001$ & $160.4167 \pm 0.0004$\\
138 & 2MASS~J03454269+0603039 & 03:45:42.712 & +06:03:03.66 & $21.11 \pm 0.12$ & $-13.801 \pm 0.091$ & $5.55 \pm 0.07$ & $15.389 \pm 0.001$ & $33.7294 \pm 0.0002$ & $83.3927 \pm 0.0002$\\
139 & 2MASS~J03454440+0603283 & 03:45:44.425 & +06:03:28.04 & $19.950 \pm 0.077$ & $-13.189 \pm 0.053$ & $5.56 \pm 0.05$ & $12.3683 \pm 0.0004$ & $65.4671 \pm 0.0002$ & $64.4250 \pm 0.0001$\\
\hline
\hline
10 & V766~Tau & 03:51:15.896 & +13:02:45.52 & $23.77 \pm 0.11$ & $-23.228 \pm 0.079$ & $6.19 \pm 0.06$ & $6.247 \pm 0.001$ & $\cdots$ & $\cdots$\\
182 & 2MASS~J03511041+1302467 & 03:51:10.454 & +13:02:46.16 & $23.10 \pm 0.17$ & $-23.14 \pm 0.12$ & $6.1 \pm 0.1$ & $16.117 \pm 0.001$ & $79.5332 \pm 0.0001$ & $270.4681 \pm 0.0001$\\
\hline
\hline
19~A & HD~23376 & 03:44:58.957 & +08:19:10.09 & $26.61 \pm 0.11$ & $-24.306 \pm 0.066$ & $6.89 \pm 0.06$ & $9.2549 \pm 0.0003$ & $\cdots$ & $\cdots$\\
19~B & TYC~658--1007--2 & 03:44:59.048 & +08:19:13.81 & $26.577 \pm 0.099$ & $-24.198 \pm 0.062$ & $6.99 \pm 0.07$ & $10.493 \pm 0.002$ & $3.9534 \pm 0.0001$ & $19.894 \pm 0.001$\\
\enddata
\tablecomments{See section~\ref{sec:comoving} for more details.}
\end{deluxetable*}
\end{longrotatetable}
\global\pdfpageattr\expandafter{\the\pdfpageattr/Rotate 0}

One notable case of a star with co-moving components is 29~Tau, the most massive member of \asso. 29~Tau (MUTA~5, Gaia~DR2 3276605295710700032) is a B3 + A7 binary star \citep{1980ApJS...44..489B}, with three co-moving systems within 70$''$: 29~Tau~B (MUTA~139; 2MASS~J03454440+0603283; Gaia~DR2 3276604922051089664), which is itself a spectral binary \citep{2001AJ....122.3466M}; 29~Tau~C (MUTA~137; 2MASS~J03454104+0602349; Gaia~DR2 3276604544094119424); and 29~Tau~D (MUTA~138; 2MASS~J03454269+0603039; Gaia~DR2 3276604544093968896). In addition to these six system components, there are two other \gaia\ entries within $\simeq$\,42$''$ of 29~Tau (Gaia~DR2 3276604509734231808 and Gaia~DR2 3276605265648475776) located within 300\,pc of the Sun with inconsistent proper motions and parallaxes. Both of them have re-normalised unit weight error (RUWE) values of $\simeq$\,1.1 which is not clearly indicative of bad parallax solutions, and indicates that they are probably unrelated to 29~Tau. For this reason, we ignored them in this analysis but we would recommend re-visiting this when further \emph{Gaia} data releases are published. Two additional MUTA candidates are within 700--715$^{\prime\prime}$ of 29~Tau: MUTA~143 (2MASS~J03460544+0553074; Gaia~DR2~3276586333432639744) and MUTA~135 (2MASS~J03450918+0612030; Gaia~DR2 3276798401738487808). Gaia~DR2 3276584478006772224 also seems co-moving with 29~Tau at a separation of 977$\farcs$5, but was not recovered in our search because its MUTA probability (89.7\%) is below our selection threshold.

Cross-matching our list of candidates with the \cite{2017AJ....153..257O} catalog of co-moving systems yielded a total of 28 matches, to Groups 39, 43, 52, 60, 124, 242, 1099 and 1109. Each of these groups have a total of members between 2 and 7. We verified that each of these groups were included in their entirety in our list of \asso\ candidates, and found 4 missing components of Group~39 and one missing component of Group~1109. We added these objects to our list of low-likelihood \asso\ candidates despite their BANYAN~$\Sigma$ membership probabilities below 90\% (ranging from 0\% to 64\%) for completion. As demonstrated by \cite{2018ApJ...863...91F}, the algorithm of \cite{2017AJ....153..257O} tends to break up nearby associations in many sub-groups because of the strong variations and correlations in direct kinematic observables (sky position, proper motion and parallax) caused by their wide distributions on the sky. The full list of matches between our candidates and \cite{2017AJ....153..257O} groups are shown in Table~\ref{tab:oh}.

\subsection{Red Giant Stars}\label{sec:giants}

One candidate member of the \asso\ association, HD~27860, is located far above the main sequence and within the red giant branch in Figure~\ref{fig:cand_cmd}. A literature search revealed that this object has a spectral type K2\,III \citep{1981ROAn...14.....W}, consistent with its position in the color-magnitude diagram. Based on the compilations of stars within 40\,pc established by \cite{2003AJ....126.2048G} and \cite{2006AJ....132..161G}, stars with the same spectral type have an average color $B-V = 1.16$ and absolute magnitude $M_V = 1.3$\footnote{See also \url{http://www.pas.rochester.edu/~emamajek/spt/K2III.txt}}.

Using the three-dimensional extinction map \emph{STructuring by Inversion of the Local InterStellar Medium} (STILISM; \citealp{2014AA...561A..91L,2017AA...606A..65C,2018AA...616A.132L})\footnote{Available at \url{https://stilism.obspm.fr}}, we can expect HD~27860 to be subject to an extinction $E(B-V) = 0.12 \pm 0.02$ based on its sky position and distance, which translates to $A_V = 0.43 \pm 0.08$ (using a total to selective extinction ratio $R = 3.54$ for this photometric band). Correcting its observed properties in the same photometric bands ($B-V = 1.41 \pm 0.01$ and $M_V = 0.05 \pm 0.02$; \citealt{1997ESASP1200.....E}) for extinction yields an intrinsic color of $B-V = 1.29 \pm 0.02$ and an absolute magnitude $M_V = -0.38 \pm 0.08$ , placing it closer in colors to the average value for K3\,III giants ($B-V = 1.37$).

Using the bolometric correction of \cite{1996ApJ...469..355F} for this color ($BC_V = -0.73$), we estimate a bolometric magnitude $M_{\rm bol} = -1.11 \pm 0.08$ and $\log L/L_\odot = 2.36 \pm 0.03$. We estimated its effective temperature at $T_{\rm eff} \approx 4400$\,K by interpolating its extinction-corrected $B-V$ color and comparing them with averages from \cite{2003AJ....126.2048G} and \cite{2006AJ....132..161G} for spectral types K2\,III and K3\,III. These physical parameters are consistent with a luminosity class III; the \cite{2009AA...508..355B} solar-metallicity isochrones predict a mass of 2.44\,\msol, a surface gravity $\log g \approx 2.0$ and an age of $\simeq$\,650\,Myr.

HD~27860 seems significantly too old to be a member of \asso\ based on the color-magnitude sequence of this young association (Figure~\ref{fig:cand_cmd}). The main-sequence turn-off of a 650\,Myr association would be located at spectral types A0 or later\footnote{See \url{http://www.pas.rochester.edu/~emamajek/EEM_dwarf_UBVIJHK_colors_Teff.txt}} (i.e., at absolute \gaia\ magnitudes $M_G \approx 1.5$). The fact that \asso\ includes several members more massive than A0 strongly suggests that HD~27860 is a chance interloper despite its high 98.6\% Bayesian membership probability, and we therefore reject it from our list of candidate members.

\subsection{White Dwarfs}\label{sec:wds}

\begin{figure}
 	\centering
 	\includegraphics[width=0.465\textwidth]{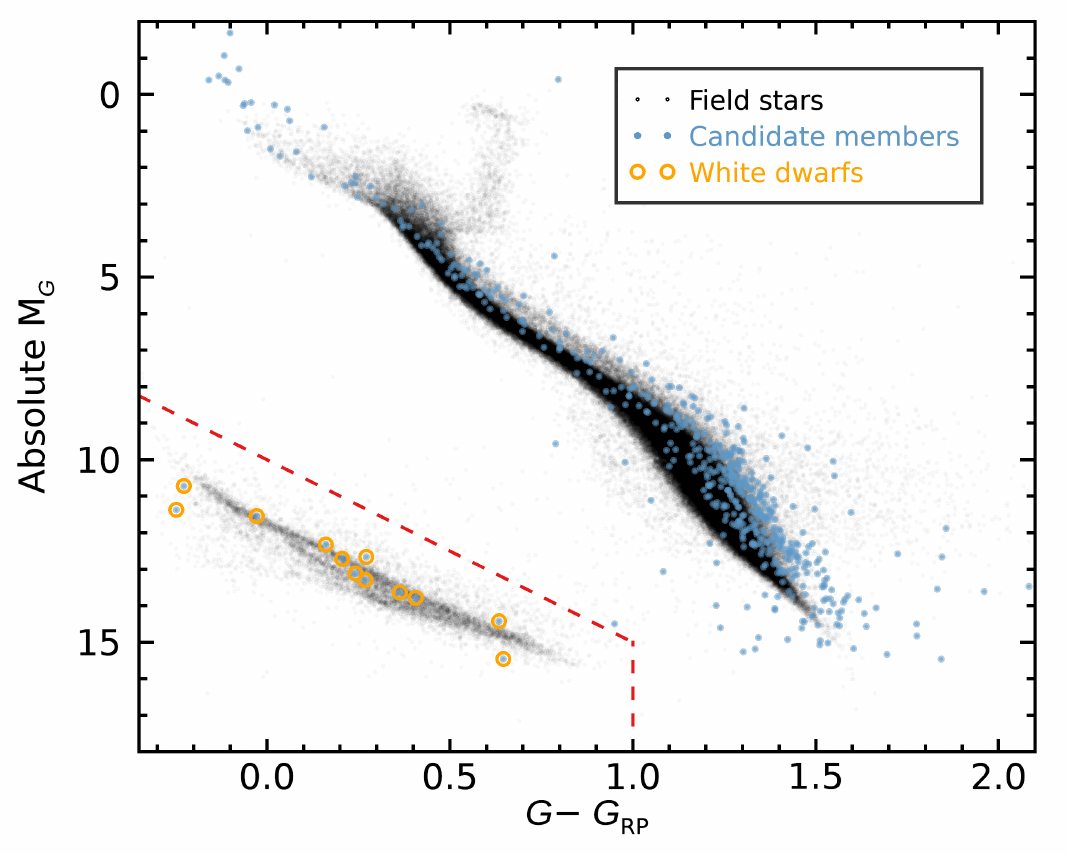}
 	\caption{Selection criterion for white dwarfs based on \gaia\ color-magnitude positions. See Section~\ref{sec:wds} for more details.}
 	\label{fig:wd_select}
\end{figure}

\begin{figure}
 	\centering
 	\includegraphics[width=0.465\textwidth]{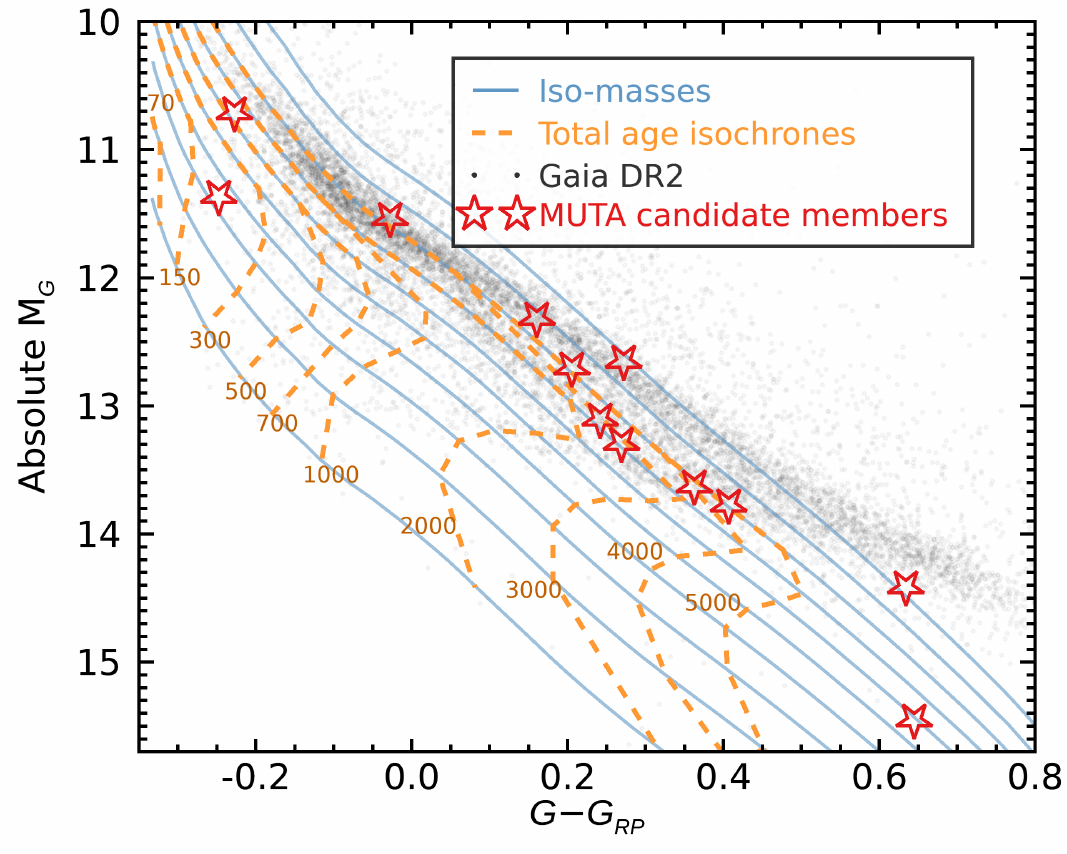}
 	\caption{\asso\ candidates recovered in \gaia\ data which color-magnitude positions are consistent with white dwarfs (red star symbols). Nearby white dwarfs in \gaia\ are indicated with black dots, and total age isochrones from 70\,Myr to 5\,Gyr are indicated with orange dashed lines. Iso-masses from 0.4\,\msol\ (top) to 1.3\,\msol\ (bottom) by steps of 0.1\,\msol\ are displayed with blue lines. Most white dwarfs recovered here are too old to be coeval with \asso. No correction for interstellar extinction was applied in this figure. See Section~\ref{sec:wds} for more details.}
 	\label{fig:wds}
\end{figure}

A subset of \asso\ members are located below the main sequence and within the color-magnitude sequence of white dwarfs in Figure~\ref{fig:initcmd}. We flagged all candidates with an absolute $G$-band magnitude fainter than $\left(G - G_{\rm RP}\right)\cdot 5 + 10$ and a color $G - G_{\rm RP} < 1.0$ {}(shown in Figure~\ref{fig:wd_select}) as likely white dwarfs, and compared them to total age isochrones obtained by combining MIST stellar main-sequence lifetimes \citep{2016ApJ...823..102C} and the the Montr\'eal white dwarf cooling tracks \citep{2001PASP..113..409F}\footnote{Available at \url{http://www.astro.umontreal.ca/~bergeron/CoolingModels/}, see also \citet{2006AJ....132.1221H,2006ApJ...651L.137K,2011ApJ...730..128T} and \cite{2011ApJ...737...28B}.} in Figure~\ref{fig:wds}. 

All but two white dwarfs in our sample are clearly much older than 150\,Myr, inconsistent with the main-sequence turn-off age of \asso\ ($\lesssim$\,80\,Myr). The two youngest and hottest white dwarfs in this figure are WD~0350+098 (MUTA~190; other designations include 1RXS~J035315.5+095700, SDSS~J035315.72+095633.7) and WD~0340+103 (MUTA~125; other designations include RBS~466, 1RXS~J034314.1+102941, and \\SDSS~J034314.35+102938.4), and are discussed further in Section~\ref{sec:wd2}.

\begin{figure}
 	\centering
 	\includegraphics[width=0.465\textwidth]{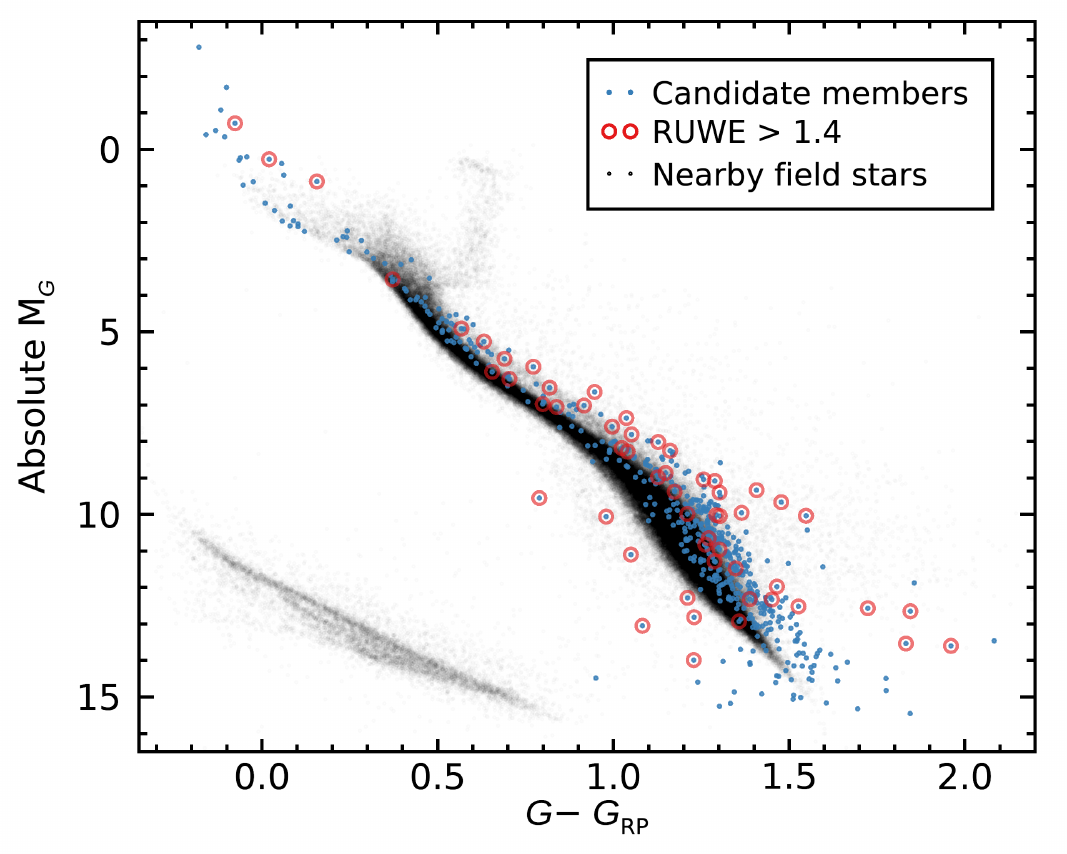}
 	\caption{\gaia\ color-magnitude diagram of \asso\ members and candidates (red circles) compared with nearby field stars (black circles). All objects flagged as problematic because of their poor \gaia\ astrometric solutions (RUWE $> 1.4$) are marked with red circles. A large fraction of these problematic solutions are located below the sequence of members, likely because of contamination by an unresolved source, or consist of possible multiple systems located above the \asso\ sequence. See Section~\ref{sec:ruwe} for more details.}
 	\label{fig:ruwe}
\end{figure}

We can estimate a false-positive rate for our list of \asso\ candidate members based on the fact that we uncovered 10 white dwarfs that are clearly too old for this young association. The number density of white dwarfs, $4.49 \pm 0.38 \times 10^{-3}$ objects\,pc$^{-3}$ \citep{2018MNRAS.480.3942H}, is small compared with that of main-sequence stars ($98.4 \pm 6.8 \times 10^{-3}$ objects\,pc$^{-3}$; \citealt{2012ApJ...753..156K}). Assuming that white dwarfs have similar kinematics to main-sequence stars, this means we could expect as many as $220_{-22}^{+25}$ stars in our sample to be contaminants if we applied no other cuts than BANYAN~$\Sigma$ probabilities based on proper motion and parallax without radial velocity measurements (none of the white dwarf contaminants have radial velocity measurements). An additional 28 \gaia\ sources would have been uncovered in our survey if we used only these observables and no other criteria, leaving our estimated number of contaminants to $192_{-22}^{+25}$ in our final list of candidates, or $34_{-4}^{+5}$\% of our full sample of \ngaiasearch\ objects. The majority of contaminants are expected to be M dwarfs. 

\subsection{Poor Astrometric Solutions}\label{sec:ruwe}

The \gaia\ team recommends placing low confidence in astrometric solutions with a RUWE larger than 1.4\footnote{As described at \url{https://www.cosmos.esa.int/web/gaia/dr2-known-issues}.}. We therefore flagged all 52 \asso\ candidates and members with RUWE~$> 1.4$ (shown in Figure~\ref{fig:ruwe}) and consider them as low-likelihood candidates; we consider that an observational follow-up of these objects will potentially be useful, but should be less prioritary. It is likely that some of these issues will be resolved in the next \gaia\ data release.

\subsection{Visual Inspection of Finder Charts}\label{sec:fcharts}

\begin{figure*}[p]
 	\centering
 	\includegraphics[width=0.98\textwidth]{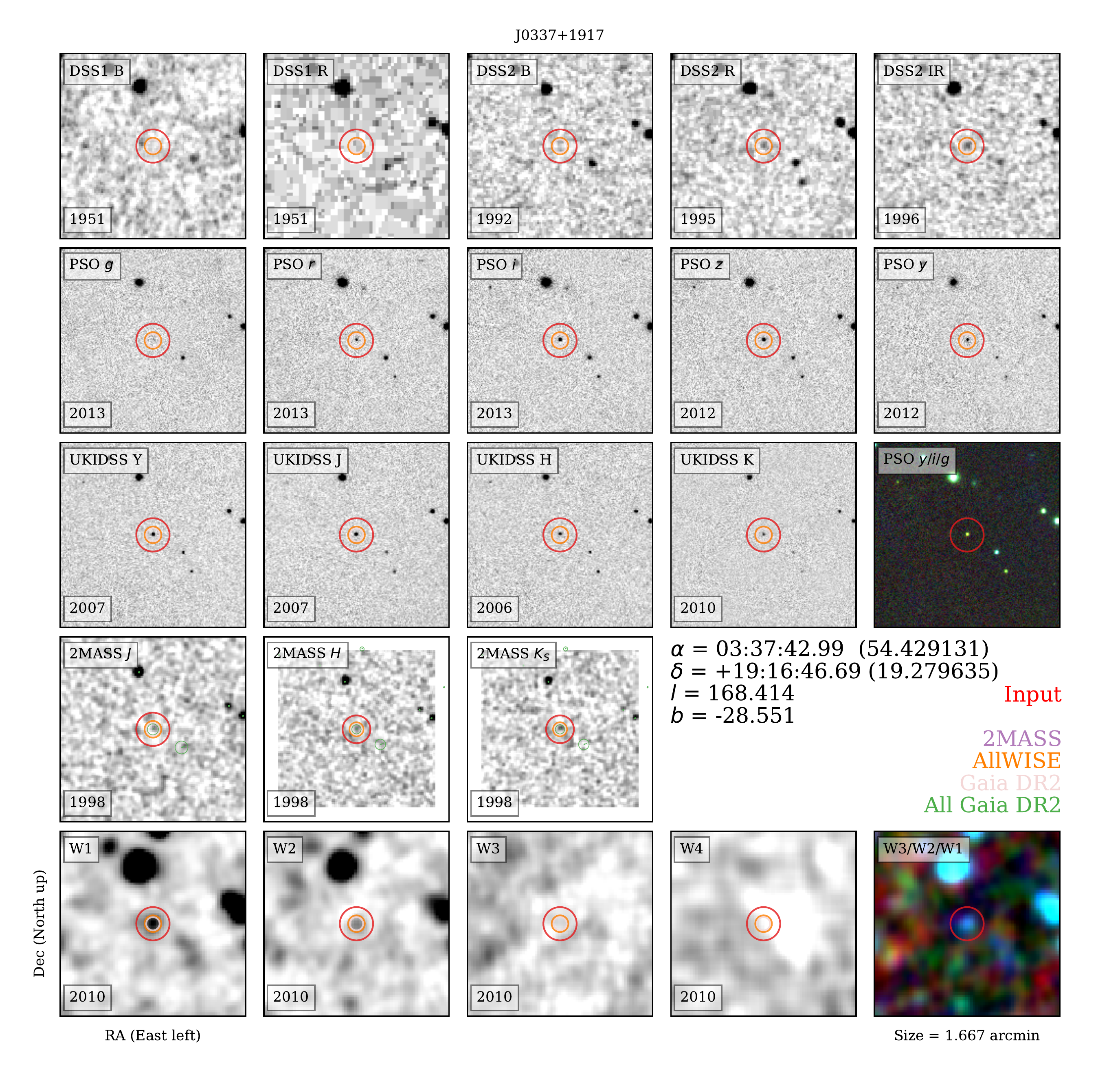}
 	\caption{Finder charts for WISEA~J033742.99+191646.7, a problematic candidate because its position in a \gaia\ color-magnitude diagram is well below the main sequence, likely because of contamination from a background source at a very small angular separation. See Section~\ref{sec:fcharts} for more details.}
 	\label{fig:fccont}
\end{figure*}

We generated finder charts for all \asso\ objects with available survey data from DSS, SDSS \citep{2015ApJS..219...12A}, UKIDSS \citep{2007MNRAS.379.1599L}, VHS \citep{2013Msngr.154...35M}, Pan-STARRS \citep{2016arXiv161205560C}, {\it WISE} \citep{2010AJ....140.1868W} and 2MASS \citep{2006AJ....131.1163S} data with the \texttt{finder\_charts.py} Python package \citep{zenodofindercharts}\footnote{Available at \url{https://github.com/jgagneastro/finder_charts}.}, on which we overlaid \gaia\ catalog entries with arrows and symbol sizes indicating their individual proper motions and distances. We used these figures to identify and correct any mismatches in our automated cross-matches to \emph{2MASS} and \emph{WISE}, which tends to happen when a target has a missing entry in either catalog.

We also verified that binaries and co-moving systems had the correct component attached to each catalog, and noted \nvisualcomovers\ stars that visually appeared co-moving with one of our targets at a similar distance, but were not recovered with our co-moving search described in Section~\ref{sec:comoving}. Those usually have \gaia\ proper motions or parallaxes that are slight mis-matches to our \asso\ candidate or member, and are listed in Table~\ref{tab:viscomovers}. It is possible that some of these systems suffer from a bad parallax solution, either because they are themselves multiple systems (e.g., 30~Tau and TYC~661--1404--1, respectively MUTA~3~A and MUTA~3~B), or contaminated by a background source (althoug they all have RUWE~$\leq 1.4$). We listed these systems that almost seem co-moving in Table~\ref{tab:viscomovers} for later follow-up, but we excluded them from the current analysis.

A number of \asso\ candidates are located well below the main sequence in a \gaia\ color-magnitude diagram (see Figure~\ref{fig:cand_cmd}), but yet not faint enough to be credible white dwarfs (see Section~\ref{sec:wds}). A fraction of these objects failed the \gaia\ RUWE~$\leq 1.4$ selection criterion for good astrometric solutions, indicating that bad parallax solutions are likely part of the explanation. Figure~\ref{fig:fccont} shows a finder chart for one such object (WISEA~J033742.99+191646.7)\footnote{All finder charts are available as online-only supplementary data.}. In this example, the finder chart shows that it is well detected at red-optical wavelengths (e.g., Pan-STARRS) and in \emph{WISE} $W1$, but too faint to be detected in 2MASS in the near-infrared. This unusual combination indicates a likely contribution from two distinct blackbodies. The presence of an accretion disks could potentially explain this, however those usually result in much redder \gaia\ \gaiagr\ colors, which would push the object far to the right of, rather than below, the main sequence. The simplest explanation seems to be that this object is a blend of two sources, maybe located at different distances, but at an angular separation small enough that they are unresolved in all the aforementioned surveys.

We assigned MUTA identifiers (31 to 372) to all candidate members that were not rejected or defined as low-likelihood candidates based either on their poor astrometric solutions, ages that are definitely too old, or problematic position in a color-magnitude diagram. We ordered these identifiers by right ascension. Stars identified in Section~\ref{sec:comoving} as co-moving with a well-behaved MUTA candidate or member which did not have a MUTA identified were assigned identifiers 373--375. Those still without identifiers that belong in one of the \cite{2017AJ....153..257O} groups associated with \asso\ were assigned identifiers 376--382, and those visually identified as co-moving with a well-behaved candidate in this section were assigned identifiers 383--386.

\clearpage
\pagebreak
\startlongtable
\tablewidth{0.985\textwidth}
\begin{deluxetable*}{lllcc}
\tablecolumns{5}
\tablecaption{MUTA objects in common with \cite{2017AJ....153..257O}.\label{tab:oh}}
\tablehead{\colhead{\asso} & \colhead{} & \colhead{\gaia} & \colhead{\cite{2017AJ....153..257O}} & \colhead{Object}\\
 \colhead{ID} & \colhead{Name} & \colhead{ID} & \colhead{Group} & \colhead{Type\tablenotemark{a}}}
\startdata
368 & HD~31125 & 3226496187146449920 & 39 & Candidates\\
369 & TYC~4745--475--1 & 3224698799168916864 & 39 & Candidates\\
372~A & TYC~4741--307--1 & 3225639289631939456 & 39 & Candidates\\
379 & BD+00~884 & 3231439080323844864 & 39 & Incomplete\\
380 & HD~32264 & 3225291882613467520 & 39 & Incomplete\\
381 & HD~32721 & 3212973572810773120 & 39 & Incomplete\\
382 & HD~33023 & 3212956839618107648 & 39 & Incomplete\\
10 & V766~Tau & 37136834159399808 & 43 & Initial\\
21 & HD~286374 & 3303308245556503296 & 43 & Initial\\
22 & PPM~119410 & 36595943156045824 & 43 & Initial\\
26 & TYC~662--217--1 & 3304906145189468416 & 43 & Initial\\
28 & TYC~664--136--1 & 39841357885932288 & 43 & Initial\\
377 & HIP~18778 & 3301831773241303552 & 43 & Initial\\
13 & HD~23990 & 3302396166303947904 & 52 & Initial\\
19~A & HD~23376 & 3278197770802258944 & 52 & Initial\\
19~B & TYC~658--1007--2 & 3278197766505583232 & 52 & Initial\\
95 & HD~22073 & 11397988505713536 & 52 & Candidates\\
140 & TYC~658--828--1 & 3278300987456845440 & 52 & Candidates\\
17 & HD~27687 & 3286590824092307200 & 60 & Initial\\
18 & HD~28356 & 3285720938596464640 & 60 & Initial\\
25 & TYC~80--202--1 & 3297372944352021120 & 60 & Initial\\
30~A & TYC~668--737--1 & 3299167141170181888 & 60 & Initial\\
290 & BD+05~638 & 3284966433101477376 & 60 & Candidates\\
33 & HD~17008 & 127148009968227584 & 124 & Candidates\\
35 & TYC~1785--155--1 & 114510012864474112 & 124 & Candidates\\
41 & TYC~1790--927--1 & 115353480017970560 & 124 & Candidates\\
11 & HD~28715 & 3285542336676520448 & 242 & Initial\\
324~A & HD~29182 & 3282435563491664896 & 242 & Candidates\\
324~B & TYC~90--953--1 & 3282434979377650176 & 242 & Incomplete\\
20 & HIP~17133 & 38088873789758720 & 1099 & Initial\\
117~A & TYC~663--362--1 & 38076641722829440 & 1099 & Candidates\\
376 & TYC~665--150--1 & 38398936068862464 & 1109 & Candidates\\
378 & HD~286412 & 3305439511410844800 & 1109 & Incomplete\\
\enddata
\tablenotetext{a}{Initial: members of \asso\ from our initial list. Candidates: candidates of \asso\ recovered in Section~\ref{sec:newmembers}. Incomplete: Targets missing from our list of \asso\ initial members and new candidates.}
\tablecomments{See section~\ref{sec:comoving} for more details.}
\end{deluxetable*}
\startlongtable
\tabletypesize{\scriptsize}
\tablewidth{0.985\textwidth}
\begin{longrotatetable}
\global\pdfpageattr\expandafter{\the\pdfpageattr/Rotate 90}
\begin{deluxetable*}{llcccccccc}
\tablecolumns{10}
\tablecaption{Wide multiple candidate systems in \asso\ visually identified but not recovered in Section~\ref{sec:comoving}.\label{tab:viscomovers}}
\tablehead{\colhead{\asso} & \colhead{} & \colhead{R.A.} & \colhead{Decl.} & \colhead{$\mu_\alpha\cos\delta$} & \colhead{$\mu_\delta$} & \colhead{Parallax} & \colhead{\gaia} & \colhead{Sep.} & \colhead{Pos. Ang.}\\
\colhead{ID} & \colhead{Name} & \colhead{(hh:mm:ss.sss)} & \colhead{(dd:mm:ss.ss)} & \colhead{(\masyr)} & \colhead{(\masyr)} & \colhead{(mas)} & \colhead{$G$ mag} & \colhead{($''$)} & \colhead{(\textdegree)}}
\startdata
3~A & 30~Tau & 03:48:16.292 & +11:08:35.52 & $25.27 \pm 0.28$ & $-23.69 \pm 0.23$ & $7.74 \pm 0.17$ & 5.040 & $\cdots$ & $\cdots$\\
3~B & TYC~661-1404-1 & 03:48:16.835 & +11:08:40.16 & $25.62 \pm 0.15$ & $-24.97 \pm 0.11$ & $7.223 \pm 0.070$ & 9.269 & $9.2389 \pm 0.0001$ & $59.8505 \pm 0.0007$\\
\hline
\hline
5 & 29~Tau & 03:45:40.466 & +06:02:59.78 & $21.88 \pm 0.29$ & $-13.65 \pm 0.26$ & $5.31 \pm 0.25$ & 5.295 & $\cdots$ & $\cdots$\\
383 & 2MASS~J03453759+0603048 & 03:45:37.587 & +06:03:04.31 & $-1.86 \pm 0.11$ & $-25.471 \pm 0.085$ & $5.427 \pm 0.070$ & 15.481 & $43.1779 \pm 0.0002$ & $276.0326 \pm 0.0001$\\
\hline
\hline
97~A & 2MASS~J03350340+1431490 & 03:35:03.438 & +14:31:48.54 & $26.99 \pm 0.14$ & $-25.93 \pm 0.11$ & $7.342 \pm 0.081$ & 15.369 & $\cdots$ & $\cdots$\\
97~B & 2MASS~J03350317+1431358 & 03:35:03.209 & +14:31:35.33 & $26.87 \pm 0.39$ & $-25.57 \pm 0.30$ & $6.84 \pm 0.24$ & 18.097 & $13.6275 \pm 0.0002$ & $194.1236 \pm 0.0008$\\
\hline
\hline
104~A & 2MASS~J03361762+2153391 & 03:36:17.665 & +21:53:38.50 & $29.492 \pm 0.088$ & $-30.262 \pm 0.068$ & $7.526 \pm 0.047$ & 10.910 & $\cdots$ & $\cdots$\\
104~B & 2MASS~J03361732+2153271 & 03:36:17.360 & +21:53:26.42 & $29.95 \pm 0.71$ & $-31.17 \pm 0.52$ & $7.98 \pm 0.31$ & 18.454 & $12.8056 \pm 0.0002$ & $199.354 \pm 0.002$\\
\hline
\hline
117~A & TYC~663--362--1 & 03:40:57.781 & +13:09:03.06 & $24.66 \pm 0.25$ & $-25.44 \pm 0.21$ & $6.749 \pm 0.098$ & 10.493 & $\cdots$ & $\cdots$\\
117~B & 2MASS~J03405723+1308577 & 03:40:57.261 & +13:08:57.23 & $27.03 \pm 0.71$ & $-24.78 \pm 0.50$ & $7.19 \pm 0.33$ & 18.437 & $9.5851 \pm 0.0003$ & $232.539 \pm 0.002$\\
\hline
\hline
153~A & TYC~1252--301--1 & 03:47:23.901 & +18:43:17.68 & $21.128 \pm 0.079$ & $-23.175 \pm 0.057$ & $5.926 \pm 0.041$ & 11.689 & $\cdots$ & $\cdots$\\
153~B & Gaia~DR2~44752086050666368 & 03:47:23.645 & +18:43:18.70 & $21.33 \pm 0.60$ & $-26.33 \pm 0.57$ & $6.36 \pm 0.34$ & 17.855 & $3.7781 \pm 0.0003$ & $285.639 \pm 0.004$\\
\hline
\hline
177~A & 2MASS~J03505694+0730565 & 03:50:56.976 & +07:30:56.18 & $30.41 \pm 0.23$ & $-22.22 \pm 0.16$ & $8.29 \pm 0.12$ & 16.916 & $\cdots$ & $\cdots$\\
177~B & Gaia~DR2~3277369048270999936 & 03:50:56.968 & +07:30:53.92 & $27.9 \pm 2.2$ & $-21.7 \pm 1.4$ & $6.3 \pm 1.3$ & 20.438 & $2.2609 \pm 0.0005$ & $183.03 \pm 0.03$\\
\hline
\hline
225~A & 2MASS~J04021281+0817400 & 04:02:12.839 & +08:17:39.75 & $23.38 \pm 0.24$ & $-22.68 \pm 0.17$ & $6.62 \pm 0.13$ & 16.635 & $\cdots$ & $\cdots$\\
225~B & 2MASS~J04021257+0817410 & 04:02:12.593 & +08:17:40.67 & $22.00 \pm 0.36$ & $-23.46 \pm 0.25$ & $6.19 \pm 0.19$ & 17.316 & $3.7653 \pm 0.0002$ & $284.200 \pm 0.002$\\
\hline
\hline
271 & 2MASS~J04181095+0934586 & 04:18:10.980 & +09:34:58.24 & $15.97 \pm 0.27$ & $-21.58 \pm 0.21$ & $5.83 \pm 0.16$ & 17.228 & $26.4111 \pm 0.0001$ & $326.8006 \pm 0.0003$\\
384 & 2MASS~J04181193+0934365 & 04:18:11.958 & +09:34:36.14 & $19.11 \pm 0.19$ & $-21.51 \pm 0.14$ & $4.74 \pm 0.11$ & 16.800 & $\cdots$ & $\cdots$\\
\hline
\hline
277~A & 2MASS~J04200165+0759584 & 04:20:01.666 & +07:59:57.72 & $22.83 \pm 0.68$ & $-23.94 \pm 0.47$ & $6.31 \pm 0.34$ & 15.336 & $\cdots$ & $\cdots$\\
277~B & Gaia~DR2~3298956138016754048 & 04:20:01.719 & +07:59:58.51 & $19.87 \pm 0.90$ & $-21.65 \pm 0.42$ & $6.69 \pm 0.14$ & 16.289 & $1.1173 \pm 0.0003$ & $44.75 \pm 0.02$\\
\hline
\hline
279 & 2MASS~J04201617+0959534 & 04:20:16.202 & +09:59:53.06 & $17.34 \pm 0.38$ & $-20.97 \pm 0.19$ & $6.14 \pm 0.17$ & 17.379 & $7.1832 \pm 0.0001$ & $18.446 \pm 0.001$\\
385 & TYC~671--129--1 & 04:20:16.048 & +09:59:46.25 & $16.91 \pm 0.14$ & $-22.140 \pm 0.066$ & $5.668 \pm 0.061$ & 10.795 & $\cdots$ & $\cdots$\\
\hline
\hline
318~A & 2MASS~J04341953+0226260 & 04:34:19.560 & +02:26:25.89 & $16.32 \pm 0.42$ & $-20.02 \pm 0.29$ & $5.77 \pm 0.22$ & 12.150 & $\cdots$ & $\cdots$\\
318~B & Gaia~DR2~3279527149078835712 & 04:34:19.467 & +02:26:25.91 & $15.62 \pm 0.75$ & $-21.63 \pm 0.44$ & $6.24 \pm 0.33$ & 15.960 & $1.4009 \pm 0.0003$ & $270.61 \pm 0.01$\\
\hline
\hline
329~A & 2MASS~J04372971--0051241 & 04:37:29.730 & --00:51:24.47 & $15.026 \pm 0.042$ & $-16.665 \pm 0.027$ & $6.050 \pm 0.025$ & 13.223 & $\cdots$ & $\cdots$\\
329~B & Gaia~DR2~3229491776511286016 & 04:37:29.780 & --00:51:25.66 & $14.70 \pm 0.36$ & $-17.42 \pm 0.19$ & $5.75 \pm 0.14$ & 16.507 & $1.4169 \pm 0.0001$ & $147.563 \pm 0.006$\\
\hline
\hline
331~A & 2MASS~J04382750-0342441 & 04:38:27.523 & --03:42:44.47 & $23.399 \pm 0.072$ & $-20.462 \pm 0.051$ & $6.076 \pm 0.041$ & 14.931 & $\cdots$ & $\cdots$\\
331~B & Gaia~DR2~3201810884087980800 & 04:38:27.437 & --03:42:46.23 & $19.59 \pm 0.85$ & $-17.76 \pm 0.49$ & $6.77 \pm 0.40$ & 18.416 & $2.1809 \pm 0.0003$ & $216.359 \pm 0.007$\\
\hline
\hline
368 & HD~31125 & 04:53:04.828 & --01:16:33.04 & $12.67 \pm 0.11$ & $-15.782 \pm 0.072$ & $5.644 \pm 0.051$ & 7.918 & $\cdots$ & $\cdots$\\
386 & HD~31124 & 04:53:04.574 & --01:15:52.17 & $19.54 \pm 0.19$ & $-17.88 \pm 0.12$ & $6.081 \pm 0.098$ & 8.046 & $41.0425 \pm 0.0001$ & $354.6787 \pm 0.0001$\\
\hline
\hline
$\cdots$ & 2MASS~J03343284+1212290 & 03:34:32.872 & +12:12:28.55 & $27.892 \pm 0.090$ & $-30.240 \pm 0.063$ & $6.740 \pm 0.044$ & 12.302 & $\cdots$ & $\cdots$\\
$\cdots$ & Gaia~DR2~40541334474313728 & 03:34:33.106 & +12:12:29.76 & $28.23 \pm 0.88$ & $-29.63 \pm 0.62$ & $7.50 \pm 0.54$ & 19.087 & $3.6272 \pm 0.0004$ & $70.436 \pm 0.005$
\enddata
\tablecomments{See section~\ref{sec:fcharts} for more details.}
\end{deluxetable*}
\end{longrotatetable}
\clearpage
\pagebreak
\global\pdfpageattr\expandafter{\the\pdfpageattr/Rotate 0}
\global\pdfpageattr\expandafter{\the\pdfpageattr/Rotate 0}
\global\pdfpageattr\expandafter{\the\pdfpageattr/Rotate 0}
\global\pdfpageattr\expandafter{\the\pdfpageattr/Rotate 0}

\section{CORRECTING EXTINCTION IN \emph{GAIA}~DR2 PHOTOMETRY}\label{sec:ext}

\begin{figure}
 	\centering
 	\includegraphics[width=0.465\textwidth]{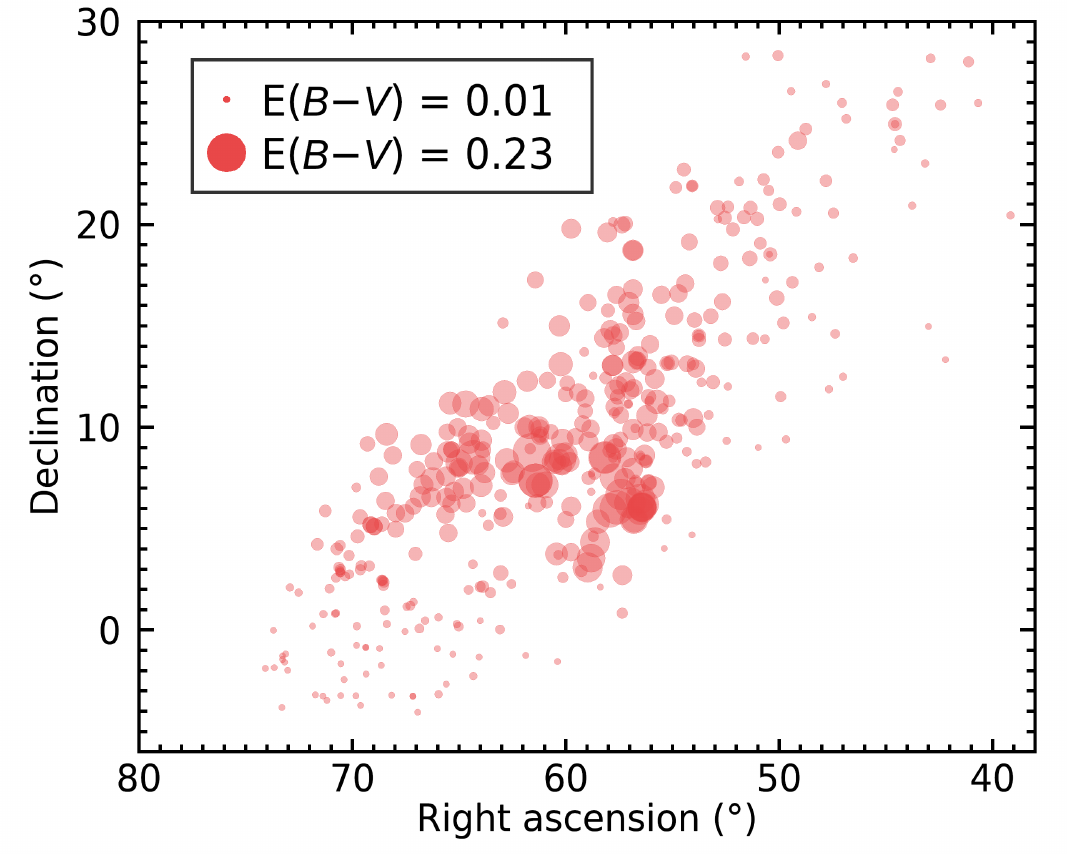}
 	\caption{Individual $E(B-V)$ extictions of \asso\ objects based on the STILISM three-dimensional extinction map combined with the sky positions and \gaia\ distances of \asso\ objects. See Section~\ref{sec:ext} for more details.}
 	\label{fig:ext}
\end{figure}

The \asso\ association is distant enough that some of its members appear slightly reddened by interstellar dust. We used STILISM (\citealp{2014AA...561A..91L,2017AA...606A..65C,2018AA...616A.132L})\footnote{Available at \url{https://stilism.obspm.fr}} to determine the individual $E(B-V)$ extinction values for individual \asso\ objects based on their sky position and \gaia\ distance. The resulting individual extinction values are displayed in Figure~\ref{fig:ext}.

\begin{figure}
 	\centering
 	\includegraphics[width=0.465\textwidth]{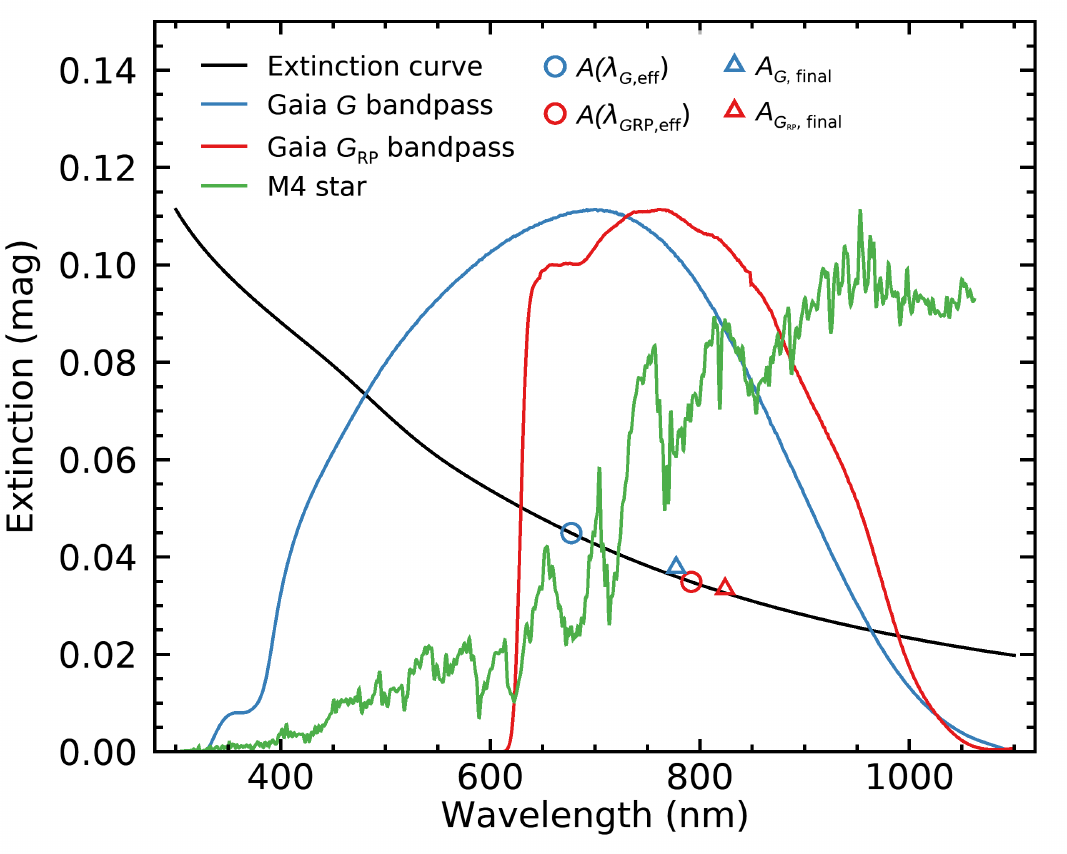}
 	\caption{\cite{1999PASP..111...63F} interstellar extinction curve (black line) compared with \gaia\ $G$ and \gaiagr\ bandpasses (blue and red, respectively) and the spectral flux density of an M4 low-mass star (green). Using only the effective wavelength of \gaia\ bandpasses to estimate extinction (blue and red circles) leads to an over estimation of de-reddening and a mistaken reddening vector angle compared with a more careful extinction correction that accounts for the stellar flux across the \gaia\ bandpasses (blue and red triangles). This effect is highly dependent on the spectral type of the star because of the wide \gaia\ bandpasses. See Section~\ref{sec:ext} for more detail.}
 	\label{fig:extcurve}
\end{figure}

\begin{figure}
 	\centering
 	\includegraphics[width=0.465\textwidth]{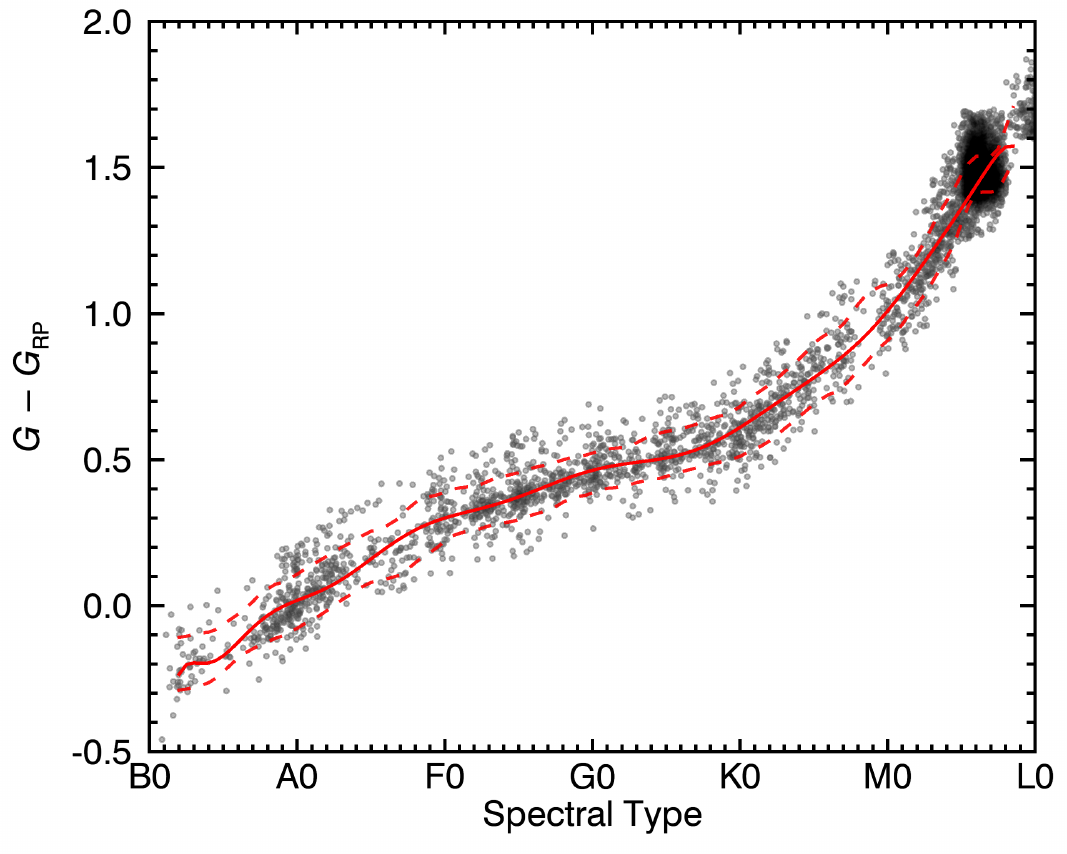}
 	\caption{\gaia\ \gaiagr\ colors as a function of spectral type for known nearby young stars and brown dwarfs (black dots). A polynomial fit is shown as a solid red line with 1$\sigma$ scatter as a dashed red line. We used this relation to estimate spectral types when no literature data was available. See Section~\ref{sec:spt} for more details. The polynomial coefficients for the red line are available as online-only material.}
 	\label{fig:sptcolor}
\end{figure}

We corrected the color-magnitude diagram position of \asso\ members and candidates with an iterative method to account for the wide \gaia\ photometric bandpasses. As shown in Figure~\ref{fig:extcurve}, even the \gaiar\ bandpass spans a significant region over which both the extinction curve of \cite{1999PASP..111...63F} and the spectral energy density of an M-type star vary significantly. As a consequence, the reddening vectors in \gaia\ color-magnitude sequences will differ significantly across spectral types.

The flux of a star with a spectral energy density $S_\lambda$ observed through an instrument with a bandpass $P_\lambda$ is given by:

\begin{align}
	F = \frac{\int_0^\infty S_\lambda P_\lambda \mathrm{d}\lambda}{\int_0^\infty P_\lambda \mathrm{d}\lambda}.
\end{align}

In the presence of interstellar extinction $E_\lambda$, the observed flux is:

\begin{align}
	F_{\rm reddened} = \frac{\int_0^\infty E_\lambda S_\lambda P_\lambda \mathrm{d}\lambda}{\int_0^\infty P_\lambda \mathrm{d}\lambda},
\end{align}

\noindent and therefore the correction factor that remains valid for wide bandpasses is:

\begin{align}
	\frac{F_{\rm reddened}}{F} = \frac{\int_0^\infty E_\lambda S_\lambda P_\lambda \mathrm{d}\lambda}{\int_0^\infty S_\lambda P_\lambda \mathrm{d}\lambda}.
\end{align}

\begin{figure*}[p]
 	\centering
 	\includegraphics[width=0.98\textwidth]{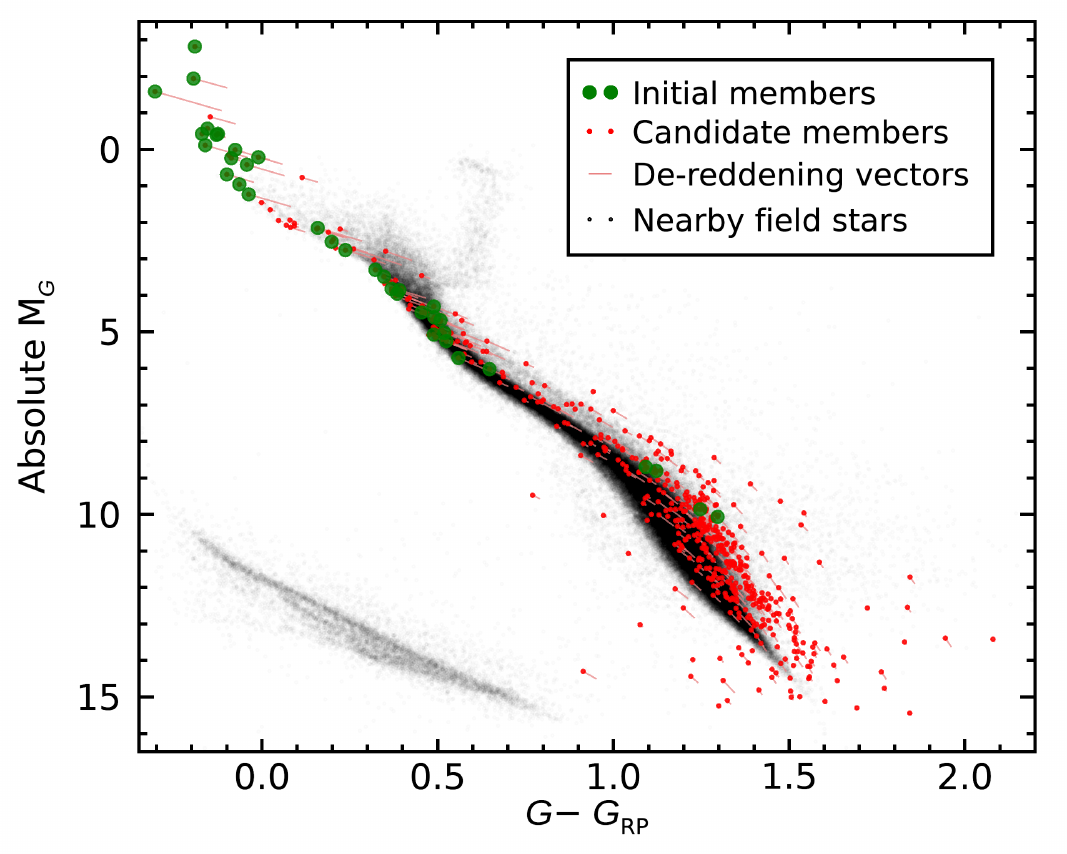}
 	\caption{\gaia\ color-magnitude diagram of our initial list of \asso\ members (green circles) and additional candidate members (red dots). De-reddening vectors are indicated with red lines (dots are located at the corrected position). A proper de-reddening correction that accounts for the wide \gaia\ bandpasses moves low-mass stars parallel to the sequence, and moves higher-mass stars mostly towards the left. See Section~\ref{sec:ext} for more details.}
 	\label{fig:cmd}
\end{figure*}

In effect, this correction is a weighted average of the extinction curve, where the weight is given by the product of the stellar spectral energy density with the instrumental bandpass. In general, the spectral energy densities of \asso\ members and candidates have not been measured, and their spectral types are unknown. We therefore used an iterative method where the photometric spectral type of each star is first estimated from its \gaiagr\ color. The \gaiagr\ versus spectral type relation for stars with spectral types B0 to L0 is shown in Figure~\ref{fig:sptcolor}. These data were drawn from the list of nearby young association members of \cite{2018ApJ...856...23G} and the List of Ultracool Dwarfs\footnote{Available at \url{http://astro.umontreal.ca/~gagne/ultracool_dwarfs.php}} that includes data from previous lists of brown dwarfs \citep{2012ApJS..201...19D,2014PhDT........56M,2015ApJS..219...33G,2016ApJ...833...96L,2016ApJS..225...10F}. A polynomial relation was fitted to the data and is also displayed in the figure; the coefficients to this polynomial sequence are available as online-only material. We preferred using a \gaia\ color to spectral type relation rather than a \gaia\ absolute magnitude to spectral type relation, because unresolved multiples would bias the latter more significantly.

We used the Pickles Atlas of spectral energy distributions for B0--M9 stars \citep{1998PASP..110..863P} and interpolated the \gaia\ instrumental bandpasses and the extinction curve of \citeauthor{1999PASP..111...63F} (\citeyear{1999PASP..111...63F}; with a nominal total to selective extinction value $R(V) = 3.1$) on the Pickles wavelength vector to determine an appropriate extinction correction.

The resulting extinction-corrected \gaiagr\ color was then used to obtain a better photometric spectral type estimate, which we used in turn to correct the raw \gaiagr\ color anew. This step was repeated until the photometric spectral type estimate of a star remained unchanged. A total of four iterations were needed for the de-reddening correction to converge for all \asso\ stars. The resulting extinction vectors and corrected color-magnitude diagram of \asso\ are shown in Figure~\ref{fig:cmd}.

In Tables~\ref{tab:ext1} and \ref{tab:ext2}, we provide reddening values $R(G)$ and $R(G_{\rm RP})$ as a function of spectral types or uncorrected \gaia\ $G - G_{\rm RP}$ colors, which can be used to de-redden the \gaia\ photometry of main-sequence or young stars with the following relations:

\begin{align}
	G_{\rm corr} &= G_{\rm uncorr} - E(B-V)\cdot R(G),\\
	G_{\rm RP,corr} &= G_{\rm RP,uncorr} - E(B-V)\cdot R(G_{\rm RP}).
\end{align}

\startlongtable
\tablewidth{0.985\textwidth}
\begin{deluxetable}{cccc}
\tablecolumns{4}
\tablecaption{\gaia\ de-reddening relations as a function of spectral type that account for its large photometric bandpasses.\label{tab:ext1}}
\tablehead{\colhead{Spectral} & \colhead{$R(G)$} & \colhead{$R(G_{\rm RP})$} & \colhead{$R(G_{\rm BP})$}\\
\colhead{Type} & \colhead{(mag)} & \colhead{(mag)} & \colhead{(mag)}}
\startdata
B3 & $3.112 \pm 0.001$ & $1.938 \pm 0.001$ & $3.670 \pm 0.001$\\
B5 & $3.10 \pm 0.02$   & $1.936 \pm 0.002$ & $3.65 \pm 0.02$  \\
B7 & $3.029 \pm 0.001$ & $1.926 \pm 0.001$ & $3.566 \pm 0.001$\\
B9 & $3.002 \pm 0.006$ & $1.925 \pm 0.001$ & $3.540 \pm 0.006$\\
A1 & $2.962 \pm 0.003$ & $1.919 \pm 0.001$ & $3.503 \pm 0.002$\\
A3 & $2.939 \pm 0.006$ & $1.916 \pm 0.001$ & $3.488 \pm 0.004$\\
A5 & $2.880 \pm 0.001$ & $1.904 \pm 0.001$ & $3.452 \pm 0.001$\\
A7 & $2.840 \pm 0.001$ & $1.901 \pm 0.001$ & $3.431 \pm 0.001$\\
A9 & $2.782 \pm 0.001$ & $1.893 \pm 0.001$ & $3.400 \pm 0.001$\\
F1 & $2.724 \pm 0.001$ & $1.885 \pm 0.001$ & $3.369 \pm 0.001$\\
F3 & $2.724 \pm 0.001$ & $1.885 \pm 0.001$ & $3.369 \pm 0.001$\\
F5 & $2.693 \pm 0.005$ & $1.883 \pm 0.001$ & $3.348 \pm 0.002$\\
F7 & $2.637 \pm 0.001$ & $1.874 \pm 0.001$ & $3.311 \pm 0.001$\\
F9 & $2.632 \pm 0.003$ & $1.873 \pm 0.001$ & $3.308 \pm 0.002$\\
G1 & $2.61 \pm 0.01$   & $1.869 \pm 0.001$ & $3.291 \pm 0.009$\\
G3 & $2.577 \pm 0.004$ & $1.865 \pm 0.001$ & $3.264 \pm 0.004$\\
G5 & $2.568 \pm 0.001$ & $1.864 \pm 0.001$ & $3.254 \pm 0.001$\\
G7 & $2.55 \pm 0.01$   & $1.864 \pm 0.001$ & $3.240 \pm 0.008$\\
G9 & $2.526 \pm 0.004$ & $1.863 \pm 0.001$ & $3.221 \pm 0.003$\\
K1 & $2.491 \pm 0.008$ & $1.858 \pm 0.001$ & $3.192 \pm 0.008$\\
K3 & $2.40 \pm 0.01$   & $1.843 \pm 0.002$ & $3.128 \pm 0.007$\\
K5 & $2.316 \pm 0.008$ & $1.829 \pm 0.001$ & $3.052 \pm 0.008$\\
K7 & $2.224 \pm 0.001$ & $1.803 \pm 0.001$ & $2.997 \pm 0.001$\\
K9 & $2.193 \pm 0.004$ & $1.786 \pm 0.002$ & $3.004 \pm 0.001$\\
M1 & $2.118 \pm 0.006$ & $1.755 \pm 0.002$ & $2.985 \pm 0.003$\\
M3 & $1.960 \pm 0.006$ & $1.699 \pm 0.002$ & $2.949 \pm 0.001$\\
M5 & $1.847 \pm 0.003$ & $1.654 \pm 0.001$ & $2.922 \pm 0.001$
\enddata
\tablecomments{See section~\ref{sec:ext} for more details.}
\end{deluxetable}

\clearpage
\pagebreak

\startlongtable
\tablewidth{0.985\textwidth}
\begin{deluxetable}{cccc}
\tablecolumns{4}
\tablecaption{\gaia\ de-reddening relations as a function of uncorrected $G - G_{\rm RP}$.\label{tab:ext2}}
\tablehead{\colhead{Uncorrected} & \colhead{$R(G)$} & \colhead{$R(G_{\rm RP})$} & \colhead{$R(G_{\rm BP})$}\\
\colhead{$G - G_{\rm RP}$} & \colhead{(mag)} & \colhead{(mag)} & \colhead{(mag)}}
\startdata
$-$0.18 & $1.938 \pm 0.001$ & $3.112 \pm 0.001$ & $3.670 \pm 0.001$ \\
$-$0.08 & $1.930 \pm 0.002$ & $3.05 \pm 0.02$   & $3.59 \pm 0.02$   \\
0.02 & $1.923 \pm 0.002$ & $2.99 \pm 0.01$    & $3.53 \pm 0.01$   \\
0.12 & $1.919 \pm 0.001$ & $2.961 \pm 0.009$  & $3.504 \pm 0.007$ \\
0.22 & $1.908 \pm 0.003$ & $2.90 \pm 0.02$    & $3.47 \pm 0.01$   \\
0.32 & $1.90 \pm 0.01$ & $2.80 \pm 0.08$      & $3.40 \pm 0.05$   \\
0.42 & $1.881 \pm 0.003$ & $2.68 \pm 0.03$    & $3.34 \pm 0.02$   \\
0.52 & $1.865 \pm 0.001$ & $2.570 \pm 0.005$  & $3.257 \pm 0.004$ \\
0.62 & $1.860 \pm 0.001$ & $2.507 \pm 0.006$  & $3.206 \pm 0.006$ \\
0.72 & $1.842 \pm 0.001$ & $2.397 \pm 0.009$  & $3.124 \pm 0.007$ \\
0.82 & $1.823 \pm 0.005$ & $2.30 \pm 0.02$    & $3.04 \pm 0.01$   \\
0.92 & $1.796 \pm 0.003$ & $2.211 \pm 0.005$  & $3.000 \pm 0.001$ \\
1.02 & $1.769 \pm 0.003$ & $2.157 \pm 0.006$  & $3.000 \pm 0.001$ \\
1.12 & $1.735 \pm 0.003$ & $2.057 \pm 0.008$  & $2.965 \pm 0.001$ \\
1.22 & $1.687 \pm 0.002$ & $1.930 \pm 0.004$  & $2.944 \pm 0.001$ \\
1.32 & $1.656 \pm 0.001$ & $1.852 \pm 0.003$  & $2.924 \pm 0.001$ \\
1.42 & $1.641 \pm 0.001$ & $1.818 \pm 0.001$  & $2.912 \pm 0.001$ 
\enddata
\tablecomments{See section~\ref{sec:ext} for more details.}
\end{deluxetable}

\section{DISCUSSION}\label{sec:discussion}

In this section, we discuss various properties of the \asso\ members and of their population as a whole. Photometric spectral type estimates and additional substellar candidates are discussed in Sections~\ref{sec:spt} and \ref{sec:substellar}. This is followed by an estimation of the isochronal age of \asso\ (Section~\ref{sec:isoage}) and a discussion of the cooling ages of the two hot white dwarf candidate members of \asso\ (Section~\ref{sec:wd2}). We discuss literature lithium absorption measurements for K- to G-type members of \asso\ in Section~\ref{sec:li}, and discuss the present-day mass function of \asso\ in Section~\ref{sec:imf}. The stellar activity of its members is assessed in Section~\ref{sec:act}. \asso\ is placed in context with the Galactic kinematic structure recently unveiled by \cite{2019AJ....158..122K} in Section~\ref{sec:kounkel}.

\subsection{Photometric Spectral Type Estimates}\label{sec:spt}

\begin{figure}
 	\centering
 	\includegraphics[width=0.465\textwidth]{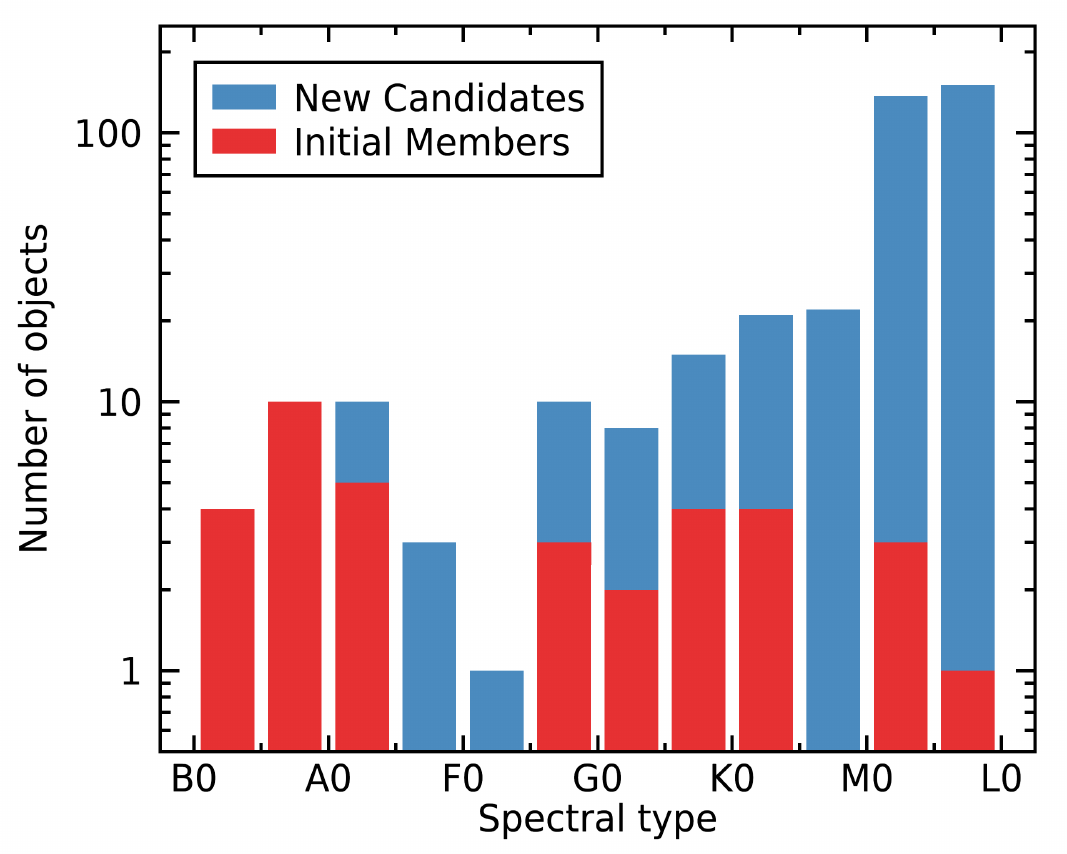}
 	\caption{Distribution of observed and estimated photometric spectral types for initial \asso\ members (red bars) and candidate members (blue bars). Data from \gaia\ allowed us to recover candidate members with photometric spectral types as late as M9. Two hot white dwarf candidates are excluded from this figure. See Section~\ref{sec:spt} for more details.}
 	\label{fig:spts}
\end{figure}

The extinction correction method described above directly provides photometric spectral type estimates for \asso\ candidates and members with no spectral type information in the literature. We used a slightly different method to estimate the photometric spectral types of objects near the substellar regime with near-infrared 2MASS--\emph{WISE} colors $J - W2 > 1.5$, corresponding to a spectral types $\simeq$\,M6 and later \citep{2015ApJS..219...33G}. For these redder objects, we used the spectral type to $J - W2$ relation of \cite{2015ApJS..219...33G} to determine a more accurate subtype given that the \gaia\ \gaiagr\ colors are more spread and based on lower-quality detections in these cases (e.g., see \citealt{2019MNRAS.485.4423S}). All photometric spectral type estimates are shown in Figure~\ref{fig:spts}.

\subsection{Substellar Objects}\label{sec:substellar}

\begin{figure}
 	\centering
 	\includegraphics[width=0.465\textwidth]{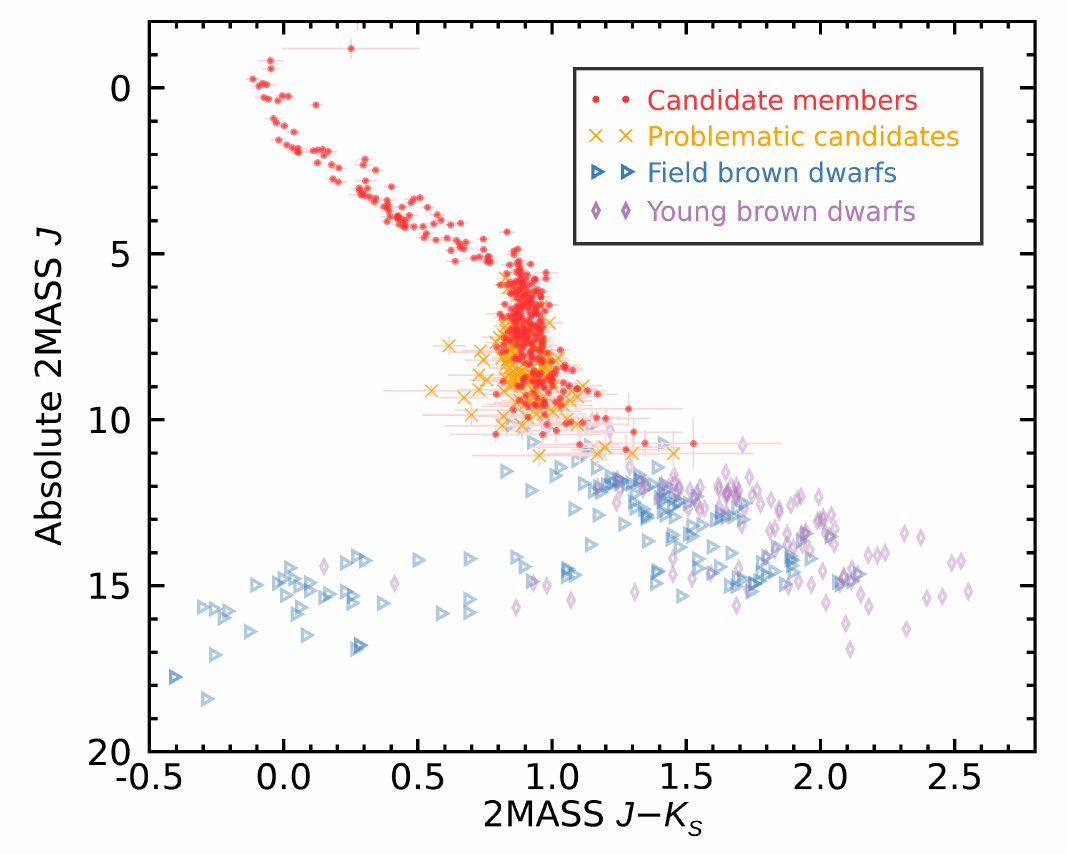}
 	\caption{Absolute 2MASS $J$-band magnitudes versus $J-K_{\rm S}$ colors for field (rightward blue triangles) and young (purple diamonds) brown dwarfs compared with all \asso\ candidates and members (filled red circles). The \asso\ candidates barely reach the sequence of young L-type brown dwarfs, and seem brighter or redder than the field brown dwarfs sequence, as expected for young objects. A fraction of the candidates with problematic \gaia\ colors (orange crosses) do not follow the \asso\ sequence, which is expected if their photometry is contaminated by background objects. Only spectral types L0 and later are shown for all brown dwarf data. See Section~\ref{sec:substellar} for more details.}
 	\label{fig:jk}
\end{figure}

\begin{figure}
 	\centering
 	\includegraphics[width=0.465\textwidth]{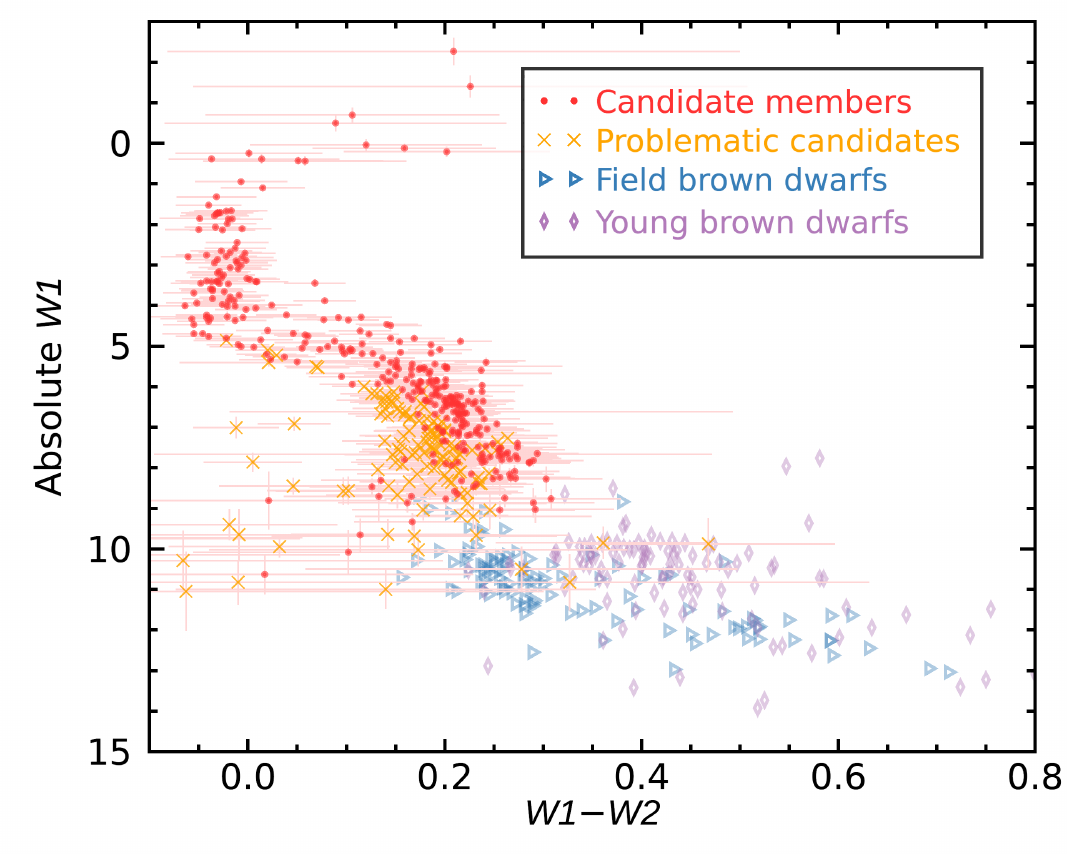}
 	\caption{Absolute \emph{WISE} $W1$-band magnitudes versus $W1-W2$ colors for field and youngbrown dwarfs compared with all \asso\ candidates and members. Color coding is the same as for Figure~\ref{fig:jk}. Only spectral types L0 and later are shown for brown dwarf data. See Section~\ref{sec:substellar} for more details.}
 	\label{fig:w1w2}
\end{figure}

In Figures~\ref{fig:jk} and \ref{fig:w1w2}, we show near-infrared color-magnitude sequences of \asso\ candidates based on 2MASS and \emph{WISE} photometry, compared with those of field-aged and young L-type or later low-mass stars and brown dwarfs. In both cases, the \asso\ sequence forms a prolongation of the young substellar sequences at brighter absolute magnitudes, and there is a small overlap indicating that a few \asso\ candidates discussed here may have spectral types as late as $\simeq$\,L0 (although at the age of \asso\ the substellar boundary is near spectral type M7; \citealp{2012RSPTA.370.2765A,2015AA...577A..42B,2015ApJ...810..158F}). \cite{2011ApJS..197...19K} devised a rejection criterion based on \emph{WISE} photometry to distinguish extragalactic sources from brown dwarfs, but our only \asso\ candidates with a sufficient $W3$-band detection were not red enough in $W1-W2$ color to apply the rejection criterion.

\subsection{Isochronal Age}\label{sec:isoage}

The locus of \asso\ candidates and members compiled in this work forms a sequence in color-magnitude space that sits between those of the Pleiades association ($112 \pm 5$\,Myr; \citealt{2015ApJ...813..108D}) and the Tucana-Horologium (see \citealp{2001ApJ...559..388Z,2000AJ....120.1410T}), Columba and Carina associations ($\simeq$\,45\,Myr; \citealp{2008hsf2.book..757T,2015MNRAS.454..593B}). We coss-matched all bona fide members of these four associations compiled by \cite{2018ApJ...856...23G} with \gaia\ for this comparison, and built an empirical isochrone for each of them by fitting their sequence with a high-order polynomial. The cross matches with \gaia\ were all inspected for spurious matches by building finder charts similar to those discussed in Section~\ref{sec:fcharts}. The color-magnitude positions of all members were corrected for extinction by interstellar dust with the method described in Section~\ref{sec:ext}. This procedure only had a noticeable but small effect on the Pleiades members.

All known unresolved binaries were removed from these lists, and their color-magnitude diagrams were visually inspected to remove the obvious sequence of unresolved binaries and triples that were shifted up by 0.75 and 1.19\,mag in \gaia\ $G$-band magnitude, respectively. The detailed lists of members used to build these isochrones will be presented in an upcoming publication, along with those of other nearby young associations.

Representing a young association's color-magnitude sequence with a polynomial curve can be complicated by the fact that they contain many more low-mass stars (e.g., \citealt{2010AJ....139.2679B}), which would cause an over-fitting of the data in the red part of the color-magnitude diagram. To avoid this, we first build a moving box average and standard deviation of the members' absolute \gaia\ $G$-band magnitudes in bins of 0.05\,mag in $G-G_{\rm RP}$ colors, and we subsequently fit a 11-order (Tucana-Horologium, Columba and Carina) or 15-order (Pleiades) polynomial, which were found to be appropriate given the number of stars and the range of colors occupied by the members of these associations. Columba, Tucana-Horologium and Carina were combined as a single $\simeq$\,45\,Myr-old population as they all share the same age \citep{2015MNRAS.454..593B}. This allowed us to build a more accurate empirical isochrone given the larger number of resulting members.

\begin{figure}
 	\centering
 	\includegraphics[width=0.465\textwidth]{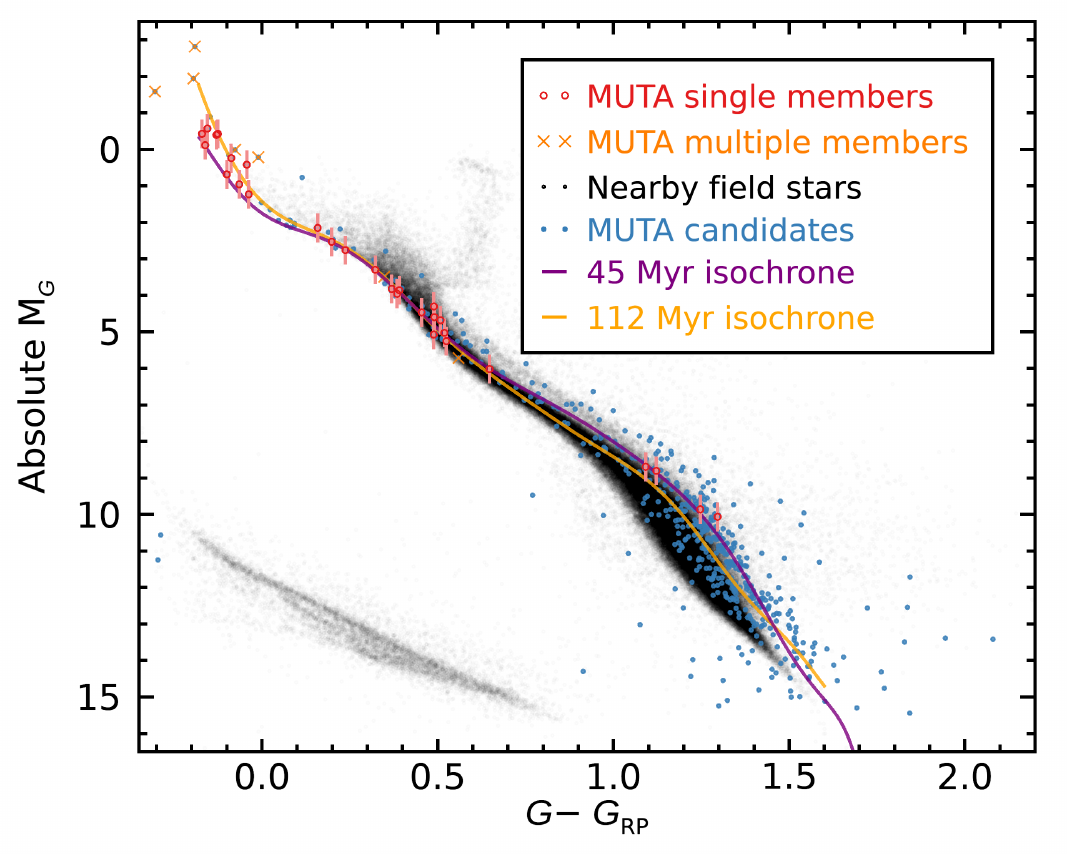}
 	\caption{\gaia\ color-magnitude diagram of \asso\ members used for isochrone fitting (red filled circles) and other candidates (blue filled circles) compared with field stars within 100\,pc of the Sun (black dots) and empirical isochrones built from the Pleiades associations (orange line) and a combination of the Tucana-Horologium, Carina and Columba associations (purple line). \asso\ objects flagged as potential unresolved or contaminated objects are identified with orange crosses. See Section~\ref{sec:isoage} for more detail.}
 	\label{fig:cmdbinrej}
\end{figure}

We used our initial list of \asso\ members (Table~\ref{tab:initialmembers}) to determine an isochronal age for the association, by comparing each member's absolute $G$-band magnitude with a hybrid isochrone built from a weighted sum of the $\simeq$\,45\,Myr and $\simeq$\,112\,Myr empirical isochrones described above. We assumed that the members are spread around the best-fitting hybrid isochrone along a Gaussian likelihood with a standard deviation of 0.35\,mag, typical of other young associations. Members that are either known binaries or have a \gaia\ RUWE above 1.4 were not used for this isochronal age determination. These latter objects are identified in Figure~\ref{fig:cmdbinrej}, along with the empirical isochrones built from the Pleiades and the Tucana-Horologium, Columba and Carina associations.

\begin{figure}
 	\centering
 	\includegraphics[width=0.465\textwidth]{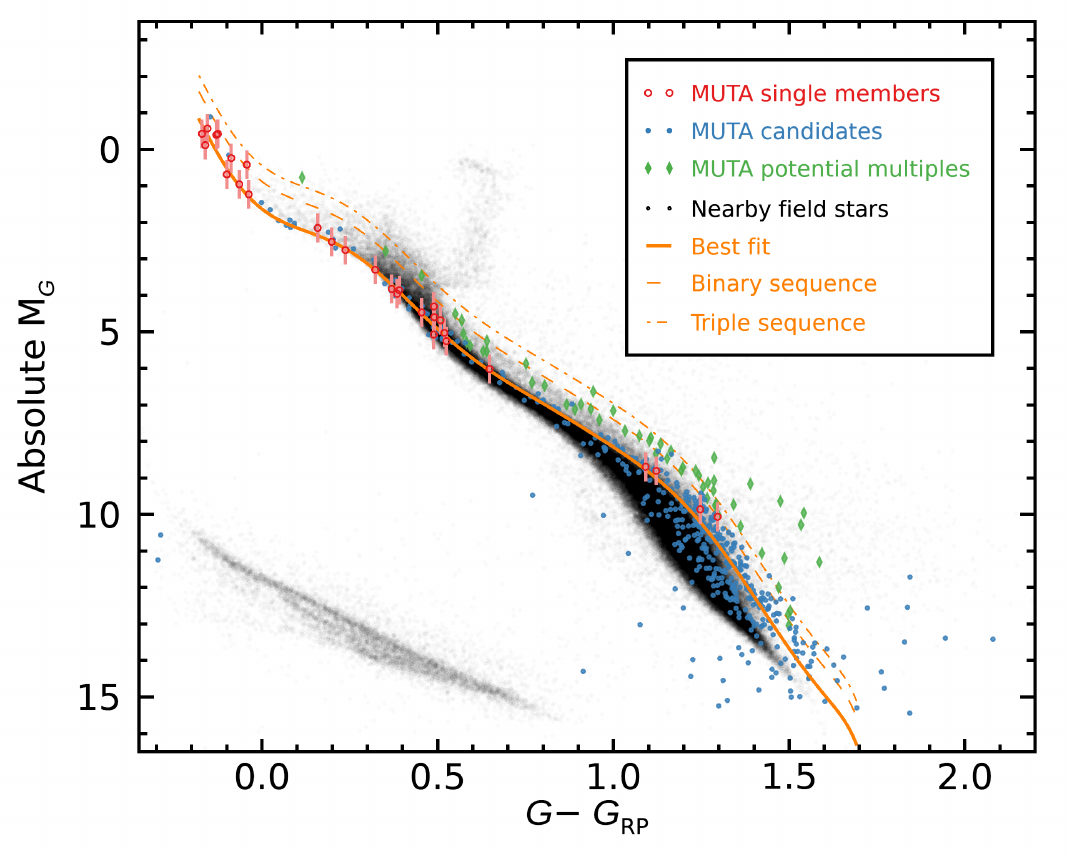}
 	\caption{\asso\ members used for isochrone fitting (red circles and error bars) fitted with a linear combination of empirical isochrones. The best fit, corresponding to an age of \isoage, is represented with an orange line. Similar isochrones shifted by 0.75\,mag and 1.19\,mag are also shown as orange dashed and dash-dotted lines, respectively, to represent the locations of unresolved equal-luminosity binaries and triples. Other candidate members of \asso\ are shown as blue circles, and those flagged as possible binaries are shown as green diamonds. See Section~\ref{sec:isoage} for more detail.}
 	\label{fig:isofit}
\end{figure}

\begin{figure}
 	\centering
 	\includegraphics[width=0.465\textwidth]{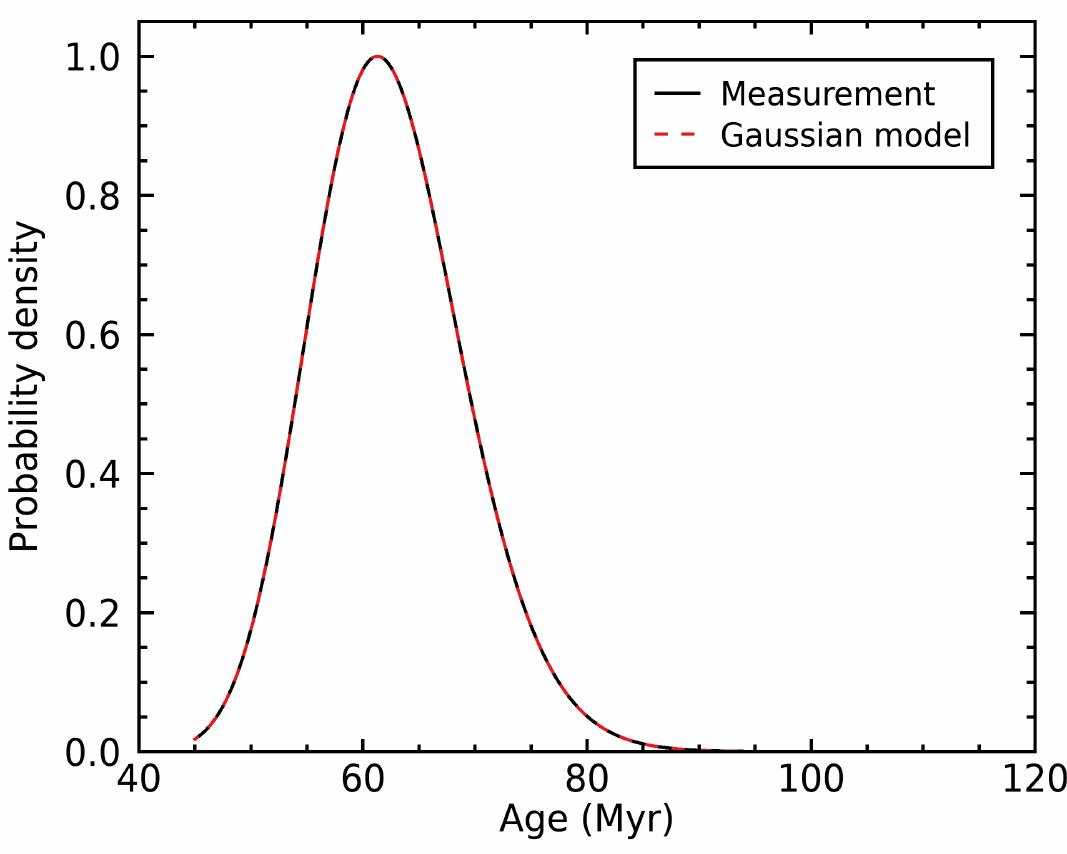}
 	\caption{Relative probability density function for the isochronal age of \asso\ determined from fitting a combination of empirical isochrones of nearby young associations (black line). A normal probability density function in logarithm age is also shown (red dashed line). The observed \asso\ age is well represented by a Gaussian distribution at \isoage.
 	See Section~\ref{sec:isoage} for more detail.}
 	\label{fig:age_pdf}
\end{figure}

\begin{figure}
 	\centering
 	\includegraphics[width=0.465\textwidth]{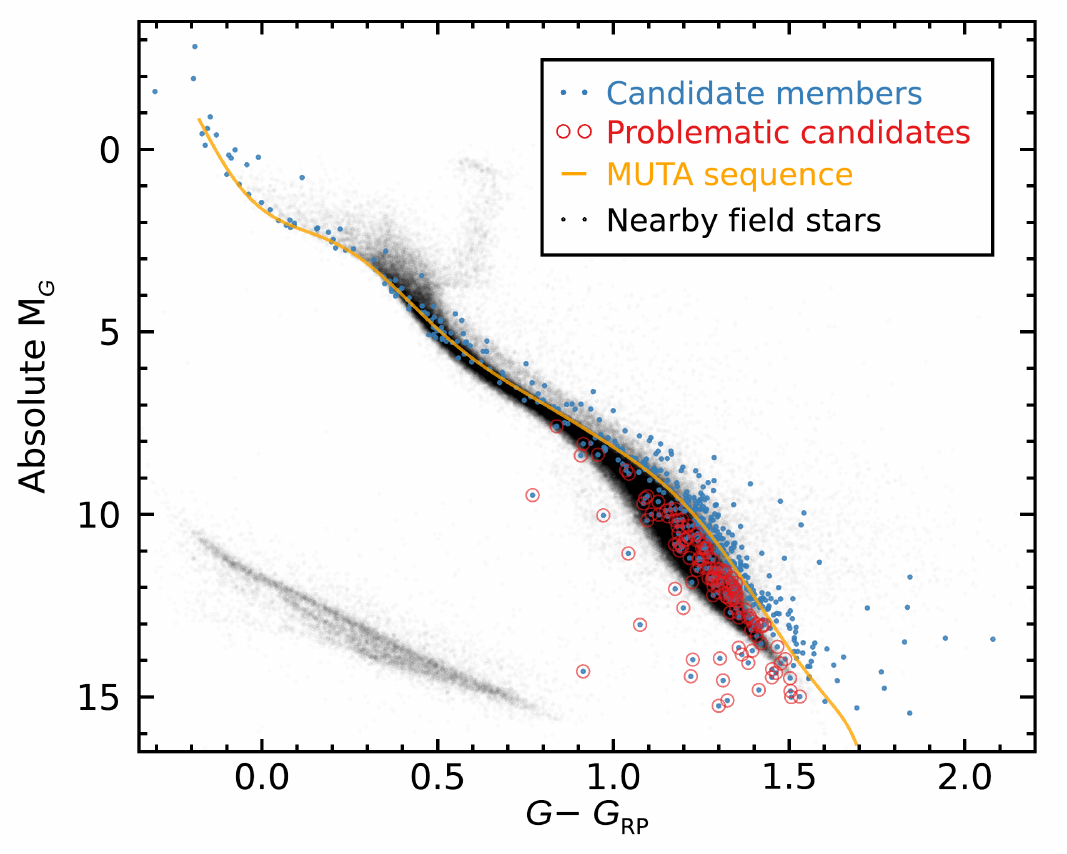}
 	\caption{\gaia\ color-magnitude diagram of \asso\ members and candidates (blue dots) compared with nearby \gaia\ entries (black dots) and the best-fitting hybrid isochrone for \asso\ members. We identified all candidates with an absolute magnitude more than 0.35\,mag fainter than this hybrid isochrone as problematic because they likely correspond to bad \gaia\ astrometric solutions, field-aged low-mass stars that have kinematics similar to \asso\ by chance, or to sources contaminated by a background object. See Section~\ref{sec:isoage} for more details.}
 	\label{fig:isocutoff}
\end{figure}

A one-dimensional grid search was performed to identify the linear combination of the $\simeq$\,45\,Myr and $\simeq$\,112\,Myr empirical isochrones that best matches the \asso\ stars. A thousand values for a linear coefficient $\alpha_i$ were chosen with $\alpha\in\left[0,1\right]$ to build a set of hybrid isochrones $I_i$ built from the $\simeq$\,45\,Myr isochrone $I_{\rm 45}$ and the $\simeq$\,112\,Myr isochrone $I_{\rm 112}$:

\begin{align}
	I_i = \alpha_i\cdot I_{\rm 45} + \left(1-\alpha_i\right)\cdot I_{\rm 112}.
\end{align}

The goodness-of-fit of each hybrid isochrone for the $10^3$ values of $\alpha_i$ were assessed by calculating the Gaussian likelihood that the \gaia\ absolute $G$-band magnitudes of \asso\ members $y_j$ and their associated standard deviations $\sigma_j$ match the model $I_{ij}$ in each color bin $j$:

\begin{align}
	\ln P_i = -0.5\cdot\sum_j \left(\frac{y_j-I_{ij}}{\sigma_j+0.35\,{\rm mag}}\right)^2.
\end{align}

The best-fitting linear combination is displayed in Figure~\ref{fig:isofit}. The ages $A_i$ corresponding to each hybrid isochrone $I_i$ were taken as a linear combination of the individual empirical isochrones in logarithm space:

\begin{align}
	\log A_i = \alpha_i\cdot \log\left({\rm 45\,Myr}\right) + \left(1-\alpha_i\right)\cdot\log\left({\rm 112\,Myr}\right).
\end{align}

The resulting probability density function $P\left(A_i\right)$ is shown in Figure~\ref{fig:age_pdf}. It is well represented by a Gaussian in logarithm of age, with an average and characteristic width that correspond to $\log A ({\rm yr}) =$ \assologage, or an age of \isoage.

We also calculated a probability density function for the relative age parameter $\alpha$ because the age estimates of both our reference populations could change in the future. For example, some recent lithium depletion boundary age estimates for the Pleiades are as old as $148 \pm 19$\,Myr \citep{2004ApJ...604..272B}, and \cite{2014AJ....147..146K} estimated a slightly younger age for Tucana-Horologium based on the lithium depletion boundary: they found ages of $38 \pm 2$\,Myr or $41 \pm 2$\,Myr, depending on the evolutionary models that they used. The age of \asso\ can thus be refined with the equation above (i.e., a simple interpolation in log age), replacing $\alpha_i$ with a Gaussian probability density function at $0.65 \pm 0.12$ for $\alpha$. Using the two extreme ends of these age estimates for the Pleiades and Tucana-Horologium would correspond to \asso\ ages of $55 \pm 7$\,Myr, or $69 \pm 10$\,Myr, placing two conservative boundaries for the possible age of \asso.

All \asso\ candidate members located more than 0.35\,mag fainter than the best-fitting hybrid isochrone were marked as problematic candidates because they likely correspond to interloping field-aged M dwarfs or contaminated \gaia\ entries. This flagging procedure is displayed in Figure~\ref{fig:isocutoff}. This step has removed 135 objects from our list of good-quality candidates; we note that this number is comparable to the number of contaminants ($192_{-22}^{+25}$) we have estimated in Section~\ref{sec:wds} based on the number of old white dwarf interlopers.

\subsection{White Dwarf Cooling Ages}\label{sec:wd2}

In Section~\ref{sec:wds}, we noted that our search for additional \asso\ candidates yielded \nwhitedwarfs\ white dwarfs seemingly co-moving with \asso, 10 of which are clearly too cold, and therefore too old, to be credible members. The only two exceptions are WD~0340+103 (MUTA~125) and WD~0350+098 (MUTA~190), which seem to be aged about 200-800\,Myr from a first comparison with total-age cooling tracks. However, both white dwarfs are so hot that a direct comparison of color-magnitude relations at visible wavelengths is imprecise, as this regime only samples the Rayleigh-Jeans end of their spectral energy distributions. Furthermore, the \gaia\ de-reddening procedure developed here cannot be applied to white dwarfs directly. For this reason, we investigated the properties of both white dwarfs in more details.

WD~0340+103 is an extremely hot white dwarf, which properties have been estimated at $\log g = 8.6$, \teff $= 42,617$\,K and a mass of 1.03\,\msol\ by \cite{2019MNRAS.482.4570G}. However, these properties were obtained by fitting models to the \gaia\ photometry of WD~0340+103, and the visible photometry of hot stars is relatively insensitive to their fundamental properties given that it only samples the Rayleigh-Jeans limit of their spectral energy distribution. For this reason, we obtained more reliable fundamental parameters by making use of spectroscopy instead of photometry.

\begin{figure}
 	\centering
 	\includegraphics[width=0.465\textwidth]{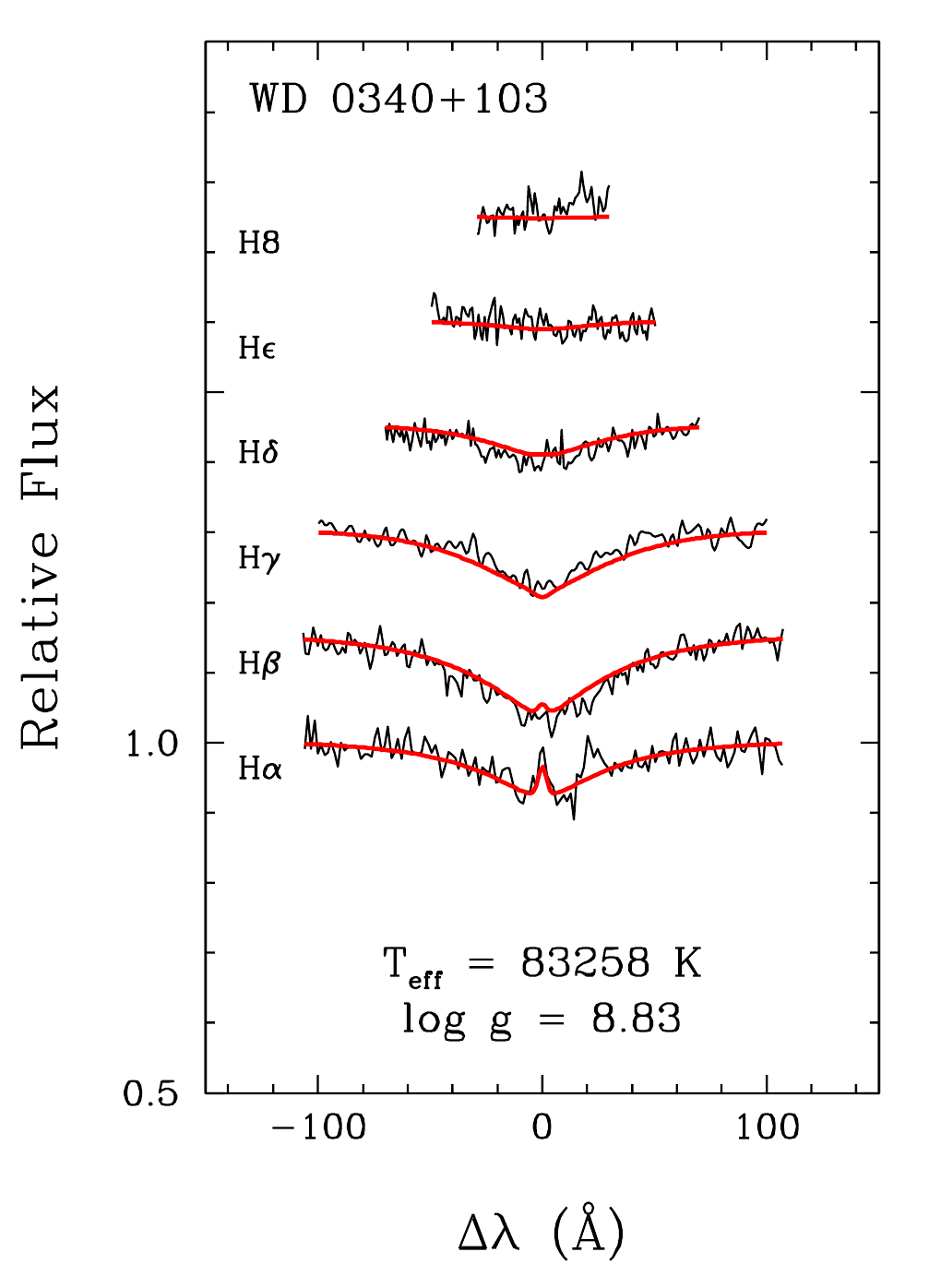}
 	\caption{Model fit to the Balmers lines of WD~0340+103 (MUTA~125). See Section~\ref{sec:wd2} for more details.}
 	\label{fig:fitspec}
\end{figure}

We first determined the effective temperature and surface gravity of WD~0340+103 by fitting its SDSS optical spectrum \citep{2012ApJS..203...21A} with the grid of non-local thermodynamic equilibrium atmosphere models of A. B\'edard (2020, in preparation). This yielded a very hot temperature of $83,000 \pm 2,000$\,K, and $\log g = 8.83 \pm 0.08$. Because WD~0340+103 only exhibits hydrogen features given its DA spectral type, we assumed a pure-hydrogen atmospheric composition. We used the fitting procedure described in \citet{1992ApJ...394..228B} and \citet{2005ApJS..156...47L}: briefly, the normalized Balmer lines are adjusted with theoretical line profiles using the Levenberg-Marquardt least-squares method. The observed spectrum of WD~0340+103 was well reproduced by this method, including the emission component at the core of the H$\alpha$ line, and as illustrated in Figure~\ref{fig:fitspec}. The positions of the lower Balmer lines (H$\alpha$, H$\beta$, and H$\gamma$) were used to measure a total redshift of $138 \pm 21$\,\kms, due in part to the gravitational redshift and radial velocity of WD~0340+103.

In a second step, we calculated the mass, radius, luminosity, and cooling age that correspond to the effective temperature and surface gravity of WD~0340+103 using the thick-hydrogen layer ($M_{\rm H}/M=10^{-4}$) cooling tracks of A.~B\'edard et al. (2020, in preparation), which are appropriate for the study of hot white dwarfs. Following \citet{2006AJ....132.1221H}, we also computed the absolute SDSS $g$-band magnitude, which we combined with the observed (dereddened) SDSS $g$-band magnitude to evaluate its spectroscopic distance. The atmospheric and stellar parameters of WD~0340+103 are summarized in Table~\ref{tab:wd0340}. Our analysis shows that WD~0340+103 is a highly unusual white dwarf: It is extremely hot, young, and massive. Furthermore, we note that the spectroscopic distance is slightly farther than its \gaia\ trigonometric distance, but the values are consistent within measurement errors.

We used the MESA Isochrones and Stellar Tracks (MIST; \citealt{2016ApJ...823..102C}) to estimate a progenitor mass of $6.7 \pm 0.4$\,\msol\ for WD~0340+103. This corresponds to a spectral type of about B2, just one subclass earlier than the earliest-type members of \asso\ (29~Tau, 30~Tau, $\mu$~Tau and $\mu$~Eri are all B3 stars). This is consistent with the extremely young cooling age of only $270,000 \pm 30,000$~years which we derived for WD~0340+103. Such a progenitor star has a main-sequence lifetime of $59_{-6}^{+8}$\,Myr, corresponding to a total age of \wdage, consistent with our isochronal age of \isoage. Combining both estimates in an error-weighted average allows us to refine our age estimate for \asso\ at \assoage. The core composition of this massive white dwarf likely does not consist of carbon and oxygen, but rather oxygen and neon \citep{2018MNRAS.480.1547L,2019AA...625A..87C}. This is expected to have a significant effect on the calculated cooling age of about 20\% (e.g., see \citealp{2018ApJ...861L..13G,2015ASPC..493..137S,ameliephd}), however, in the present scenario the age estimate of WD~0340+103 is completely dominated by its main-sequence lifetime, and its core composition will therefore not have any significant effect on our total age estimation.

The detailed properties of WD~0350+098 are harder to determine because of its lack of spectral lines, likely due to extreme Zeeman broadening caused by a strong magnetic field. Much like WD~0340+103, the age estimate based on \gaia\ photometry alone may be unreliable given its extremely blue colors and hot temperature. Adding UV photometry from \emph{GALEX} \citep{2005ApJ...619L...1M} to better constrain its temperature yielded an estimate of $31,000 \pm 1,000$\,K with a radius of $0.0073_{-0.0005}^{+0.0006}$\,\rsol, however, these uncertainties are likely underestimated because the models we used do not include magnetic fields. These parameters would correspond to a mass of $1.09_{-0.05}^{+0.04}$\,\msol\ and a surface gravity of $\log g = 8.75 \pm 0.09$. Using non-magnetic cooling tracks yields a cooling age estimate of $79_{-10}^{+20}$\,Myr. The main-sequence lifetime that corresponds to the $6.1 \pm 0.5$\,\msol\ progenitor is $74_{-12}^{+15}$\,Myr, making WD~0350+098 too old for \asso\ membership if we take our analysis at face value. However, the lack of magnetic fields in our treatment could have introduced a significant bias in the determination of its cooling age and mass (and therefore its main-sequence lifetime), and for this reason we keep it as a candidate member of \asso.

\begin{deluxetable}{lcc}
\renewcommand\arraystretch{0.9}
\tabletypesize{\small}
\tablecaption{Properties of WD~0340+103 (MUTA~125) \label{tab:wd0340}}
\tablehead{\colhead{Property} & \colhead{Value} & \colhead{Ref.}}
\vspace{-0.2cm}\startdata
\sidehead{\textbf{Position and Kinematics}\vspace{-0.1cm}}
\emph{Gaia}~DR2 Source ID & 36321786805002880 & 1\\
R.A. ep. 2015.5\tablenotemark{a} & 03:43:14.370 $\pm 0.09$ & 1\\
Decl. ep. 2015.5\tablenotemark{a}  & +10:29:38.15 $\pm 0.06$ & 1\\
$\mu_\alpha\cos\delta$ (\masyr) & $31.51 \pm 0.18$ & 1\\
$\mu_\delta$ (\masyr) & $-22.55 \pm 0.12$ & 1\\
Parallax (mas) & $6.8 \pm 0.1$ & 1\\
Trigonometric distance (pc) & $145.7 \pm 2.3$ & 1\\
Spectroscopic distance (pc) & $163.4_{-15}^{+16}$ & 2\\
RV$_{\rm opt}$\tablenotemark{b} (\kms) & $14.3 \pm 3.4$ & 2\\
RV$_{\rm mes}$ (\kms) & $27 \pm 21$ & 2\\
\sidehead{\textbf{Photometric Properties}\vspace{-0.1cm}}
$G_{\rm BP}$ (\emph{Gaia}~DR2) & $16.307 \pm 0.009$ & 1\\
$G$ (\emph{Gaia}~DR2) & $16.539 \pm 0.001$ & 1\\
$G_{\rm RP}$ (\emph{Gaia}~DR2) & $16.766 \pm 0.005$ & 1\\
$u_{\rm AB}$ (SDSS~DR12) & $15.946 \pm 0.005$ & 3\\
$g_{\rm AB}$ (SDSS~DR12) & $16.298 \pm 0.003$ & 3\\
$r_{\rm AB}$ (SDSS~DR12) & $16.748 \pm 0.004$ & 3\\
$i_{\rm AB}$ (SDSS~DR12) & $17.090 \pm 0.005$ & 3\\
$z_{\rm AB}$ (SDSS~DR12) & $17.392 \pm 0.016$ & 3\\
\sidehead{\textbf{Fundamental Properties}\vspace{-0.1cm}}
Spectral type & DA & 4\\
\teff\ (K) & $83,000 \pm 2,000$ & 2\\
$\log g$ & $8.83 \pm 0.08$ & 2\\
Mass (\msol) & $1.16 \pm 0.04$ & 2\\
Radius (\rsol) & $0.0069_{-0.0005}^{+0.0006}$ & 2\\
$\log{L/L_\odot}$ & $0.31 \pm 0.08$ & 2\\
Cooling age (Myr) & $0.27 \pm 0.03$ & 2\\
Progenitor mass (\msol) & $6.7 \pm 0.4$ & 2\\
Progenitor spectral type & B2 & 2\\
Total age (Myr) & $60_{-6}^{+8}$ & 2\\
\enddata
\tablenotetext{a}{J2000 position at epoch 2015.5 from the \emph{Gaia}~DR2 catalog. Measurement errors are given in units of milliarcseconds.}
\tablenotetext{b}{Optimal radial velocity predicted by BANYAN~$\Sigma$ that assumes membership in \asso.}
\tablerefs{(1)~\citealt{GaiaCollaboration:2018io}, (2)~This work, (3)~\citealt{2015ApJS..219...12A}, (4)~\citealt{2013ApJS..204....5K}.
}
\end{deluxetable}

\subsection{Lithium}\label{sec:li}

The equivalent width of the \ion{Li}{1}~$\lambda$6708\,\AA\ spectral line is a well-established age indicator. Because lithium burns at lower temperatures than hydrogen, it is relatively fragile and will disappear over time if it is allowed to be transported in layers deep enough in a star to reach the threshold temperature for lithium burning. The temperature profile of a star, combined with the location of its convective layers, will determine whether lithium gets burned at all, and how fast it does so. Lower-mass stars (late-K or early-M spectral types) have deep convective layers that allow them to burn through all lithium within only $\simeq$\,30\,Myr \citep{2001AA...377..512R}, whereas higher-mass stars, with their shallower convective layers, burn lithium more gradually. It takes more than a billion years for stars with spectral types G0 and earlier to burn lithium in their photospheres such that the \ion{Li}{1}~$\lambda$6708\,\AA\ absorption line disappears completely \citep{1999AJ....117..330J}. As a result, the sequence in temperature versus \ion{Li}{1} absorption line for K-type or earlier stars evolves slowly with time, and makes it possible to place weak constraints on the ages of such early-type stars (e.g., \citealp{2001AA...371..652B,1993AJ....106.1059S}). Similarly, the K-type lithium depletion boundary, where stars below a given temperature stop displaying the lithium absorption line, can be used to place constraints on the age of a stellar population. The location of this boundary is, however, not very sensitive to age for populations $\simeq$\,10\,Myr and older \citep{2014AJ....147..146K}.

Brown dwarfs with masses below $\simeq$\,60\,\mjup\ do not burn lithium despite their fully convective structure, because they do not reach temperatures sufficient to do so even at their core (e.g., \citealt{2015AA...577A..42B}). Low-mass stars and brown dwarfs with masses above 60\,\mjup\ burn their photosphere lithium slowly, causing the appearance of a second, age-dependent boundary where the lithium absorption line begins appearing again below a threshold in effective temperature. The effective temperatures, spectral types and bolometric luminosities at which this second, M-type lithium depletion boundary occurs, is a strong function of age over the first hundreds of millions of years that follow stellar formation. The lithium depletion boundary has therefore become a popular diagnostic tool to determine precise ages for stellar populations with known M-type stars (e.g., \citealp{2014AJ....147..146K,2014ApJ...792...37M,2017AJ....154...69S}).

Measuring the equivalent width of the lithium absorption line accurately requires high-resolution spectroscopy, ideally with a resolving power $\lambda/\Delta\lambda$\,$>$\,10,000 to avoid contamination from otherwise blended spectral lines such as \ion{Fe}{1} \citep{2010ApJ...723.1542X}. Such measurements require long exposure times and they have thus typically only been obtained for known populations of nearby associations or open clusters. However, a literature search revealed that \ion{Li}{1} equivalent width measurements have been obtained by \cite{1997AAS..124..449M} for nine members or candidate members (and one low-likelihood candidate) of \asso\ in a follow-up of \emph{ROSAT} X-ray bright sources \citep{1995AA...295L...5N} in the vicinity of Taurus-Auriga. These measurements were obtained at a relatively low resolving power ($\lambda/\Delta\lambda \simeq$\,8,400),\footnote{\cite{1997AAS..124..449M} also obtained measurements at $\lambda/\Delta\lambda \simeq$\,4200, but inspecting the Isaac Newton Group Archive at \url{http://casu.ast.cam.ac.uk/casuadc/ingarch/query} indicated that none of these lower-resolution observations have been obtained for \asso\ objects.} meaning that the equivalent widths may be slightly overestimated because of line blending. We obtained effective temperatures for these ten stars from \cite{2010ApJ...723.1542X}, \cite{2018AA...616A...1G} and \cite{2019AJ....158...93B}, where available, listed in Table~\ref{tab:li} along with the lithium equivalent width measurements of \cite{1997AAS..124..449M}.

\begin{figure*}
	\centering
	\subfigure[20--25\,Myr]{\includegraphics[width=0.49\textwidth]{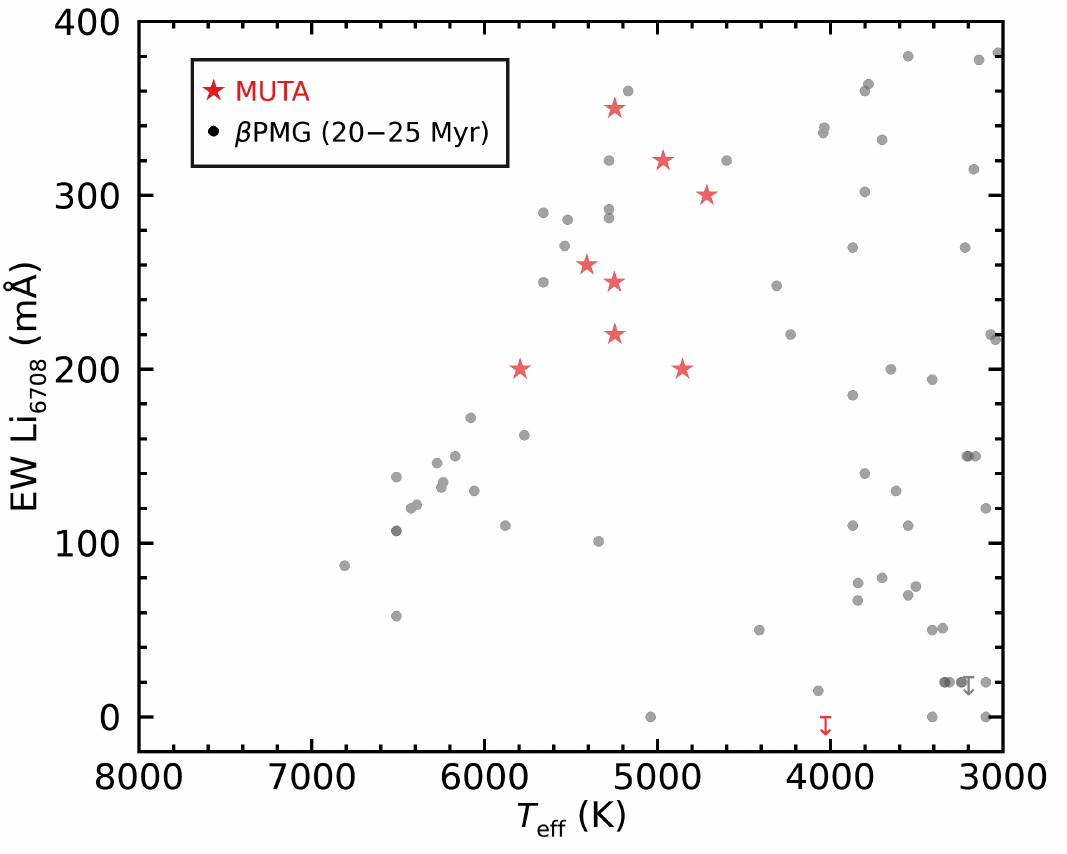}\label{fig:li_20_25}}
	\subfigure[45--50\,Myr]{\includegraphics[width=0.49\textwidth]{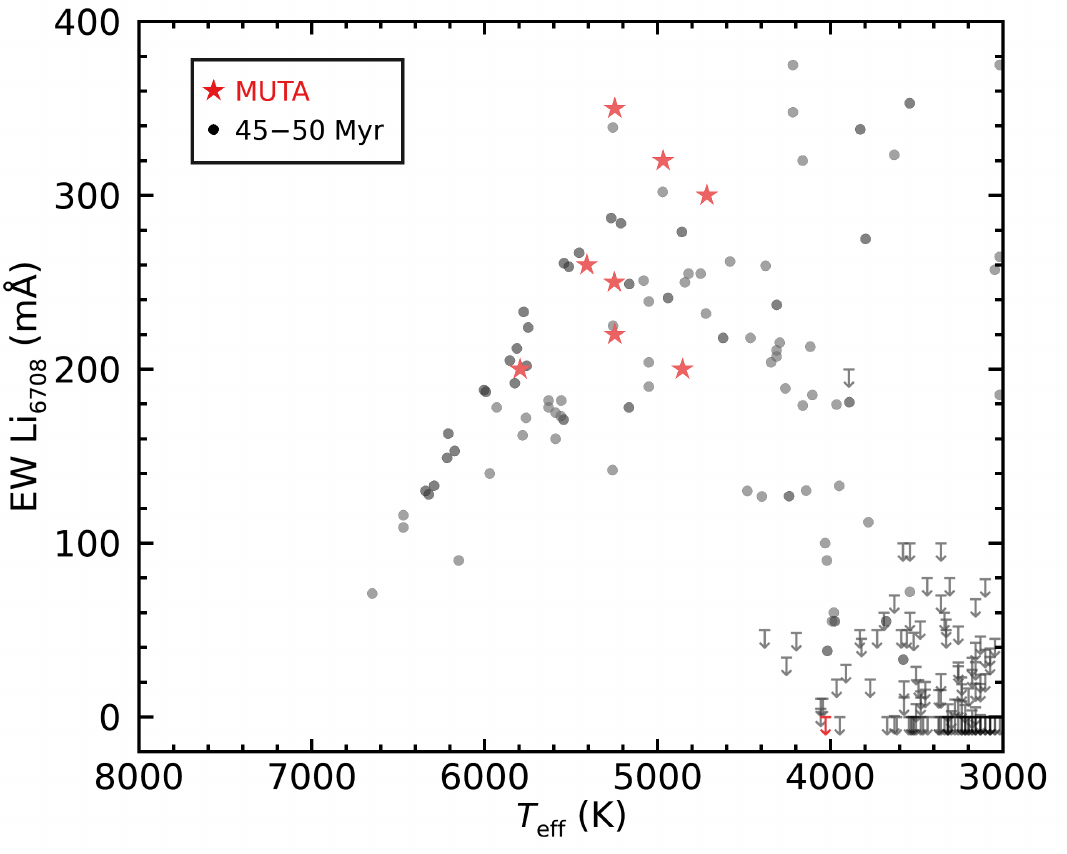}\label{fig:li_45_50}}
	\subfigure[110--125\,Myr]{\includegraphics[width=0.49\textwidth]{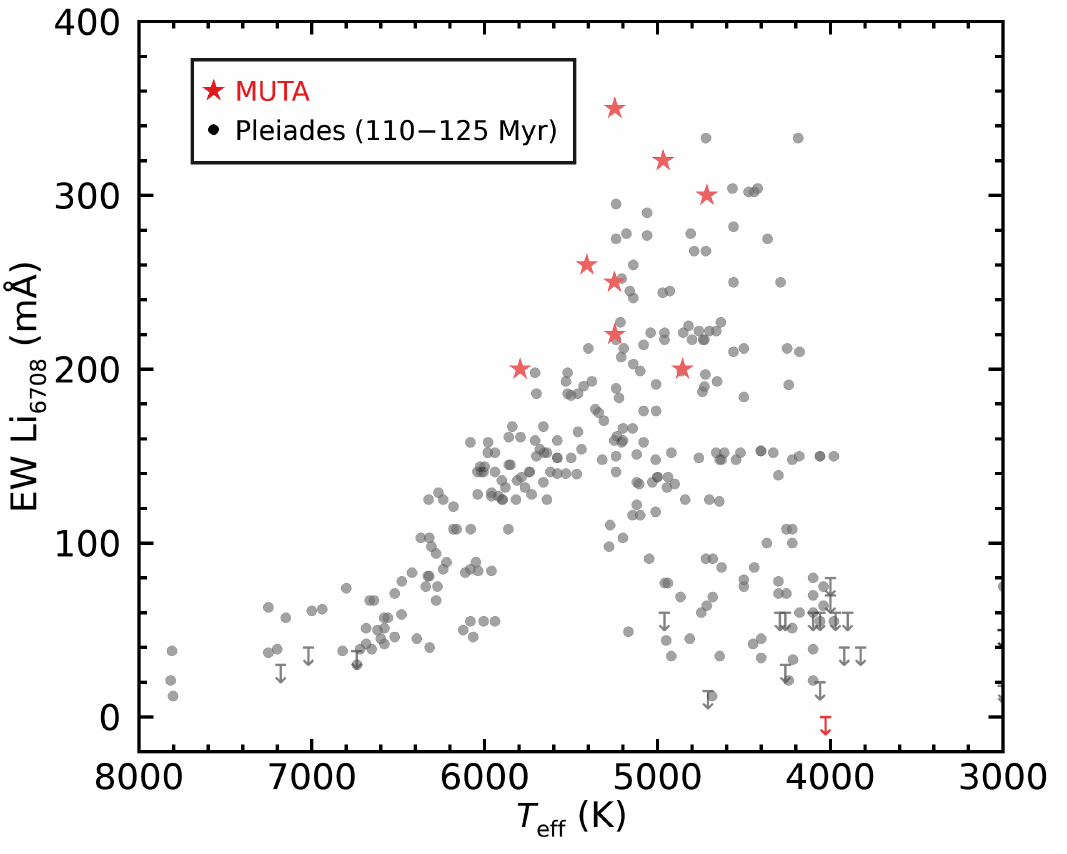}\label{fig:li_110_125}}
	\subfigure[150--175\,Myr]{\includegraphics[width=0.49\textwidth]{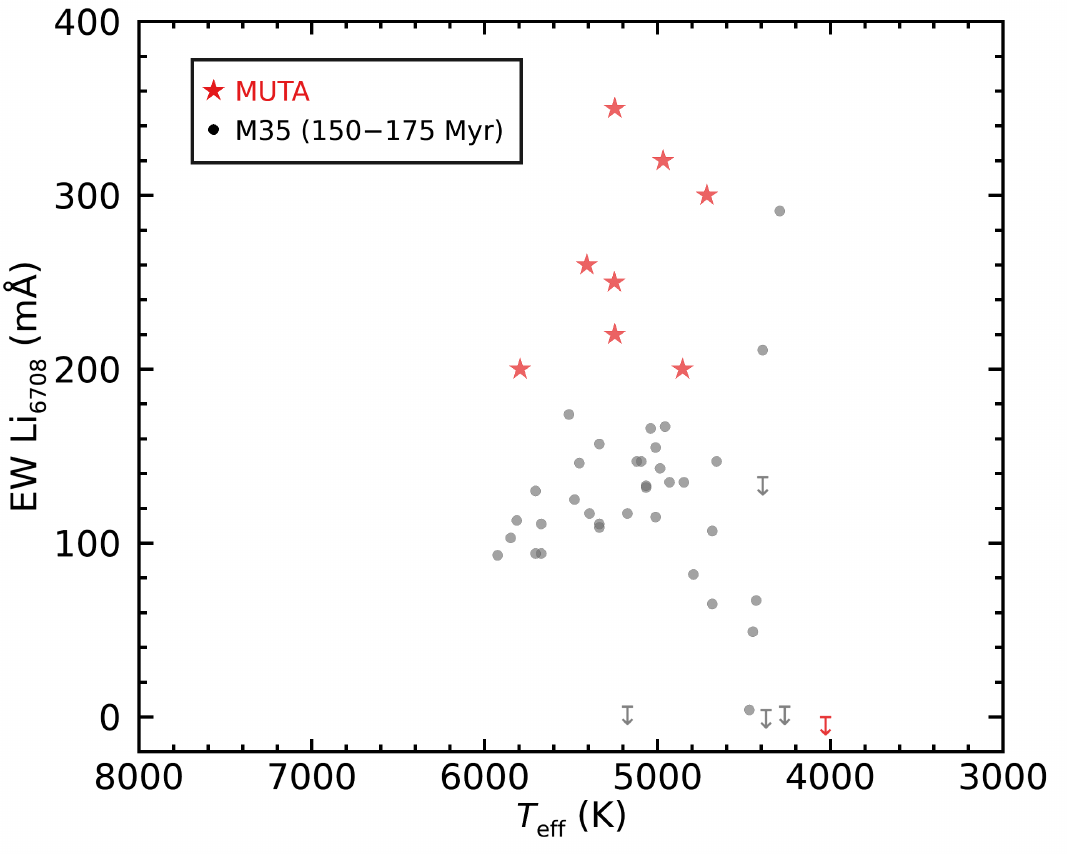}\label{fig:li_150_175}}
	\caption{Effective temperature versus the equivalent width of the \ion{Li}{1}~$\lambda$6708\,\AA\ absorption line for \asso\ members and candidates (red stars), compared with other known, coeval populations (grey circles). The 45--50\,Myr sequence was built from members of the Tucana-Horologium association and IC~2602 and IC~2391 open clusters. Upper limits are indicated with downward arrows. Although all measurements for \asso\ members are based on a lower resolving power ($\lambda/\Delta\lambda$\,$\approx$\,8,400) compared with the reference sequences ($\lambda/\Delta\lambda$\,$>$\,10,000), they indicate that \asso\ seems roughly consistent with an age of 20--125\,Myr. $\beta$PMG indicates the $\beta$~Pictoris moving group. See Section~\ref{sec:li} for more detail.}
	\label{fig:liall}
\end{figure*}

In Figure~\ref{fig:liall}, we compare these available \asso\ temperature versus lithium measurements with other literature data for stellar populations across a range of ages. The 20--25\,Myr sequence was built from the $\beta$~Pictoris moving group ($\beta$PMG, e.g., see \citealp{2001ApJ...562L..87Z,2004ARAA..42..685Z,2015MNRAS.454..593B}, measurements are from \citealp{2008ApJ...689.1127M,2014ApJ...792...37M,2017AJ....154...69S}). The 40--50\,Myr sequence was built from the stellar populations of the Tucana-Horologium association discussed earlier (lithium equivalent width measurements are by \citealt{2014AJ....147..146K}) and the IC~2602 and IC~2391 open clusters \citep{2001AA...377..512R,2004ApJ...614..386B,2010MNRAS.409.1002D}. The 110--125\,Myr sequence was built from the Pleiades association \citep{1993AJ....106.1059S,1996AJ....112..186J,2018AA...613A..63B}, and the 150--175\,Myr was built from the M35 open cluster \citep{2001AA...371..652B,2015AA...575A.120B}.

\startlongtable
\tablewidth{0.985\textwidth}
\begin{deluxetable*}{lllccl}
\tablecolumns{6}
\tablecaption{Lithium equivalent width measurements for \asso\ .\label{tab:li}}
\tablehead{\colhead{\asso} &\colhead{Common} & \colhead{ROSAT} & \colhead{EW(Li)} & \colhead{\teff} & \colhead{\teff}\\
\colhead{ID} & \colhead{Name} & \colhead{Name} & \colhead{(m\AA)} & \colhead{(K)} & \colhead{Ref.} }
\startdata
24 & RX~J0348.5+0832 & RX J0348.5+0832 & 260 & 5409 & 2\\
27 & RX~J0338.3+1020 & RX J0338.3+1020 & 250 & 5250 & 2\\
29 & RX~J0358.2+0932 & RX J0358.1+0932 & 200 & 4855 & 1\\
94 & V1267~Tau & RX J0333.1+1036 & 320 & 4967 & 1\\
159 & TYC~71--542--1 & RX J0347.9+0616 & 200 & 5794 & 2\\
195 & 2MASS~J03545074+1232061 & RX J0354.8+1232 & 0   & 4028 & 3\\
318~A & 2MASS~J04341953+0226260 & RX J0434.3+0226 & 300 & 4714 & 1\\
350 & TYC~91--702--1 & RX J0442.9+0400 & 220 & 5247 & 2\\
362 & V1831~Ori & RX J0450.0+0151 & 350 & 5247 & 1\\
376 & TYC~665--150--1 & RX J0357.3+1258 & 250 & 5943 & 2
\enddata
\tablecomments{All lithium equivalent width measurements are from \cite{1997AAS..124..449M}. TYC~665--150--1 was excluded from Figure~\ref{fig:liall} because it is a low-likelihood candidate member of \asso\ (its separation from the \asso\ model in $UVW$ space is 8.1\,\kms). See section~\ref{sec:li} for more details.}
\tablerefs{(1)~\citealt{2019AJ....158...93B}; (2)~\citealt{2010ApJ...723.1542X}; (3)~\citealt{2018AA...616A...1G}.}
\end{deluxetable*}

Although the available \asso\ measurements do not span either of the lithium depletion boundaries, they seem consistent with an age in the range 20--125\,Myr, with the caveat that our comparison sequences were built from higher-resolution spectra compared with \asso\ measurements. This likely biases our range slightly towards young ages, but this result seems consistent with our previous age assessments based on empirical isochrones and white dwarf cooling ages. Obtaining higher-resolution optical spectra for \asso\ members, as well as extending the range of spectral types over which lithium equivalent widths are measured, will allow us to further constrain the age of \asso.

\subsection{Present-Day Mass Function}\label{sec:imf}

We used the empirically corrected MIST solar-metallicity model isochrones of \cite{2016ApJ...823..102C} as described by \cite{2018ApJ...865..136G}\footnote{We used the models based on the revised \emph{Gaia}~DR2 photometric zero points of \cite{2018arXiv180409368E} available at \url{http://waps.cfa.harvard.edu/MIST/model_grids.html}} with a nominal stellar rotation of $v/v_{\rm crit} = 0$ to estimate the masses of \asso\ members and candidates based on their position in a \gaia\ absolute $G$ versus \gaiagr\ color-magnitude diagram. This method uses the differences between the empirical Pleiades sequence and the 112\,Myr MIST isochrone to correct for systematic effects such as the increased stellar activity and strong magnetic fields of low-mass stars.

The masses for \asso\ candidates with very red colors ($J - W2 > 1.5$) were estimated with the method of \cite{2014ApJ...783..121G}, which is more reliable than extrapolating MIST isochrones or using lower quality \gaia\ photometry, but potentially suffers from different systematics. The method is based on a comparison of the absolute 2MASS $J$, $H$, $K_{\rm S}$ and \emph{WISE} $W1$ and $W2$ photometry of \asso\ candidates with BT-Settl models \citep{2012RSPTA.370.2765A} in the same respective bandpasses, and combining the individual estimates in a likelihood analysis. These model-dependent mass estimates range from $\simeq$\,35\,\mjup\ to 0.2\,\msol, covering the substellar-to-stellar transition and overlapping slightly with the range of masses (0.1--6.0\,\msol) obtained with MIST isochrones for bluer targets.

\begin{figure}
 	\centering
 	\includegraphics[width=0.465\textwidth]{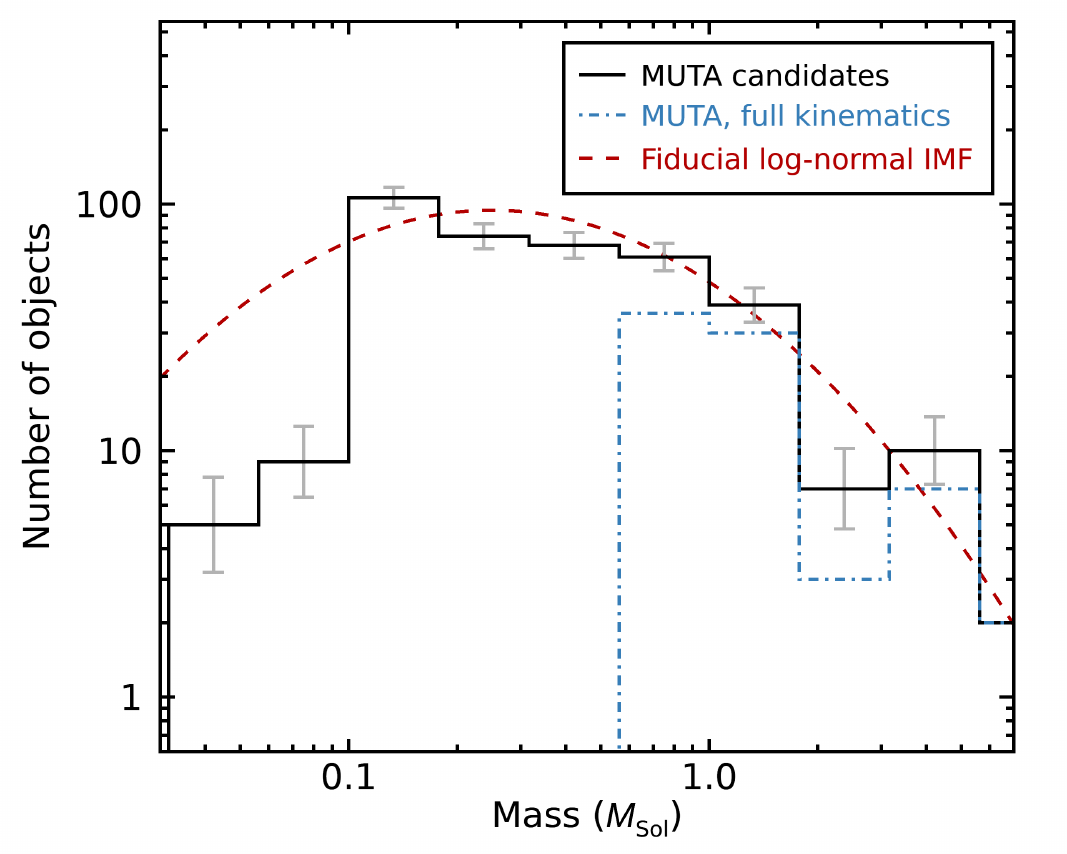}
 	\caption{Present-day mass function of \asso\ (thick black bars) compared with a fiducial log-normal initial mass function with a peak mass 0.25\,\msol\ and a logarithm characteristic width of 0.5, anchored on the $>$\,0.2\,\msol\ population of \asso. Gray error bars represent uncertainties associated with Poisson statistics. The subset of members with full kinematics and therefore a more reliable membership are shown with a dash-dotted blue line. Our set of candidates is consistent with a complete population down to $\simeq$\,0.1\msol\ if a log-normal mass function is realistic for \asso, but the brown dwarfs population still seems mostly incomplete. We did not include the progenitor masses of the two white dwarf candidates in this figure. See Section~\ref{sec:imf} for more details.}
 	\label{fig:imf}
\end{figure}

The resulting present-day mass function of \asso\ members and candidates is displayed in Figure~\ref{fig:imf} along with a fiducial log-normal mass function ($\sigma = 0.5$\,dex, $m_c = 0.25$\,\msol). We fitted its amplitude to our \asso\ members with masses above 1\,\msol, but the width and central position were not fitted. This particular mass function was shown to be a good fit to other nearby young associations by \cite{2012EAS....57...45J}. The log-normal mass function is a good match to our distribution of \asso\ members and candidates down to 0.1\,\msol, indicating that its present-day mass function may be similar to other young associations of the Solar neighborhood. Assuming that the population of \asso\ is complete above 0.2\,\msol\ indicates that about 65 brown dwarf members would remain to be found, for a total stellar and substellar population of $\simeq$\,450 members.

\subsection{Stellar Rotation and Activity}\label{sec:act}

Young stars lose angular momentum as they age, and their rotation periods consequently slow down with time. Because the rate of angular momentum loss depends on the rotation period, members of stellar associations with a wide range of rotation periods will eventually converge to a tight sequence as a function of their mass \citep{Barnes2003,vanSaders2016}. The timescale for this convergence for Sun-like stars is $<650$\,Myr and decreases with increasing stellar mass \citep{Delorme2011,2016ApJ...822...47D,2019AJ....158...77C}, but a partial sequence is apparent even at $\simeq$\,112\,Myr for higher-mass stars \citep{Rebull2016}.

Depending on the mass, this trend of longer rotation periods for older ages {\it reverses} for the youngest (pre-main-sequence) stars, as they spin up while contracting onto the main sequence. The youngest stars therefore also have longer rotation periods. The scatter at these younger ages is also larger because of a large spread in the initial rotation periods. Thus, the rotation period versus color sequence of \asso\ can still be used as an additional test of our assigned age by comparing with similarly-aged groups.

As bounds for the expected age of \asso, we used members the Pleiades \citep[$\simeq$112\,Myr;][]{2015ApJ...813..108D} and Praesepe clusters ($\simeq$\,800\,Myr; \citealt{2015ApJ...807...24B}), in addition to members of the Columba, Carina, and Tucana-Horologium associations discussed earlier ($\simeq$45\,Myr). We included the older Praesepe cluster as an example of a clearly older population in the color-rotation period diagram, because the differences between \asso\ and the Pleiades are subtle. We collected the rotation period measurements of the Pleiades and Praesepe members from \citet{Rebull2016} and \citet{2017ApJ...842...83D}, respectively. We obtained light curves for each member of the younger three associations from the {\it TESS} or {\it K2} missions, where available. We restricted our sample to targets with \gaia\ $G-G_{\rm RP}>0.2$, as the variability period in bluer stars may be impacted by pulsations as much as rotation. For those observed by {\it K2} (16 stars), we used K2SFF processed light curves \citep{Vanderburg2014}. For {\it TESS} targets with short-cadence data, we used light curves from the Science Processing Operations Center \citep[SPOC, ][]{SPOC_LC} and for others we extracted light curves from the full-frame images using Eleanor\footnote{\url{https://github.com/afeinstein20/eleanor}} \citep{Eleanor}. We excluded targets with flux contamination ratios above 1, even when rotation consistent with youth was present in the curve.

\clearpage
\pagebreak
\startlongtable
\tablewidth{0.985\textwidth}
\begin{deluxetable*}{lcccccl}
\tablecolumns{7}
\tablecaption{{\it TESS} and {\it K2} potation periods.\label{tab:rots}}
\tablehead{\colhead{} & \colhead{R.A.} & \colhead{Decl.} & \colhead{Period~1} & \colhead{Period~2\tablenotemark{a}} & \colhead{Young} & \colhead{}\\
\colhead{Name} & \colhead{(hh:mm:ss.sss)} & \colhead{(dd:mm:ss.ss)} & \colhead{(days)} & \colhead{(days)} & \colhead{Association\tablenotemark{b}} & \colhead{Source} }
\startdata
2MASS~J03303685+1610599 & 03:30:36.887 & +16:10:59.58 & 1.70 & $\cdots$ & MUTA & 2\\
2MASS~J03350134+1418016 & 03:35:01.376 & +14:18:01.14 & 4.38 & $\cdots$ & MUTA & 2\\
2MASS~J03361762+2153391 & 03:36:17.665 & +21:53:38.50 & 4.38 & $\cdots$ & MUTA & 2\\
2MASS~J03371337+1307315 & 03:37:13.411 & +13:07:30.93 & 0.66 & $\cdots$ & MUTA & 2\\
2MASS~J03373508+1705162 & 03:37:35.111 & +17:05:15.93 & 6.19 & $\cdots$ & MUTA & 2\\
RX~J0338.3+1020 & 03:38:18.266 & +10:20:16.32 & 3.24 & $\cdots$ & MUTA & 1\\
2MASS~J03385230+1635406 & 03:38:52.328 & +16:35:40.21 & 0.34 & $\cdots$ & MUTA & 2\\
TYC~1235--156--1 & 03:39:39.516 & +15:29:54.47 & 4.43 & $\cdots$ & MUTA & 2\\
TYC~663--362--1 & 03:40:57.781 & +13:09:03.06 & 2.59 & $\cdots$ & MUTA & 1\\
TYC~660--135--1 & 03:41:45.000 & +10:54:27.46 & 5.12 & $\cdots$ & MUTA & 1\\
2MASS~J03420359+1631392 & 03:42:03.617 & +16:31:38.80 & 4.92 & $\cdots$ & MUTA & 2\\
HD~23376 & 03:44:58.957 & +08:19:10.09 & 0.81 & $\cdots$ & MUTA & 1\\
BD+04~589 & 03:47:13.551 & +05:26:23.49 & 4.75 & $\cdots$ & MUTA & 1\\
TYC~1252--301--1 & 03:47:23.901 & +18:43:17.68 & 4.09 & $\cdots$ & MUTA & 2\\
BD+07~543 & 03:47:31.345 & +07:57:26.39 & 3.57 & $\cdots$ & MUTA & 1\\
TYC~661--560--1 & 03:47:53.694 & +11:48:57.98 & 1.55 & $\cdots$ & MUTA & 1\\
RX~J0348.5+0832 & 03:48:31.461 & +08:31:36.43 & 0.41 & $\cdots$ & MUTA & 1\\
EPIC~210811401 & 03:48:50.333 & +20:02:27.56 & 5.10 & $\cdots$ & MUTA & 2\\
2MASS~J03495031+1440552 & 03:49:50.345 & +14:40:54.73 & 1.02 & $\cdots$ & MUTA & 2\\
PPM~119410 & 03:50:50.558 & +11:00:05.12 & 1.80 & $\cdots$ & MUTA & 1\\
2MASS~J03511041+1302467 & 03:51:10.454 & +13:02:46.16 & 2.54 & $\cdots$ & MUTA & 1\\
EPIC~210361663 & 03:54:50.776 & +12:32:05.61 & 3.26 & $\cdots$ & MUTA & 2\\
HD~286374 & 03:56:19.224 & +11:25:10.84 & 1.57 & $\cdots$ & MUTA & 2\\
HD~286380 & 03:56:20.741 & +10:47:47.24 & 2.51 & $\cdots$ & MUTA & 1\\
TYC~665--150--1 & 03:57:21.412 & +12:58:16.37 & 0.86 & $\cdots$ & MUTA & 2\\
2MASS~J03573875+1142322 & 03:57:38.786 & +11:42:31.85 & 3.92 & $\cdots$ & MUTA & 2\\
RX~J0358.2+0932 & 03:58:12.749 & +09:32:21.97 & 1.44 & $\cdots$ & MUTA & 1\\
TYC~662--217--1 & 03:59:42.158 & +12:10:08.14 & 4.80 & $\cdots$ & MUTA & 2\\
2MASS~J04072953--0115000 & 04:07:29.559 & --01:15:00.13 & 1.04 & $\cdots$ & MUTA & 1\\
TYC~74--1393--1 & 04:12:18.449 & +00:01:31.28 & 2.94 & $\cdots$ & MUTA & 1\\
2MASS~J04210781--0111328 & 04:21:07.848 & --01:11:33.15 & 4.29 & $\cdots$ & MUTA & 1\\
TYC~668--737--1 & 04:21:24.386 & +08:53:54.34 & 5.31 & $\cdots$ & MUTA & 1\\
BD--03~753 & 04:22:23.528 & --02:40:04.13 & 0.90 & $\cdots$ & MUTA & 1\\
HD~27687 & 04:22:24.213 & +06:31:45.14 & 0.53 & 0.39 & MUTA & 1\\
BD+05~638 & 04:22:33.022 & +05:41:38.82 & 3.37 & 6.32 & MUTA & 1\\
HD~28356 & 04:28:32.733 & +06:05:52.07 & 0.81 & $\cdots$ & MUTA & 1\\
BD--03~789 & 04:28:37.716 & --03:15:44.58 & 1.72 & $\cdots$ & MUTA & 1\\
2MASS~J04372578--0210117 & 04:37:25.800 & --02:10:12.12 & 1.13 & $\cdots$ & MUTA & 1\\
2MASS~J04372971--0051241 & 04:37:29.730 & --00:51:24.47 & 3.59 & $\cdots$ & MUTA & 1\\
2MASS~J04391308--0045039 & 04:39:13.102 & --00:45:04.39 & 0.42 & $\cdots$ & MUTA & 1\\
BD+06~731 & 04:39:15.500 & +07:01:43.92 & 6.08 & 8.66 & MUTA & 1\\
TYC~4739--1225--1 & 04:39:20.251 & --03:14:21.79 & 3.61 & $\cdots$ & MUTA & 1\\
BD--02~1047 & 04:52:07.364 & --01:58:57.43 & 1.47 & 0.69 & MUTA & 1\\
TYC~4741--307--1 & 04:56:18.287 & --01:53:33.04 & 2.44 & $\cdots$ & MUTA & 1\\
HD~37402 & 05:34:26.201 & --60:06:14.58 & 1.93 & $\cdots$ & CAR & 1\\
HD~42270 & 05:53:29.503 & --81:56:52.20 & 1.87 & $\cdots$ & CAR & 1\\
HD~43199 & 06:10:52.922 & --61:29:58.79 & 0.52 & $\cdots$ & CAR & 1\\
AL~442 & 06:11:30.043 & --72:13:37.79 & 0.85 & $\cdots$ & CAR & 1\\
AB~Pic & 06:19:12.941 & --58:03:14.83 & 3.81 & $\cdots$ & CAR & 1\\
HIP~32235 & 06:43:46.270 & --71:58:34.45 & 3.94 & $\cdots$ & CAR & 1\\
HIP~33737 & 07:00:30.501 & --79:41:45.06 & 5.21 & $\cdots$ & CAR & 1\\
2MASS~J07013884--6236059 & 07:01:38.844 & --62:36:05.98 & 3.93 & 5.58 & CAR & 1\\
2MASS~J07065772--5353463 & 07:06:57.714 & --53:53:45.75 & 2.87 & $\cdots$ & CAR & 1\\
2MASS~J08040534--6316396 & 08:04:05.300 & --63:16:39.11 & 2.01 & $\cdots$ & CAR & 1\\
2MASS~J08194309--7401232 & 08:19:43.099 & --74:01:23.22 & 0.42 & $\cdots$ & CAR & 1\\
2MASS~J09032434--6348330 & 09:03:24.265 & --63:48:32.65 & 4.42 & $\cdots$ & CAR & 1\\
2MASS~J09180165--5452332 & 09:18:01.547 & --54:52:32.85 & 0.37 & $\cdots$ & CAR & 1\\
HIP~46063 & 09:23:34.921 & --61:11:35.61 & 3.92 & $\cdots$ & CAR & 1\\
2MASS~J09315840--6209258 & 09:31:58.328 & --62:09:25.46 & 1.93 & $\cdots$ & CAR & 1\\
TWA~21 & 10:13:14.666 & --52:30:53.85 & 4.43 & $\cdots$ & CAR & 1\\
HD~14691 & 02:22:01.693 & --10:46:40.40 & 0.46 & $\cdots$ & COL & 1\\
2MASS~J03083950--3844363 & 03:08:39.597 & --38:44:36.32 & 0.69 & $\cdots$ & COL & 1\\
2MASS~J03320347--5139550 & 03:32:03.559 & --51:39:54.87 & 5.56 & $\cdots$ & COL & 1\\
HIP~17248 & 03:41:37.453 & +55:13:05.02 & 4.70 & $\cdots$ & COL & 1\\
2MASS~J04091413--4008019 & 04:09:14.199 & --40:08:01.98 & 3.28 & $\cdots$ & COL & 1\\
HIP~19775 & 04:14:22.624 & --38:19:01.54 & 1.74 & $\cdots$ & COL & 1\\
CD--36~1785 & 04:34:50.821 & --35:47:21.13 & 2.31 & $\cdots$ & COL & 1\\
HD~29329 & 04:46:00.914 & +76:36:37.66 & 0.92 & $\cdots$ & COL & 1\\
HIP~22226 & 04:46:49.568 & --26:18:08.93 & 0.89 & $\cdots$ & COL & 1\\
HD~31242 & 04:51:53.585 & --46:47:13.11 & 3.01 & $\cdots$ & COL & 1\\
HD~272836 & 04:53:05.246 & --48:44:38.49 & 4.60 & $\cdots$ & COL & 1\\
HIP~23316 & 05:00:51.910 & --41:01:06.56 & 2.29 & $\cdots$ & COL & 1\\
2MASS~J05195695--1124440 & 05:19:56.985 & --11:24:44.48 & 4.38 & $\cdots$ & COL & 1\\
2MASS~J05241317--2104427 & 05:24:13.213 & --21:04:43.14 & 4.17 & $\cdots$ & COL & 1\\
HIP~25709 & 05:29:24.132 & --34:30:55.43 & 6.14 & 2.75 & COL & 1\\
AH~Lep & 05:34:09.189 & --15:17:03.54 & 2.10 & $\cdots$ & COL & 1\\
HD~37484 & 05:37:39.655 & --28:37:34.70 & 0.82 & $\cdots$ & COL & 1\\
2MASS~J05395494--1307598 & 05:39:54.968 & --13:08:00.11 & 1.89 & $\cdots$ & COL & 1\\
AI~Lep & 05:40:20.753 & --19:40:11.12 & 1.69 & $\cdots$ & COL & 1\\
HD~38397 & 05:43:35.843 & --39:55:24.50 & 2.27 & $\cdots$ & COL & 1\\
HIP~28036 & 05:55:43.189 & --38:06:16.10 & 0.95 & $\cdots$ & COL & 1\\
HD~41071 & 06:00:41.325 & --44:53:49.75 & 5.46 & $\cdots$ & COL & 1\\
HIP~30030 & 06:19:08.069 & --03:26:21.01 & 1.36 & $\cdots$ & COL & 1\\
CD--40~2458 & 06:26:06.918 & --41:02:53.59 & 4.21 & $\cdots$ & COL & 1\\
HIP~490 & 00:05:52.680 & --41:45:12.23 & 3.00 & $\cdots$ & THA & 1\\
2MASS~J00125703--7952073 & 00:12:57.525 & --79:52:08.03 & 0.99 & $\cdots$ & THA & 1\\
HIP~1113 & 00:13:53.335 & --74:41:18.61 & 3.62 & $\cdots$ & THA & 1\\
2MASS~J00144767--6003477 & 00:14:47.860 & --60:03:48.67 & 0.49 & $\cdots$ & THA & 1\\
2MASS~J00152752--6414545 & 00:15:27.705 & --64:14:55.61 & 3.69 & $\cdots$ & THA & 1\\
GJ~3017 & 00:15:36.842 & --29:46:01.77 & 0.84 & $\cdots$ & THA & 1\\
HIP~1481 & 00:18:26.332 & --63:28:39.90 & 2.31 & 2.59 & THA & 1\\
2MASS~J00235732--5531435 & 00:23:57.506 & --55:31:44.58 & 2.44 & 2.99 & THA & 1\\
HIP~1993 & 00:25:14.853 & --61:30:49.12 & 4.37 & $\cdots$ & THA & 1\\
UPM~J0027--6157 & 00:27:33.500 & --61:57:17.79 & 0.55 & $\cdots$ & THA & 1\\
2MASS~J00284683--6751446 & 00:28:47.106 & --67:51:45.46 & 0.33 & $\cdots$ & THA & 1\\
2MASS~J00332438--5116433 & 00:33:24.551 & --51:16:44.33 & 0.35 & $\cdots$ & THA & 1\\
HIP~2729 & 00:34:51.397 & --61:54:58.95 & 0.38 & $\cdots$ & THA & 1\\
2MASS~J00393579--3816584 & 00:39:35.930 & --38:16:59.55 & 6.41 & $\cdots$ & THA & 1\\
2MASS~J00394063--6224125 & 00:39:40.905 & --62:24:13.39 & 0.38 & $\cdots$ & THA & 1\\
UPM~J0042--5444 & 00:42:10.272 & --54:44:44.09 & 1.78 & $\cdots$ & THA & 1\\
CD--78~24 & 00:42:20.705 & --77:47:40.20 & 2.59 & $\cdots$ & THA & 1\\
2MASS~J00425349--6117384 & 00:42:53.702 & --61:17:39.23 & 1.04 & $\cdots$ & THA & 1\\
HIP~3556 & 00:45:28.320 & --51:37:34.85 & 5.97 & 9.13 & THA & 1\\
2MASS~J00485254--6526330 & 00:48:52.746 & --65:26:33.71 & 1.01 & $\cdots$ & THA & 1\\
2MASS~J00493566--6347416 & 00:49:35.887 & --63:47:42.33 & 4.95 & $\cdots$ & THA & 1\\
UPM~J0113--5939 & 01:13:40.523 & --59:39:35.06 & 0.32 & $\cdots$ & THA & 1\\
2MASS~J01180670--6258591 & 01:18:06.926 & --62:58:59.85 & 0.35 & $\cdots$ & THA & 1\\
2MASS~J01211297--6117281 & 01:21:13.152 & --61:17:28.86 & 0.43 & $\cdots$ & THA & 1\\
CD--34~521 & 01:22:04.571 & --33:37:04.47 & 9.61 & $\cdots$ & THA & 1\\
UPM~J0122--6318 & 01:22:45.334 & --63:18:45.24 & 0.46 & $\cdots$ & THA & 1\\
HIP~6485 & 01:23:21.433 & --57:28:51.25 & 3.47 & $\cdots$ & THA & 1\\
2MASS~J01233280--4113110 & 01:23:32.961 & --41:13:11.76 & 0.74 & $\cdots$ & THA & 1\\
2MASS~J01275875--6032243 & 01:27:58.956 & --60:32:24.76 & 0.34 & $\cdots$ & THA & 1\\
HIP~6856 & 01:28:08.842 & --52:38:19.81 & 6.36 & $\cdots$ & THA & 1\\
2MASS~J01375879--5645447 & 01:37:58.967 & --56:45:45.33 & 0.71 & $\cdots$ & THA & 1\\
HD~10863 & 01:46:01.170 & --27:20:56.49 & 0.44 & $\cdots$ & THA & 1\\
2MASS~J01504543--5716488 & 01:50:45.620 & --57:16:49.23 & 0.71 & $\cdots$ & THA & 1\\
2MASS~J01505688--5844032 & 01:50:57.087 & --58:44:03.63 & 1.65 & $\cdots$ & THA & 1\\
2MASS~J01532494--6833226 & 01:53:25.212 & --68:33:22.92 & 0.60 & $\cdots$ & THA & 1\\
HIP~9141 & 01:57:49.093 & --21:54:06.12 & 3.04 & $\cdots$ & THA & 1\\
2MASS~J02001992--6614017 & 02:00:20.178 & --66:14:02.14 & 0.63 & $\cdots$ & THA & 1\\
HIP~9685 & 02:04:35.299 & --54:52:54.45 & 0.44 & $\cdots$ & THA & 1\\
2MASS~J02045317--5346162 & 02:04:53.332 & --53:46:16.75 & 0.70 & $\cdots$ & THA & 1\\
UCAC3~92--4597 & 02:07:01.904 & --44:06:38.51 & 0.39 & 3.00 & THA & 1\\
HIP~9892 & 02:07:18.209 & --53:11:56.88 & 2.39 & $\cdots$ & THA & 1\\
HIP~9902 & 02:07:26.315 & --59:40:46.23 & 1.71 & $\cdots$ & THA & 1\\
2MASS~J02125819--5851182 & 02:12:58.366 & --58:51:18.42 & 1.60 & $\cdots$ & THA & 1\\
2MASS~J02205139--5823411 & 02:20:51.580 & --58:23:41.37 & 1.28 & $\cdots$ & THA & 1\\
2MASS~J02242453--7033211 & 02:24:24.829 & --70:33:21.25 & 0.52 & $\cdots$ & THA & 1\\
2MASS~J02294869--6906044 & 02:29:48.952 & --69:06:04.36 & 0.46 & $\cdots$ & THA & 1\\
2MASS~J02321934--5746117 & 02:32:19.520 & --57:46:11.93 & 0.86 & $\cdots$ & THA & 1\\
UPM~J0234--5128 & 02:34:18.835 & --51:28:46.44 & 0.46 & $\cdots$ & THA & 1\\
2MASS~J02383255--7528065 & 02:38:32.880 & --75:28:06.41 & 0.63 & $\cdots$ & THA & 1\\
CD--53~544 & 02:41:47.002 & --52:59:52.61 & 0.52 & $\cdots$ & THA & 1\\
2MASS~J02420204--5359147 & 02:42:02.231 & --53:59:14.88 & 0.57 & 2.47 & THA & 1\\
2MASS~J02420404--5359000 & 02:42:04.237 & --53:59:00.22 & 0.57 & $\cdots$ & THA & 1\\
CD--58~553 & 02:42:33.187 & --57:39:36.95 & 7.40 & $\cdots$ & THA & 1\\
HD~17250 & 02:46:14.687 & +05:35:32.64 & 1.21 & $\cdots$ & THA & 1\\
2MASS~J02474639--5804272 & 02:47:46.569 & --58:04:27.47 & 9.45 & $\cdots$ & THA & 1\\
2MASS~J02502222--6545552 & 02:50:22.440 & --65:45:55.27 & 1.29 & $\cdots$ & THA & 1\\
2MASS~J02523550--7831183 & 02:52:35.919 & --78:31:18.08 & 0.71 & $\cdots$ & THA & 1\\
2MASS~J02553178--5702522 & 02:55:31.954 & --57:02:52.41 & 0.49 & $\cdots$ & THA & 1\\
2MASS~J03050556--5317182 & 03:05:05.712 & --53:17:18.46 & 0.44 & $\cdots$ & THA & 1\\
2MASS~J03104941--3616471 & 03:10:49.532 & --36:16:47.39 & 0.66 & $\cdots$ & THA & 1\\
2MASS~J03114544--4719501 & 03:11:45.581 & --47:19:50.27 & 4.81 & $\cdots$ & THA & 1\\
HIP~15247 & 03:16:40.753 & --03:31:49.69 & 1.03 & $\cdots$ & THA & 1\\
2MASS~J03244056--3904227 & 03:24:40.680 & --39:04:22.95 & 0.34 & 8.05 & THA & 1\\
2MASS~J03291649--3702502 & 03:29:16.628 & --37:02:50.37 & 0.54 & $\cdots$ & THA & 1\\
CD--46~1064 & 03:30:49.233 & --45:55:57.44 & 3.92 & $\cdots$ & THA & 1\\
CD--44~1173 & 03:31:55.768 & --43:59:13.61 & 2.93 & $\cdots$ & THA & 1\\
2MASS~J03454058--7509121 & 03:45:40.854 & --75:09:11.91 & 0.70 & $\cdots$ & THA & 1\\
HD~24636 & 03:48:11.720 & --74:41:38.44 & 5.72 & 0.84 & THA & 1\\
2MASS~J03512287--5154582 & 03:51:23.001 & --51:54:58.01 & 0.43 & $\cdots$ & THA & 1\\
HD~25284 & 04:00:03.918 & --29:02:16.64 & 0.31 & 2.27 & THA & 1\\
HD~25402 & 04:00:32.079 & --41:44:54.40 & 3.56 & $\cdots$ & THA & 1\\
2MASS~J04013874--3127472 & 04:01:38.846 & --31:27:47.35 & 0.46 & $\cdots$ & THA & 1\\
BD--15~705 & 04:02:16.556 & --15:21:30.22 & 3.85 & $\cdots$ & THA & 1\\
2MASS~J04074372--6825111 & 04:07:43.905 & --68:25:10.85 & 1.02 & $\cdots$ & THA & 1\\
2MASS~J04133609--4413325 & 04:13:36.194 & --44:13:32.40 & 0.78 & $\cdots$ & THA & 1\\
WOH~S~6 & 04:21:39.275 & --72:33:55.53 & 4.55 & $\cdots$ & THA & 1\\
2MASS~J04274963--3327010 & 04:27:49.718 & --33:27:01.17 & 0.67 & $\cdots$ & THA & 1\\
HIP~21632 & 04:38:44.007 & --27:02:01.97 & 2.40 & $\cdots$ & THA & 1\\
2MASS~J04435860--3643188 & 04:43:58.683 & --36:43:18.81 & 1.61 & $\cdots$ & THA & 1\\
2MASS~J04440099--6624036 & 04:44:01.138 & --66:24:03.15 & 7.94 & $\cdots$ & THA & 1\\
2MASS~J04470041--5134405 & 04:47:00.509 & --51:34:40.23 & 6.25 & $\cdots$ & THA & 1\\
TYC~8083--45--5 & 04:48:00.760 & --50:41:25.40 & 8.13 & $\cdots$ & THA & 1\\
HIP~22295 & 04:48:05.485 & --80:46:44.64 & 1.24 & $\cdots$ & THA & 1\\
CD--30~2310 & 05:18:29.092 & --30:01:32.16 & 1.70 & $\cdots$ & THA & 1\\
TYC~8098--414--1 & 05:33:25.647 & --51:17:12.77 & 5.21 & $\cdots$ & THA & 1\\
HIP~32435 & 06:46:13.720 & --83:59:28.55 & 1.59 & $\cdots$ & THA & 1\\
HIP~84642 & 17:18:14.645 & --60:27:27.57 & 4.17 & $\cdots$ & THA & 1\\
2MASS~J19225071--6310581 & 19:22:50.700 & --63:10:59.23 & 0.86 & $\cdots$ & THA & 1\\
2MASS~J20291446--5456116 & 20:29:14.491 & --54:56:13.29 & 0.88 & $\cdots$ & THA & 1\\
2MASS~J21100614--5811483 & 21:10:06.195 & --58:11:49.78 & 0.56 & $\cdots$ & THA & 1\\
2MASS~J21163528--6005124 & 21:16:35.368 & --60:05:14.13 & 0.99 & $\cdots$ & THA & 1\\
HIP~105388 & 21:20:50.012 & --53:02:04.64 & 3.45 & $\cdots$ & THA & 1\\
UPM~J2127--6841 & 21:27:50.634 & --68:41:04.64 & 0.34 & $\cdots$ & THA & 1\\
2MASS~J21370885--6036054 & 21:37:08.927 & --60:36:07.04 & 2.00 & $\cdots$ & THA & 1\\
2MASS~J21380269--5744583 & 21:38:02.765 & --57:44:59.89 & 0.68 & $\cdots$ & THA & 1\\
HIP~107345 & 21:44:30.211 & --60:58:40.34 & 4.53 & $\cdots$ & THA & 1\\
2MASS~J21504048--5113380 & 21:50:40.563 & --51:13:39.59 & 1.05 & $\cdots$ & THA & 1\\
HIP~107947 & 21:52:09.822 & --62:03:09.92 & 0.96 & $\cdots$ & THA & 1\\
2MASS~J22021626--4210329 & 22:02:16.331 & --42:10:34.73 & 4.51 & $\cdots$ & THA & 1\\
2MASS~J22025453--6440441 & 22:02:54.624 & --64:40:45.70 & 0.43 & $\cdots$ & THA & 1\\
UPM~J2222--6303 & 22:22:39.816 & --63:03:27.22 & 1.11 & $\cdots$ & THA & 1\\
2MASS~J22244102--7724036 & 22:24:41.287 & --77:24:04.82 & 0.67 & $\cdots$ & THA & 1\\
2MASS~J22444835--6650032 & 22:44:48.534 & --66:50:04.47 & 0.73 & $\cdots$ & THA & 1\\
2MASS~J22463471--7353504 & 22:46:34.912 & --73:53:51.52 & 1.65 & $\cdots$ & THA & 1\\
2MASS~J23131671--4933154 & 23:13:16.833 & --49:33:16.85 & 1.23 & $\cdots$ & THA & 1\\
2MASS~J23170011--7432095 & 23:17:00.401 & --74:32:10.53 & 0.83 & $\cdots$ & THA & 1\\
TYC~9344--293--1 & 23:26:10.958 & --73:23:50.88 & 0.57 & $\cdots$ & THA & 1\\
2MASS~J23273447--8512364 & 23:27:35.285 & --85:12:37.17 & 0.90 & $\cdots$ & THA & 1\\
CD--86~147 & 23:27:50.213 & --86:13:19.36 & 0.70 & $\cdots$ & THA & 1\\
2MASS~J23285763--6802338 & 23:28:57.841 & --68:02:35.08 & 0.37 & $\cdots$ & THA & 1\\
2MASS~J23291752--6749598 & 23:29:17.728 & --67:50:01.14 & 1.02 & $\cdots$ & THA & 1\\
2MASS~J23382851--6749025 & 23:38:28.714 & --67:49:03.52 & 0.44 & $\cdots$ & THA & 1\\
HIP~116748 & 23:39:39.712 & --69:11:45.75 & 2.86 & $\cdots$ & THA & 1\\
2MASS~J23424333--6224564 & 23:42:43.528 & --62:24:57.60 & 0.52 & $\cdots$ & THA & 1\\
2MASS~J23452225--7126505 & 23:45:22.521 & --71:26:51.46 & 1.61 & $\cdots$ & THA & 1\\
2MASS~J23474694--6517249 & 23:47:47.152 & --65:17:25.79 & 4.84 & $\cdots$ & THA & 1\\
2MASS~J23524562--5229593 & 23:52:45.779 & --52:30:00.51 & 0.91 & $\cdots$ & THA & 1\\
\enddata
\tablenotetext{a}{Second rotation period candidate.}
\tablenotetext{b}{The full names of young associations are: Carina (CAR), Columba (COL), the Tucana-Horologium association (THA) and the $\mu$~Tau Association (MUTA).}
\tablecomments{Rotation periods are accurate to approximately 2\%. Only a portion of the table is shown here. The full table is available as online-only additional material. See Section~\ref{sec:act} for more details.}
\tablerefs{(1)~{\it TESS} \citep{SPOC_LC,2015JATIS...1a4003R}; (2)~{\it K2} \citep{Vanderburg2014,2014PASP..126..398H,2010Sci...327..977B}.}
\end{deluxetable*}

\clearpage
\pagebreak

\begin{figure*}
	\centering
	\includegraphics[width=0.98\textwidth]{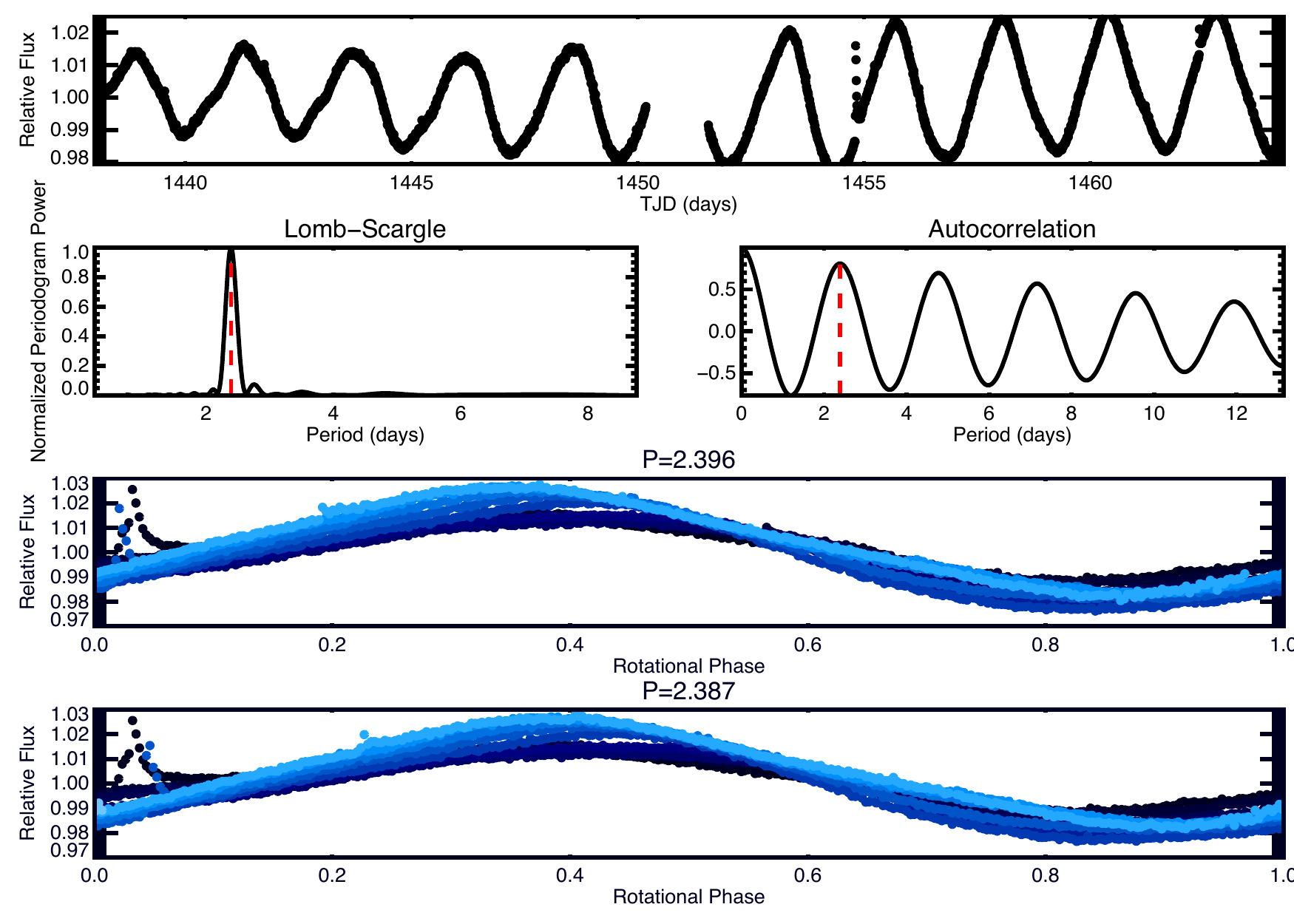}\\
	\caption{{\it TESS} light curve and rotation diagnostics of TIC178969585 (HD~29615), a G-type dwarf in the Tucana-Horologium association. The top panel shows the SPOC light curve, with the Lomb-Scargle power and autocorrelation function just below (the assigned period is marked with a red dashed line). The bottom two panels show the light curve phased to the period derived from the Lomb-Scargle (top) and autocorrelation function (bottom), color-coded by chronological order (lighter is later).}
	\label{fig:LC}
\end{figure*}

We estimated the rotation periods for each star using two methods: a modified version of the Lomb-Scargle periodogram as described in \citet{LombScargle}, and the autocorrelation function as described in \citet{McQuillan2013}. In both cases, we searched for periodic signals down to twice the Nyquist-sampling limit, and as long as a third of the total data coverage. Below the lower limit, we found that both algorithms are biased by the data sampling, particularly for long-cadence (30\,min) data. We set the lower limit for a significant detection at three full rotations. We then flagged the peak in the periodogram and the second peak in the autocorrelation function as the likely period (see Figure~\ref{fig:LC} for an example). We only considered periodic signals with false-alarm probabilities $<1$\% and for which autocorrelation and Lomb-Scargle periods agreed within 10\%. For six stars, the autocorrelation and Lomb-Scargle disagreed by an integer factor (alias), which we retained, provided the true rotation period was clear. Across all clusters, 14 out of 201 stars showed evidence of a second period, which we excluded from our sample as they are likely binaries \citep{Douglas2017}. As a final check, we visually inspected all phased light curves.

We created synthetic data sets, with random subsamples of half the data and each point perturbed by a random number following the measurement errors, to investigate the accuracy of our period determinations. We found that, when the correct period is identified, our assigned periods are accurate within 2\%, with a fail rate of $\simeq$\,5\% where the measured period is wrong by 20\% or more (usually off by an integer multiple). This assumes that all detected periods are associated with stellar rotation and not other phenomena. Periodic signals caused by binary systems, pulsations or flares could cause further false positives, if they passed our visual inspection.

\begin{figure*}
	\centering
	\includegraphics[width=0.98\textwidth]{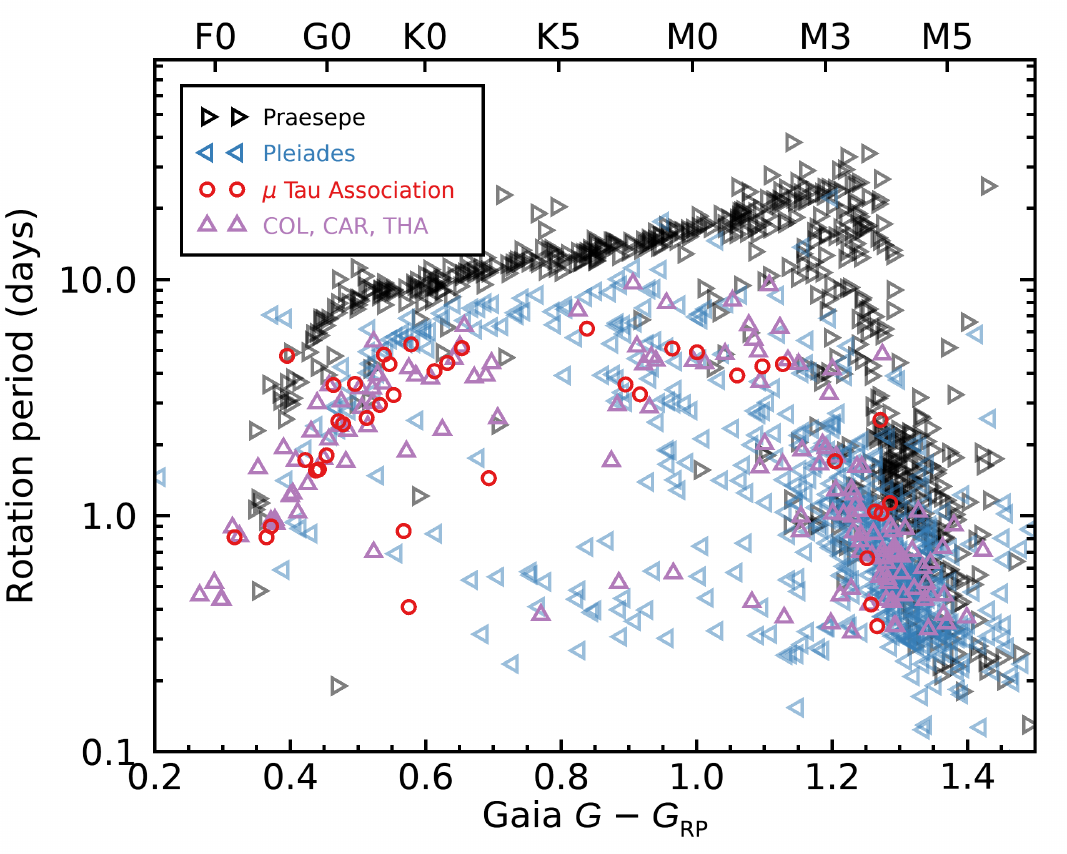}\\
	\caption{Rotation periods for stars in the Praesepe (black, $\simeq$800\,Myr) and Pleiades clusters (blue, $\simeq$112\,Myr), Columba (COL), Carina (CAR), or Tucana-Horologium (THA) associations (violet; $\simeq$45\,Myr), and \asso\ (red; $\simeq$60\,Myr) as a function of Gaia $G-G_{RP}$ color.}
	\label{fig:rot}
\end{figure*}

The resulting rotation periods are shown in Figure~\ref{fig:rot} and listed in Table~\ref{tab:rots}. While there is significant scatter in the sequence, Praesepe and Pleiades members have the longest typical rotation period at $G-G_{\rm RP}<0.8$, while members of young moving groups have the shortest periods, and \asso\ members are located in between. On the cool end ($G-G_{\rm RP}\gtrsim1.1$), Pleiades rotations are the fastest, as the $\simeq$45\,Myr stars are still contracting, although we have fewer period measurements in \asso\ in this regime. The overall trend is consistent with our assigned \assoage\ age of \asso\ based on empirical isochrones and the total age of the white dwarf WD~0340+103, though additional rotation period measurements would be useful to better map out its sequence.

\begin{figure}
 	\centering
 	\includegraphics[width=0.465\textwidth]{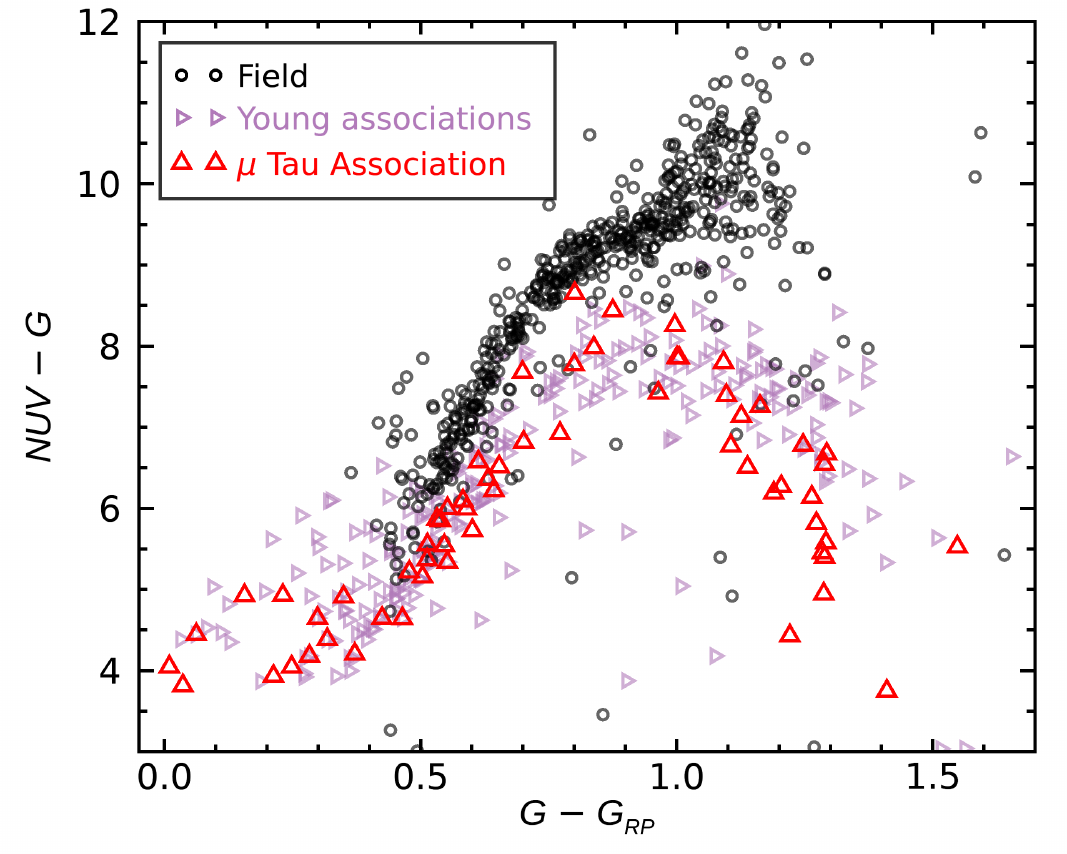}
 	\caption{\emph{GALEX} to \gaia\ $NUV - G$ color versus \gaiagr\ for field stars (black circles), members of nearby young associations (rightward purple triangles) and \asso\ candidates studied in this paper (upward red triangles). Our candidates are consistent with the young stellar population displaying a $NUV$ excess compared with field stars of the same \gaiagr\ color. See Section~\ref{sec:act} for more details.}
 	\label{fig:galex}
\end{figure}

\begin{figure}
 	\centering
 	\includegraphics[width=0.465\textwidth]{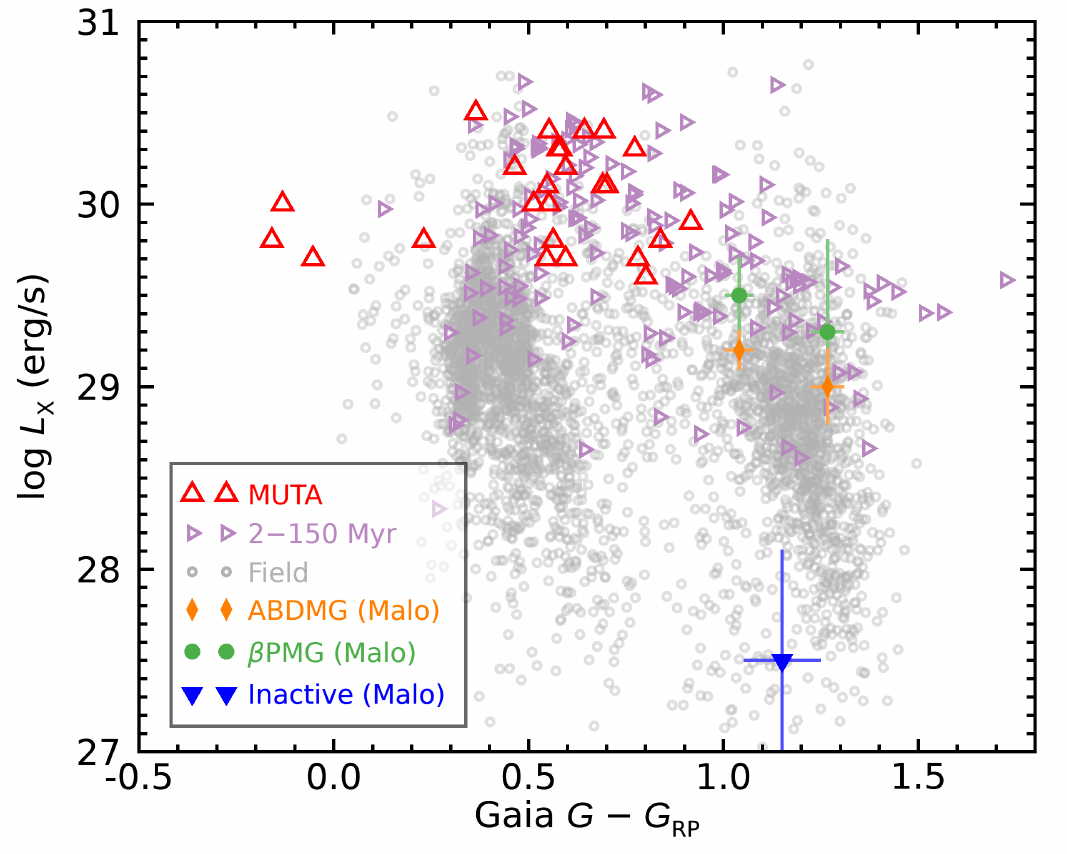}
 	\caption{Absolute X-ray luminosity for field stars (grey circles), nearby young stars (rightward purple triangles) and our \asso\ candidates (upward red triangles). The young M dwarf distributions of \cite{2014ApJ...788...81M} are also shown for comparison. Young stars tend to emit more X-ray because they are more active. In the case of low-mass stars, this effect is compounded by the larger radius of younger M dwarfs. Field stars tend to be more active at both ends of the mass spectrum, consistent with their faster average rotation rates. $\beta$PMG indicates the $\beta$~Pictoris moving group, and ABDMG indicates the AB~Doradus moving group. See Section~\ref{sec:act} for more details.}
 	\label{fig:rosat}
\end{figure}

Stellar rotation serves as a driver of magnetic activity through the dynamo effect \citep{2012AJ....143...93R}, and causes young stars to display enhanced UV and X-ray emission among other effects associated with an enhanced stellar activity \citep{2003ApJ...585..878K,2013ApJ...774..101R,2014ApJ...788...81M}. We used data from the \emph{ROSAT} all-sky survey \citep{2016AA...588A.103B} and the \emph{GALEX} catalog \citep{2005ApJ...619L...1M} to verify that our population of \asso\ members and candidates display this expected enhanced activity in a way that is consistent with other young asociations of similar ages ($\simeq$\,10--150\,Myr) in the Solar neighborood, including $\beta$PMG and the AB~Doradus moving group (ABDMG, \citealt{2004ApJ...613L..65Z}; see \citealt{2018ApJ...856...23G} for a discussion of these associations). The resulting distributions are shown in Figures~\ref{fig:galex} and \ref{fig:rosat}, and provide more evidence that \asso\ consists of a coeval and young association.

\subsection{$\mu$~Tau in the Context of the Galactic Structure}\label{sec:kounkel}

An unprecedented view of the local spatial and kinematic structure of the Galaxy was enabled with the advent of \gaia. Using these new data, \cite{2019AJ....158..122K} identified 1,901 groups of stars that appear co-moving and coeval, located within 30\textdegree\ of the Galactic plane and 1\,kpc of the Sun. Their method used the HDBSCAN unsupervised clustering algorithm\footnote{See \url{https://hdbscan.readthedocs.io}.} directly in the 5-dimensional parameter space of \gaia\ observables (sky position, proper motion and parallax) to identify over-densities; this did not allow them to efficiently recover the structure within about 70\,pc of the Sun because the large spread of nearby associations on the sky introduces strong variation and correlations in the \gaia\ 5-dimensional kinematic space of the members within a specific young association. \cite{2019AJ....158..122K} separated the over-densities among clusters and strings, the latter consisting of much larger structures with typical physical sizes of about 200\,pc and some of which also have extended kinematic distributions.

\begin{figure*}[p]
	\centering
	\subfigure[Galactic positions $X$ versus $Y$]{\includegraphics[width=0.49\textwidth]{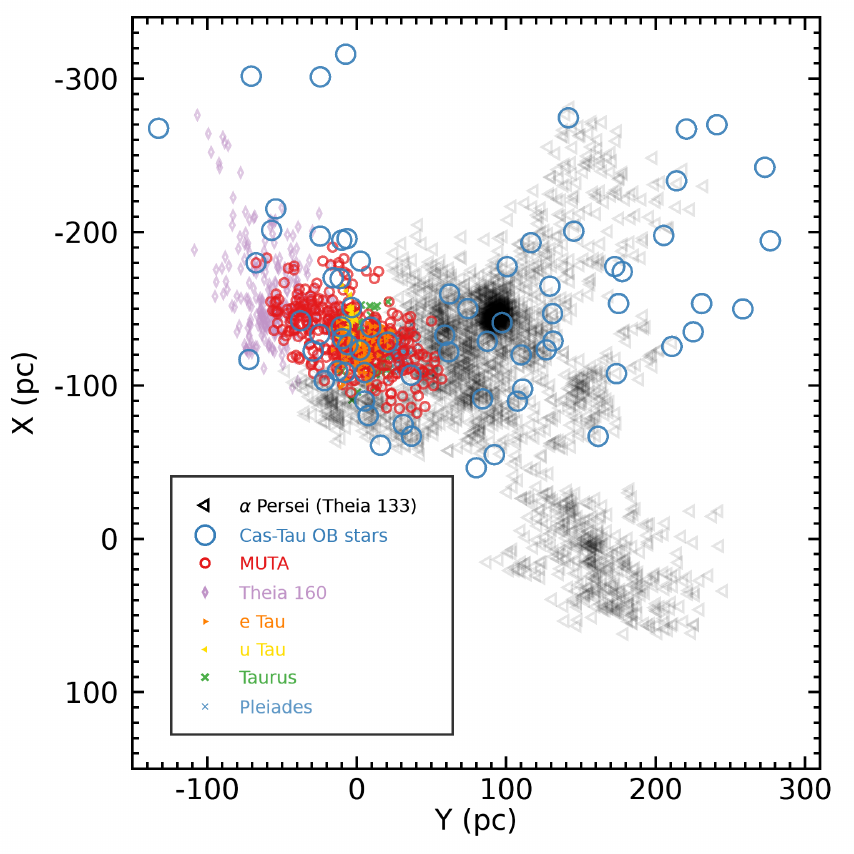}\label{fig:theia_xy}}
	\subfigure[Galactic positions $X$ versus $Z$]{\includegraphics[width=0.49\textwidth]{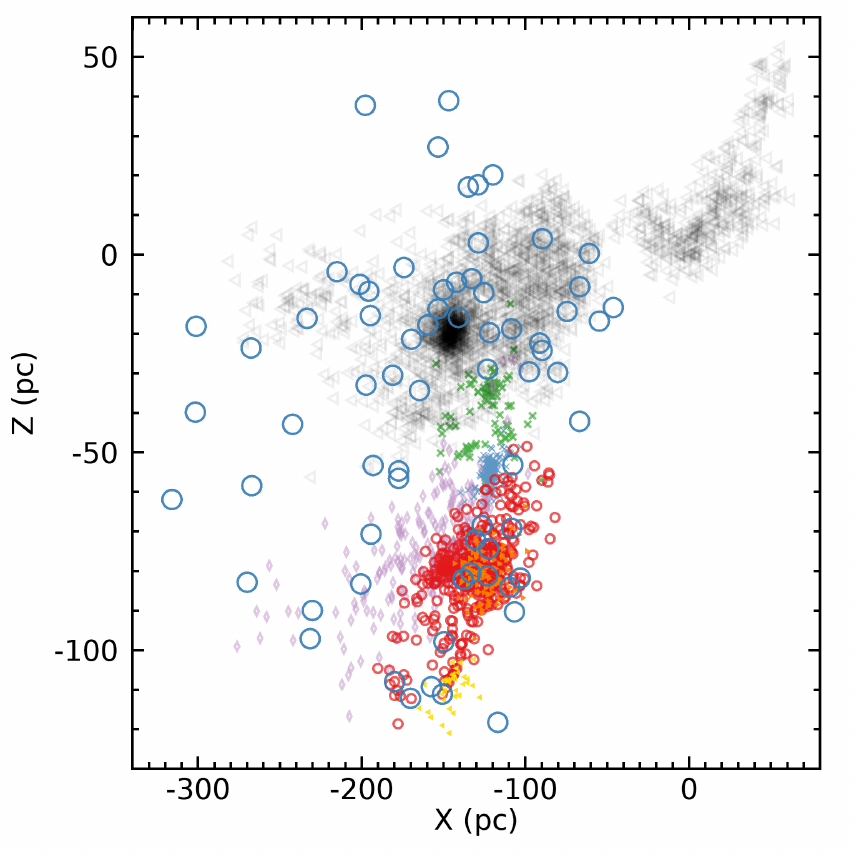}\label{fig:theia_xz}}
	\subfigure[Space velocities $U$ versus $V$]{\includegraphics[width=0.49\textwidth]{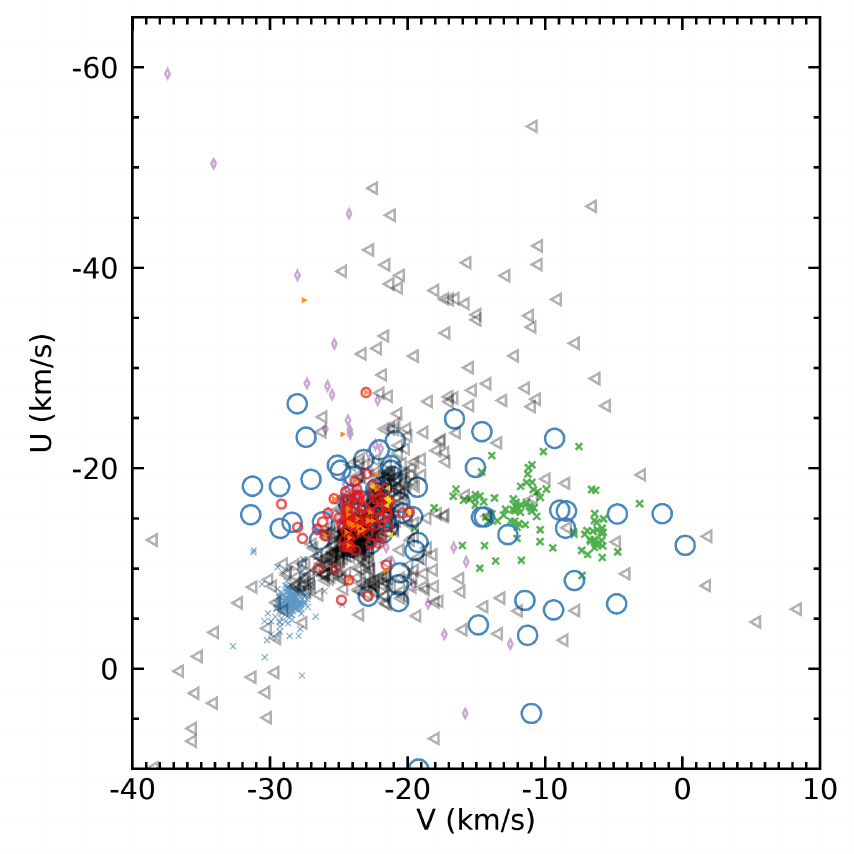}\label{fig:theia_uv}}
	\subfigure[Space velocities $U$ versus $W$]{\includegraphics[width=0.49\textwidth]{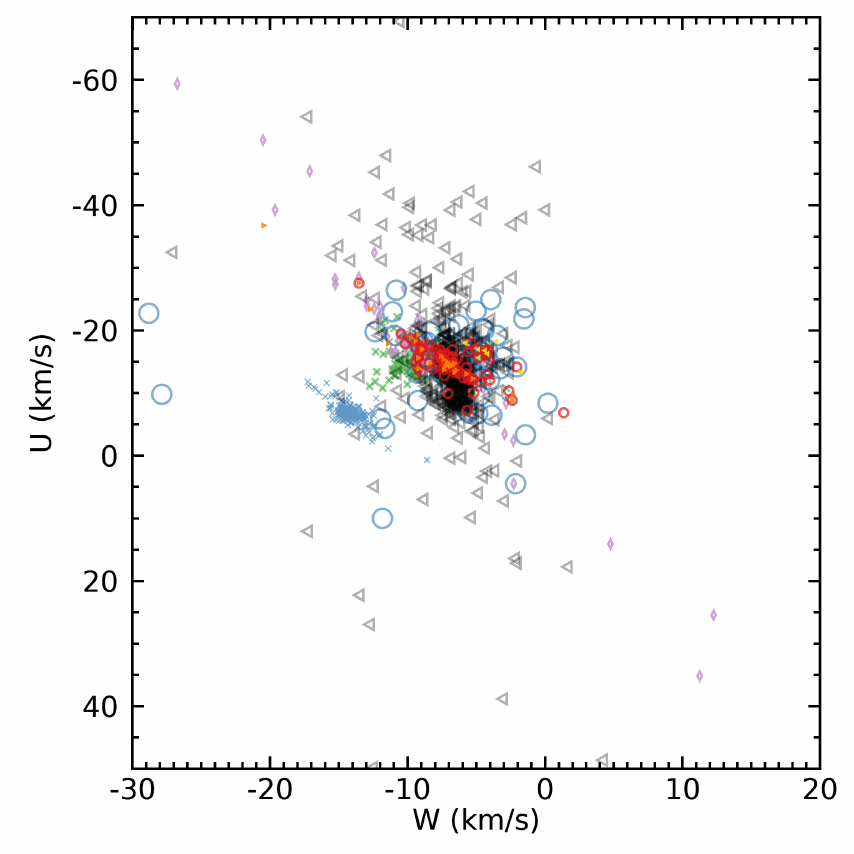}\label{fig:theia_vw}}
	\caption{Spatial and kinematic distribution of \asso\ candidates and members discussed in this work (red circles), compared with the neighbor Taurus association (rightward green triangles) and the Theia~160 kinematic string (blue diamonds). The similar kinematics and isochronal ages of Theia~160 and \asso\ indicate that these two Galactic structures may be related to one another. See Section~\ref{sec:kounkel} for more detail.}
	\label{fig:theia}
\end{figure*}

We cross-matched our sample of \asso\ candidates and members with the full \cite{2019AJ....158..122K} catalog of clustered sources to determine whether \asso\ had been recovered by their study. We found a total of 72 matches with our list, all with a single \cite{2019AJ....158..122K} string named Theia~160 that contains a total of 300 stars. Only 4 of these stars are matches to our initial list of \asso\ members (HD~28715, HD~27687, HD~28356, and TYC~668--737--1; respectively, MUTA~11, 17, 18, and 30~A). One likely explanation for the partial overlap is the $|b| < 30$\textdegree\ cut-off in Galactic latitude that they imposed, as approximately half of \asso\ falls at $b < -30$\textdegree. We show a comparison of Theia~160, \asso\ and Taurus in Figure~\ref{fig:theia}. Theia~160 is spatially more extended, but also shows a much larger spread in space velocities compared with \asso, although they are centered at similar average velocities; \asso\ members have a spread of (2.8, 2.1, 1.6)\,\kms\ in $UVW$ space, whereas the spread of Theia~160 members is (21.1, 1.7, 8.9)\,\kms. This indicates that some interlopers may contaminate the sample of Theia~160 stars, and further investigation will be required to confirm this.

In addition to the similar kinematics between \asso\ and Theia~160, \cite{2019AJ....158..122K} determined a model-dependent isochronal age of $\simeq$\,80\,Myr for Theia~160, which is close to our estimated age of \assoage. It seems likely that \asso\ and Theia~160 are related to each other; perhaps Theia~160 represents a stream or tidal tail around the more closely packed core of \asso\ (analoguous to the tidal tail around the Hyades cluster although the latter is much older; \citealt{2019AA...621L...2R}), or it is simply a fragment of \asso\ with some contaminating field stars that have more spread-out space velocities. Investigating this further will require a spectroscopic follow-up of candidates in both \asso\ and Theia~160 to complete the $UVW$ measurements of all members in both groups--although the next data release of the \gaia\ mission will likely allow to complete the $UVW$ velocities of most \asso\ members,--and determine spectroscopic signs of young ages. It is possible that our method did not recover the full spatial structure of \asso, especially regions that would lack massive stars, because BANYAN~$\Sigma$ requires an initial kinematic model to work with, which we obtained from the initial collection of young or active stars described in Section~\ref{sec:initsample}. In addition to this, \cite{2019AJ....158..122K} uncovered a large kinematic structure (Theia~133) that encompasses the $\alpha$~Persei cluster, likely related to Cas-Tau and \asso, as discussed in Section~\ref{sec:initsample}. This structure is also shown in Figure~\ref{fig:theia}.

 \cite{2020AJ....159..105L} recently published the discovery of two new associations physically nearby (but unrelated to) the Taurus-Auriga star-forming region; e~Tau and u~Tau. The group that they identified as e~Tau has significant overlap with our definition of \asso; 104 of their 119 members are in also in our list (18 in our initial members, 79 in our candidate members, 6 in our low-likelihood candidate members, and 1 in our list of rejected members). The 15 remaining objects not in our catalogs that they list as e~Tau members either have a Bayesian membership probability below 90\% or a best-case scenario separation above 5\,\kms\ with our kinematic model, which explains why we have not recovered them. We identified in this paper a total of 444 candidate members that \cite{2020AJ....159..105L} did not discuss: 18 in our initial members, 277 candidate members and 149 in our low-likelihood candidate members. An additional 12 objects in our \asso\ lists (4 initial members, 6 candidate members and 2 low-likelihood candidates) are listed as u~Tau members by \cite{2020AJ....159..105L}. The isochrone age of $\simeq$50\,Myr determined by \cite{2020AJ....159..105L} is similar to our \assoage, but is based on model isochrones rather than empirical ones.

The fact that \cite{2019AJ....158..122K} and \cite{2020AJ....159..105L} may have uncovered spatial extensions of \asso, and the presence of a large structure of additional stars coeval with \asso\ and $\alpha$~Persei hints that it would be valuable to parse the local Solar neighborood with an overdensity detection algorithm that is not hindered by the lack of radial velocity measurements or the large spread and correlations of sky positions, proper motions and parallaxes of nearby cluster members. Such a study would have the potential to uncover extended structures and connections between the \cite{2019AJ....158..122K} groups and the known nearby young associations in the Solar neighborhood, as well as new nearby associations entirely.

\section{CONCLUSIONS}\label{sec:conclusion}

We presented and characterized the \assofull\ Association, a young stellar population consisting of hundreds of members at about 150\,pc from the Sun. We built a BANYAN~$\Sigma$ spatial-kinematic model for this association to identify additional candidate members with \gaia\ and to allow other teams to search for new members. The \gaia\ photometry and parallaxes of \asso\ members allowed us to make a comparison with empirical sequences of the Pleiades, Tucana-Horologium, Carina and Columba members to determine an isochronal age relative to these other young associations. This resulted in an age estimate of \isoage\ for \asso. We identified a white dwarf (WD~0340+103) that is the remnant of a B2 \asso\ member that left its planetary nebula phase 270,000~years ago, and used its total age to further constrain the age of \asso\ at \assoage. We found literature measurements of the lithium equivalent width for K-type to G-type members of \asso\ and showed that they are consistent with our age determination. The members of this new association have a \gaia\ colors versus {\it TESS} rotation periods sequence consistent with a young age, and display an enhanced level of stellar activity compared with the field population based on UV and X-ray, consistent with a young coeval population. We also showed that its present-day mass function is similar to other known young associations. \asso\ is likely part of an extended network of stars coeval and co-moving with the $\alpha$~Persei cluster that are currently dissolving. A master table with all candidates and members of the \asso\ association is also provided here (Table~\ref{tab:master}).

The \asso\ association is a new laboratory to study stellar and exoplanet evolution at an age which was not well sampled by other associations within the Solar neighborhood. Its distance of $\simeq$\,150\,pc will make it harder to identify its substellar population, but upcoming wide area surveys such as Pan-STARRS~3$\pi$ \citep{2010HiA....15..818M} and CatWISE \citep{2019arXiv190808902E} may be able to do so in the near future. The extended ROentgen Survey with an Imaging Telescope Array (eROSITA; \citealt{2014SPIE.9144E..1TP}) on the Spektrum-Roentgen-Gamma (SRG) space telescope will also likely allow us to better study the activity of the low-mass stars in \asso.

\acknowledgments

We thank the anonymous reviewer for thoughtful and constructive comments. We thank Patrick Dufour, Aaron Rizzuto and Benjamin Tofflemire for useful comments. We thank Za Gur\={e}to Muta for guidance in choosing an acronym for the $\mu$~Tau Association. This work was partially carried under a Banting grant from the Natural Sciences and Engineering Research Council of Canada (NSERC). This research made use of: the SIMBAD database and VizieR catalog access tool, operated at the Centre de Donn\'ees astronomiques de Strasbourg, France \citep{2000AAS..143...23O}; data products from the Two Micron All Sky Survey (\emph{2MASS}; \citealp{2006AJ....131.1163S}), which is a joint project of the University of Massachusetts and the Infrared Processing and Analysis Center (IPAC)/California Institute of Technology (Caltech), funded by the National Aeronautics and Space Administration (NASA) and the National Science Foundation \citep{2006AJ....131.1163S}; data products from the \emph{Wide-field Infrared Survey Explorer} (\emph{WISE}; and \citealp{2010AJ....140.1868W}), which is a joint project of the University of California, Los Angeles, and the Jet Propulsion Laboratory (JPL)/Caltech, funded by NASA. The Digitized Sky Surveys (DSS) were produced at the Space Telescope Science Institute under U.S. Government grant NAG W-2166. The images of these surveys are based on photographic data obtained using the Oschin Schmidt Telescope on Palomar Mountain and the UK Schmidt Telescope. The plates were processed into the present compressed digital form with the permission of these institutions. The Second Palomar Observatory Sky Survey (POSS-II) was made by the California Institute of Technology with funds from the National Science Foundation, the National Geographic Society, the Sloan Foundation, the Samuel Oschin Foundation, and the Eastman Kodak Corporation. The Oschin Schmidt Telescope is operated by the California Institute of Technology and Palomar Observatory. This work presents results from the European Space Agency (ESA) space mission Gaia. Gaia data are being processed by the Gaia Data Processing and Analysis Consortium (DPAC). Funding for the DPAC is provided by national institutions, in particular the institutions participating in the Gaia MultiLateral Agreement (MLA). The Gaia mission website is https://www.cosmos.esa.int/gaia. The Gaia archive website is https://archives.esac.esa.int/gaia. The Digitized Sky Surveys were produced at the Space Telescope Science Institute under U.S. Government grant NAG W-2166. Part of this research was carried out at the Jet Propulsion Laboratory, California Institute of Technology, under a contract with the National Aeronautics and Space Administration (80NM0018D0004). TJD and EEM and gratefully acknowledge support from the Jet Propulsion Laboratory Exoplanetary Science Initiative and NASA award 17-K2GO6-0030. EEM acknowledges support from NASA grant NNX15AD53G.

\startlongtable
\tabletypesize{\footnotesize}
\tablewidth{0.985\textwidth}
\clearpage
\pagebreak
\begin{longrotatetable}
\global\pdfpageattr\expandafter{\the\pdfpageattr/Rotate 90}
\begin{deluxetable*}{llllL}
\tablecolumns{5}
\tablecaption{Main list of all systems of interest to \asso\ identified in this work.\label{tab:master}}
\tablehead{
\colhead{Name} &
\colhead{Units} & 
\colhead{Type} &
\colhead{Format} & 
\colhead{Description}
}
\startdata
muta\_id & \nodata & char & a6 & $\mu$~Tau Association (MUTA) identification number.\\
main\_name & \nodata & char & a25 & Main target name. SIMBAD-resolvable names are preferred; short names in the format J0236+2026 are given otherwise.\\
gaiadr2\_id & \nodata & char & a25 & \gaia\ identification number.\\
tm\_name & \nodata & char & a25 & 2MASS designation.\\
aw\_name & \nodata & char & a25 & AllWISE designation.\\
rosat\_name & \nodata & char & a25 & {\it ROSAT} designation.\\
tyc\_name & \nodata & char & a25 & Tycho catalog designation.\\
hip\_name & \nodata & char & a25 & Hipparcos catalog designation.\\
simbad\_id & \nodata & char & a25 & Principal SIMBAD identifier.\\
spt & \nodata & char & a10 & Literature spectral type. Spectral type estimates based on \gaia\ colors are given between parentheses.
\newline (WD) indicates likely white dwarfs.\\
spt\_ref & \nodata & char & a25 & Reference for literature spectral type.\\
member\_type & \nodata & char & a2 & Membership type. IM: Member from our initial list. CM: Candidate member. LM: Low-priority candidate member.
\newline R: Rejected candidate member.\\
source & \nodata & char & a6 & Source from which the target was obtained. INIT: Initial list described in Section~\ref{sec:initsample}. GAIA: Originates from our \gaia-based search for additional candidate members described in Section~\ref{sec:newmembers}. COM: Originates from our comover search described in Section~\ref{sec:comoving}. VIS: Originates from our visual identification of comover candidates described in Section~\ref{sec:fcharts}. OH2017: Originates from a \cite{2017AJ....153..257O} group with a partial match to our \asso\ members and candidates.\\
mem\_prob & \% & R*4 & f7.1 & BANYAN~$\Sigma$ probability for membership in \asso.\\
uvw\_sep & \kms & R*4 & f7.1 & Smallest possible separation from the center of the BANYAN~$\Sigma$ model in $UVW$ space.\\
xyz\_sep & \kms & R*4 & f7.1 & Smallest possible separation from the center of the BANYAN~$\Sigma$ model in $XYZ$ space.\\
ra & deg & R*8 & f21.16 & \gaia\ right ascension (J2000) at epoch 2015.5 in the ICRS reference frame.\\
dec & deg & R*8 & f21.16 & \gaia\ declination (J2000) at epoch 2015.5 in the ICRS reference frame.\\
pmra & \masyr & R*4 & f10.5 & \gaia\ proper motion in right ascension, including the $\cos\delta$ jacobian term.\\
pmdec & \masyr & R*4 & f10.5 & \gaia\ proper motion in declination.\\
epmra & \masyr & R*4 & f10.5 & Measurement error for \gaia\ proper motion in right ascension.\\
epmdec & \masyr & R*4 & f10.5 & Measurement error for \gaia\ proper motion in declination.\\
plx & pc & R*4 & f10.5 & \gaia\ parallax.\\
eplx & pc & R*4 & f10.5 & Measurement error for \gaia\ parallax.\\
ruwe & \nodata & R*4 & f7.1 & Re-normalised unit weight error of the \gaia\ astrometric solution. See Section~\ref{sec:ruwe} for more details.\\
rv & \kms & R*4 & f7.1 & Radial velocity measurement from the literature.\\
erv & \kms & R*4 & f7.1 & Measurement error for radial velocity measurement.\\
rv\_ref & \nodata & char & a25 & Reference for literature radial velocity measurement.\\
pred\_rv & \kms & R*4 & f7.1 & Predicted radial velocity that maximizes \asso\ membership probability obtained from BANYAN~$\Sigma$, only listed for targets without a radial velocity measurement.\\
epred\_rv & \kms & R*4 & f7.1 & $1\sigma$ confidence range on predicted radial velocity that maximizes \asso\ membership probability.\\
gaia\_g & mag & R*8 & f12.5 & \gaia\ $G$-band magnitude.\\
egaia\_g & mag & R*8 & f12.5 & Measurement error for \gaia\ $G$-band magnitude.\\
gaia\_grp & mag & R*4 & f12.5 & \gaia\ $G_{\rm RP}$-band magnitude.\\
egaia\_grp & mag & R*4 & f12.5 & Measurement error for \gaia\ $G_{\rm RP}$-band magnitude.\\
gaia\_brp & mag & R*4 & f12.5 & \gaia\ $G_{\rm BP}$-band magnitude.\\
egaia\_brp & mag & R*4 & f12.5 & Measurement error for \gaia\ $G_{\rm BP}$-band magnitude.\\
tmass\_j & mag & R*4 & f10.3 & 2MASS $J$-band magnitude.\\
etmass\_j & mag & R*4 & f10.3 & Measurement error for 2MASS $J$-band magnitude.\\
tmass\_h & mag & R*4 & f10.3 & 2MASS $H$-band magnitude.\\
etmass\_h & mag & R*4 & f10.3 & Measurement error for 2MASS $H$-band magnitude.\\
tmass\_k & mag & R*4 & f10.3 & 2MASS $K_S$-band magnitude.\\
etmass\_k & mag & R*4 & f10.3 & Measurement error for 2MASS $K_S$-band magnitude.\\
aw\_w1 & mag & R*4 & f10.3 & AllWISE $W1$-band magnitude, W1MPRO entry in the original catalog.\\
eaw\_w1 & mag & R*4 & f10.3 & Measurement error for AllWISE $W1$-band magnitude, W1SIGMPRO entry in the original catalog.\\
aw\_w2 & mag & R*4 & f10.3 & AllWISE $W2$-band magnitude, W2MPRO entry in the original catalog.\\
eaw\_w2 & mag & R*4 & f10.3 & Measurement error for AllWISE $W2$-band magnitude, W2SIGMPRO entry in the original catalog.\\
aw\_w3 & mag & R*4 & f10.3 & AllWISE $W3$-band magnitude, W3MPRO entry in the original catalog.\\
eaw\_w3 & mag & R*4 & f10.3 & Measurement error for AllWISE $W3$-band magnitude, W3SIGMPRO entry in the original catalog.\\
ebv & mag & R*4 & f7.1 & $E(B-V)$ reddening based on the STILISM reddening map combined with \gaia\ distance and sky position. See Section~\ref{sec:ext} for more details.\\
eebv & mag & R*4 & f7.1 & Measurement error for $E(B-V)$ reddening.\\
galex\_nuv & mag & R*4 & f7.1 & {\it GALEX} $NUV$-band magnitude.\\
egalex\_nuv & mag & R*4 & f7.1 & Measurement error for {\it GALEX} $NUV$-band magnitude.\\
galex\_fuv & mag & R*4 & f7.1 & {\it GALEX} $FUV$-band magnitude.\\
egalex\_fuv & mag & R*4 & f7.1 & Measurement error for {\it GALEX} $FUV$-band magnitude.\\
rosat\_hr1 & \nodata & R*4 & f10.3 & {\it ROSAT} hardness ratio HR1.\\
rosat\_hr2 & \nodata & R*4 & f10.3 & {\it ROSAT} hardness ratio HR2.\\
rosat\_counts & ct/s & R*4 & f10.3 & {\it ROSAT} X-ray counts.\\
erosat\_counts & ct/s & R*4 & f10.3 & Measurement error for {\it ROSAT} X-ray counts.\\
rosat\_lx & \nodata & R*4 & f10.3 & Absolute X-ray luminosity $\log L_X/L_\odot$ calculated from {\it ROSAT} X-ray data and \gaia\ trigonometric distance.\\
erosat\_lx & \nodata & R*4 & f10.3 & Measurement error for absolute X-ray luminosity.\\
li\_ew & m\AA & R*4 & f7.1 & Lithium absorption line equivalent width.\\
spt\_ref & \nodata & char & a25 & Reference for lithium absorption line equivalent width.\\
teff & K & R*4 & f7.1 & Effective temperature.\\
teff\_ref & \nodata & char & a25 & Reference for effective temperature.\\
is\_primary & \nodata & int & i3 & 1: Single stars or primary (brightest) star in a multiple system. 0: Companion star in a multiple system.\\
mult\_letter & \nodata & char & a3 & Identifier letter for multiple system components.\\
sep\_parent & asec & R*8 & f12.5 & Separation from parent star calculated from \gaia\ positions.\\
esep\_parent & asec & R*8 & f12.5 & Measurement error for separation.\\
pa\_parent & deg & R*8 & f12.5 & Position angle with respect to parent star calculated from \gaia\ positions.\\
epa\_parent & deg & R*8 & f12.5 & Measurement error for position angle.\\
comover\_gaiadr2\_id & \nodata & char & a40 & \gaia\ identification number for comoving star (parent or companion). Multiple entries are separated by a semicolon.\\
oh2017\_group\_id & \nodata & char & a4 & Comoving group identification number from \cite{2017AJ....153..257O}.\\
\enddata
\tablecomments{The full table data are available as online-only additional material.}
\end{deluxetable*}
\end{longrotatetable}
\global\pdfpageattr\expandafter{\the\pdfpageattr/Rotate 0}
\global\pdfpageattr\expandafter{\the\pdfpageattr/Rotate 0}
\global\pdfpageattr\expandafter{\the\pdfpageattr/Rotate 0}
\global\pdfpageattr\expandafter{\the\pdfpageattr/Rotate 0}

\emph{JG} wrote the codes, manuscript, generated figures and led the analysis; \emph{TJD} compiled an initial list of new candidates and generated Figure~\ref{fig:ohfig}; \emph{EEM} first identified the over-density associated with \asso, led the turnoff age analysis, the investigation of HD~27860 and provided the initial members list; \emph{AWM} led the rotation periods analysis, wrote part o f Section~\ref{sec:act} and built Figures~\ref{fig:LC} and \ref{fig:rot}, \emph{JKF} provided help with parsing the \emph{Gaia}~DR2 data and general comments, and \emph{AB} provided the atmosphere analysis of WD~0340+103 and Figure~\ref{fig:fitspec}.

\software{BANYAN~$\Sigma$ \citep{2018ApJ...856...23G}, Eleanor \citep{Eleanor}.}

\global\pdfpageattr\expandafter{\the\pdfpageattr/Rotate 0}

\clearpage
\pagebreak

\bibliographystyle{apj}
\bibliography{../ApJ_Library_AAfix,../ADS_Library,../Zenodo_Library,./mutaumann}

\end{document}